\documentclass[number,final]{elsarticle}

\usepackage{graphicx}
\usepackage{amssymb}

\def\deg{^\circ}
\def\gtorder{\mathrel{\raise.3ex\hbox{$>$}\mkern-14mu
 \lower0.6ex\hbox{$\sim$}}}
\def\ltorder{\mathrel{\raise.3ex\hbox{$<$}\mkern-14mu
 \lower0.6ex\hbox{$\sim$}}}

\journal{Progress in Particle and Nuclear Physics}

\begin{document}

\begin{frontmatter}

\title{Hard probes of short-range nucleon-nucleon correlations}

\author[anl]{J. Arrington}
\address[anl]{Physics Division, Argonne National Lab, Argonne, IL 60649}

\author[jlab]{D. W. Higinbotham}
\address[jlab]{Thomas Jefferson National Accelerator Facility, Newport News, VA 23606}

\author[glasgow]{G. Rosner}
\address[glasgow]{University of Glasgow, Glasgow, G12 8QQ, Scotland, UK}

\author[fiu]{M. Sargsian}
\address[fiu]{Florida International University, Miami, FL 33199}

\begin{abstract}

One of the primary goals of nuclear physics is providing a complete
description of the structure of atomic nuclei.  While mean-field calculations
provide detailed information on the nuclear shell structure for a wide range
of nuclei, they do not capture the complete structure of nuclei, in particular
the impact of small, dense structures in nuclei.  The strong, short-range
component of the nucleon-nucleon potential yields hard interactions between
nucleons which are close together, generating a high-momentum tail to the
nucleon momentum distribution, with momenta well in excess of the Fermi
momentum.  This high-momentum component of the nuclear wave-function is
one of the most poorly understood parts of nuclear structure.

Utilizing high-energy probes, we can isolate scattering from high-momentum
nucleons, and use these measurements to examine the structure and impact of
short-range nucleon-nucleon correlations. Over the last decade we have moved
from looking for evidence of such short-range structures to mapping out their
strength in nuclei and examining their isospin structure. This has been made
possible by high-luminosity and high-energy accelerators, coupled with an
improved understanding of the reaction mechanism issues involved in studying
these structures. We review the general issues related to short-range
correlations, survey recent experiments aimed at probing these short-range
structures, and lay out future possibilities to further these studies.

\end{abstract}

\today{}


\end{frontmatter}

\section{Introduction}
\label{sec:introduction}

Despite a fairly detailed understanding of the rich structure of the
nucleon-nucleon strong interaction, its implications for the dynamics of
atomic nuclei are not yet fully understood. Bulk properties of medium and heavy
nuclei follow from the rather general characteristics of nuclear forces such
as nucleons being fermions and nuclear forces being short-range and having a
repulsive core. Because of these characteristics, nuclei
demonstrate degeneracy of the Fermi system with clear identification of the
Fermi momentum, $k_{Fermi}$, for quantities such as momentum or kinetic energy
distributions of bound nucleons~\cite{Moniz:1971mt}. Refinements of such
approximations within the framework of Bethe-Goldstone~\cite{Bethe:1971xm} or
Nuclear Shell~\cite{Mayer:1955yy} models require, but are not very sensitive
to, the short-range properties of the nucleon--nucleon (NN) interaction.
Therefore, these models provide limited ability to investigate the
structure of nuclear matter beyond saturation density, where one
approaches the expected transition from nucleonic to quark-gluon degrees of
freedom. As such, the experimental evidence for short-range correlations
within the context of these models is rather indirect.

Simple shell model calculations make predictions for the momentum distribution
and occupation number (or spectroscopic factor)
for each nuclear shell.  Early measurements found that while such
mean-field calculations provide extremely successful descriptions of the
energy and momentum distribution of nucleons in these shells, they do not
correctly predict the occupancy of the shells.
In the early 70s, with the advent of high-quality,
medium-energy A(e,e$^\prime$p)X experiments on various nuclei, several
observations were made which suggested significant strength of the nuclear
spectral function beyond the single shell excitations.  For example, the
apparent violation of the Koltun sum rule~\cite{Koltun:1972kh} in proton
knock-out experiments~\cite{Bernheim:1974aa} was clearly attributed to the high
energy excitation part of the nuclear spectral function~\cite{Faessler:1975zz}
while no correlation effects were found in the momentum distributions
corresponding to the fixed nuclear shells~\cite{Royer:1975zz}. More
comprehensive measurements of the proton knock-out
reactions~\cite{Kelly:1996hd, Lapikas:1003zz} show that the spectroscopic
factor, basically the ratio of the observed strength within a given shell to
the expected strength, is less than one.  The observed strength is typically
$\sim$30--40\% below the shell-model expectation for measurements in many nuclei
and looking at several nuclear shells.  Even the most advanced Hartree-Fock
calculations, which include long-range correlations, significantly
overestimate the strength observed in the nuclear shells. A very plausible
explanation for this discrepancy is the presence of strong short-range NN
interactions. The repulsive core and tensor components of the NN force
yield hard interactions that can
excite nucleons outside of the low-lying shells.  Thus, the nucleons are
removed from the relatively low excitations associated with the nuclear shells
and moved to higher energies and momenta, where there is little or no strength
in the mean-field calculations.

\begin{figure}[htp]
\begin{center}
\includegraphics[width=0.9\textwidth]{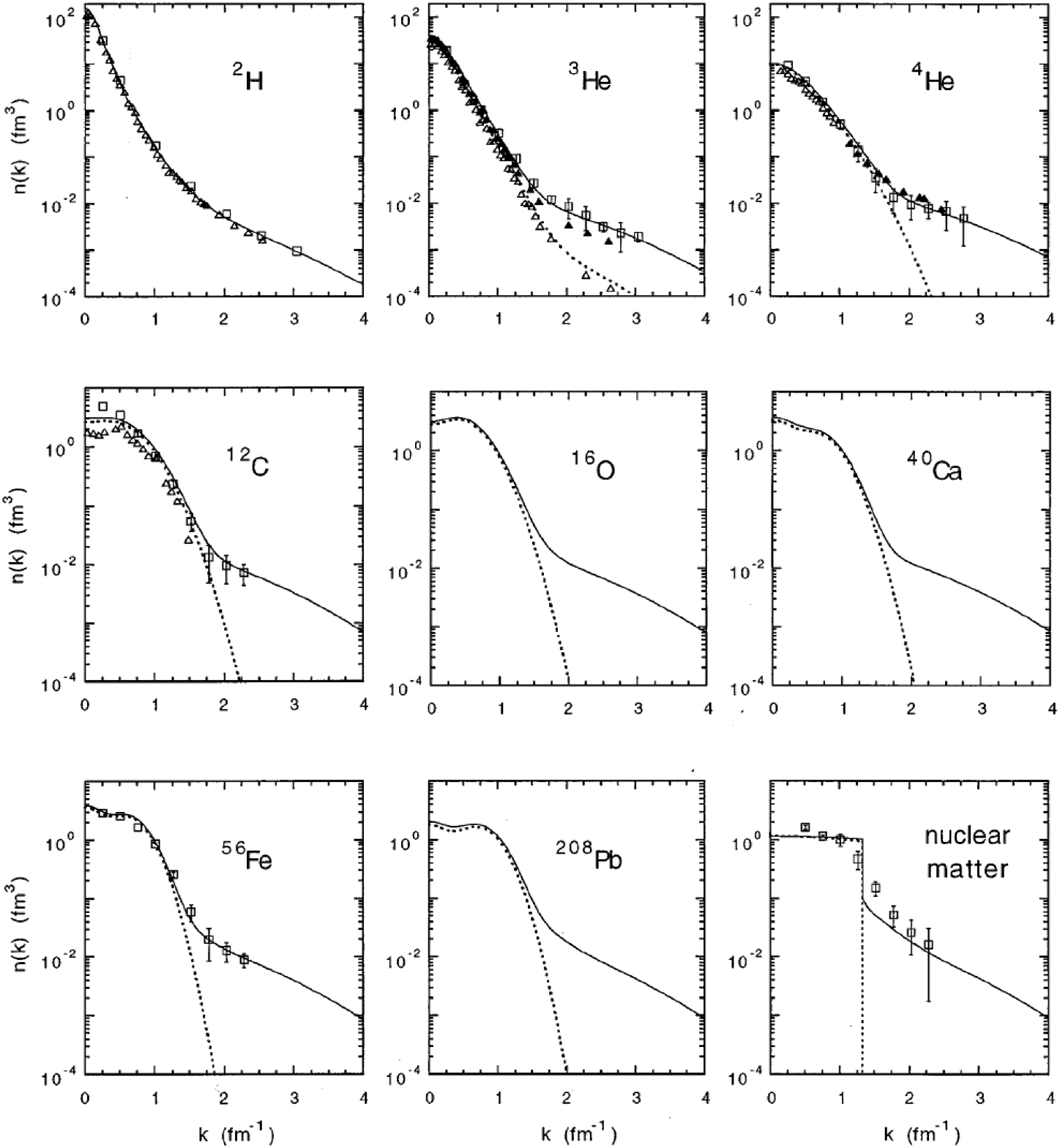}
\caption{
Nucleon momentum distributions $n(k)$ (solid lines) along with the momentum
distribution for nucleons in an average potential (dotted lines) for
various nuclei are shown. The deuteron distribution was calculated
in~\cite{CiofidegliAtti:1995qe} using the Paris potential~\cite{Lacombe:1980dr}.
Also shown are distributions for
$^3$He~\cite{degliAtti198414},
$^4$He~\cite{Schiavilla:1985gb}
$^{12}$C~\cite{Benhar1986135},
$^{16}$O~\cite{Pieper:1992gr},
$^{40}$Ca~\cite{Benhar1986135},
$^{56}$Fe~\cite{Ji:1989nr},
$^{208}$Pb~\cite{Jaminon1986445}, and
infinite nuclear matter~\cite{Fantoni1984473}.
The open squares represent the results obtained within the $y$-scaling
analysis of inclusive data from Ref.~\cite{CiofidegliAtti:1990rw} and the
open and full triangles represent the values obtained from exclusive
experiments on $^2$H~\cite{Bussiere:1981mv, TurckChieze:1984fy},
$^3$He~\cite{Jans:1982aw, Jans:1987px, Marchand:1987hd},
$^4$He~\cite{VanDenBrand:1988pv, LeGoff:1994cg} and
$^{12}$C~\cite{Frullani:1984nn}.
Figure reprinted with permission from Ref.~\cite{CiofidegliAtti:1995qe}
}
\label{all-momentum}
\end{center}
\end{figure}

To understand why the independent particle shell model can yield detailed and
precise predictions for the nucleon distributions yet fail to predict the
absolute strengths, one needs only to consider modern nucleon momentum
distributions.  At lower nucleon momenta, the distributions shown in
Fig.~\ref{all-momentum} clearly show the characteristics of a degenerate Fermi
system with broad momentum distribution, falling off rapidly at momenta
approaching $k_{Fermi}$.  In stark contrast to this behavior, the
high-momentum tail has a much less rapid falloff which is similar for all
nuclei from deuterium to nuclear matter. This universality of the high
momentum tails strongly argues against the role of the collective or
mean-field effects. Indeed, the mean-field calculations dramatically
underestimate the strength at $k> k_{Fermi}$, falling short by several orders
of magnitude at high momenta, as shown in Fig.~\ref{all-momentum}. This
universality of the high momentum tails can be easily understood if they are
generated via the short-range part of the two-nucleon potential, and are thus
independent of the shell model component of the momentum distribution.

Note that we often describe these two-body short-range correlations as
excitations in the nucleus where two nucleons undergo a hard
interaction and end up in a configuration with large relative momenta but a
small total momentum.  However, these are excitations relative to a
theoretical mean-field ground state of the nucleus are not related to any real
excited states of the nucleus; they are a contribution to the true nuclear
ground state.  When trying to probe these configurations in
high-energy reactions, the goal is to take a ``snapshot'' of the configuration
of the nucleons, and so the SRCs are often described as static configurations,
treating the nucleons as though they were simply an isolated pair of 
high-momentum nucleons, analogous to the high-momentum part of the deuteron
momentum distribution.  However, such configurations are components of the
nuclear ground wave function, with these \textit{virtual} excitations
responsible for generating most of the high-momentum nucleons.

The above scenario naturally suggests that if the two-body NN interaction is
responsible for the high-momentum tail of the nuclear momentum distribution,
then one expects the shape of the distribution beyond the Fermi momentum to be
essentially identical for all nuclei,
\begin{equation}
n_A(k) = a_2(A,Z)\cdot n_2(k)~~~~\mbox{for}~~ k > k_{Fermi}~,
\label{nk_src}
\end{equation}
neglecting the small center-of-mass momentum of the SRC.  In this case,
$n_2(k)$ represents the momentum distribution generated by the two-body
interaction and $a_2(A,Z)$ yields the relative strength in the high-momentum
tails which is related to the probability of finding these
high-momentum two-nucleon configurations, or two-nucleon short-range
correlations (2N SRCs) in the nucleus relative to the two-nucleon system.

In this scenario, the NN SRC represents a pair of nucleons where each nucleon
has a large momentum (exceeding $k_{Fermi}$) but the total momentum of the
pair is very small, i.e. a pair of nucleons with large, back-to-back momenta.
In the case of an iso-singlet SRC, one expects the
high-momentum part of the distribution to look much like the high-momentum
tails in the deuteron, which is an iso-singlet nucleon pair with zero
total momentum.

The short-range NN attraction is dominated by the tensor interaction, which
yields high momentum iso-singlet (np)$_{I=0}$ pairs but does not contribute to
the iso-triplet channel (pp, nn, np)$_{I=1}$.  Therefore, one expects the
two-body distribution to be identical to the deuteron distribution, 
$n_2(k)=n_D(k)$, and the ratio of scattering cross sections between a heavy
nucleus $A$ and the deuteron to yield $a_2(A,Z)$. The value of $a_2$ can then
be interpreted as the relative probability of finding NN SRCs in the nucleus
$A$ compared to the deuteron.

This simple picture should break down at momenta where the central repulsive
core of the NN potential dominates, as the SRCs will no longer be dominated 
by deuteron-like configurations.  Additionally, as one goes to extremely
high momenta, inclusion of three-nucleon configurations may become important.
At even higher momenta, where the
nucleon kinetic energy is comparable with the excitation energies of nucleon,
non-nucleonic degrees of freedom may also need to be taken into
account.

Extensive theoretical investigations of the impact of nucleon correlations
in the structure of nuclei have been performed using a variety of methods. 
See, for example the review of Ref.~\cite{Muther:2000qx} as well as several
more recent works~\cite{Fabrocini:1999mz, Dickhoff:2004xx, Alvioli:2005cz,
Alvioli:2007zz, Suzuki:2008cy, Vanhalst:2011es, Feldmeier:2011qy}.  Detailed
calculations employing the Green's function Monte Carlo method using various 2N
and 3N interactions have been used to study the structure and impact of
correlations in light nuclei~\cite{Forest:1996kp, Schiavilla:2006xx,
Wiringa:2008dn}.

To probe the high momentum tail experimentally one needs to deal with several
issues which follow mainly from the fact that the momentum distribution of the
nucleon is not an experimental observable.  As a result, one needs to identify
the appropriate observables which are most relevant for characterization of 
bound nucleons in the high momentum tail.

The second problem one faces is finding a probe that is able to distinguish
genuine $j$-Nucleon ($j=2, 3...$) SRCs from $j$-body processes involving long
range interactions.  The most efficient solution of this problem is to study
the SRC using a probe with large momentum transfer, $q$, and energy transfer,
$q_0$.  Most of these measurements utilize single nucleon knock-out reactions,
so one must choose kinematics which minimize contributions from more inelastic
scattering processes. The optimal approach depends on the reaction mechanism,
and the kinematic requirements are detailed in Section~\ref{sec:kinematics}.

Prior to the high-energy experiments reviewed herein, the effects of
short-range and tensor correlations were seen in valence knock-out experiments
where, as mentioned above,
the strength of the A(e,e$^\prime$p)(A-1) cross sections is
over-predicted by independent particle models by 30-40\%~\cite{Kelly:1996hd}.
Also, experiments which probed the continuum of the A(e,e$^\prime$p) reaction
found peaks in the missing momentum spectra~\cite{Marchand:1987hd,
LeGoff:1994cg}. The most straightforward interpretation of those peaks is that
a large fraction of the continuum strength arises from the breaking of an
initial-state correlated pair with large relative and small center-of-mass
momenta~\cite{Kelly:1996hd}. Such an interpretation is consistent with
calculations that include nucleon-nucleon short-range correlations such as
Muther and Dickhoff~\cite{Muther:1993yg}.

These initial indications of the importance of short-range correlations,
missing strength relative to mean-field expectations and peaks in the
missing momentum distribution, did not cleanly isolate SRCs or provide
direct information that could elucidate the structure of SRCs.  High energy
scattering experiments, aimed at isolating scattering from high-momentum
nucleons, provide a more direct way to probe SRCs in nuclei. Over the last
decade, several such high energy nucleon knock-out experiments were performed
which provided significant advances in our understanding of short-range nucleon
correlations in nuclei. In this work we review these experiments, emphasizing
their impact on our understanding the dynamics of short-range correlations.

\section{The Origin and the Features of Short-Range Correlations in Nuclei}

Already in the 1950s, it was observed the nucleons in nuclei exhibit collective
behavior in response to the absorption of specific probes such as real
photons or pions. These effects were understood on the basis of the simple
observation that a free nucleon cannot absorb a real photon below the pion
production threshold and must couple to other nuclear constituents for
photon absorption to take place. Based on the experimental fact that the real
photon has a large absorption cross section on the iso-triplet pn system
it was observed in Refs.~\cite{Levinger:1951vp,Gottfried:1958vp} that the
photo-absorption cross section will scale as $L{NZ\over A}\sigma_{\gamma d}$,
where constant $L$ was called as Levinger factor and $\sigma_{\gamma D}$
represents the photon-deuteron absorption cross section. However this
two-nucleon or ``quasi-deuteron'' regularity of the photo-absorption cross
section was a property of the nuclear response rather than its ground state
wave function. These ``quasi-deuterons'' will not be observed, for example,
in scattering by virtual photons or by real photon above the pion-production
threshold.

When we discuss short-range correlations in this work, we are referring to
properties of the ground state wave function of the nucleus which should
contribute to scattering by any probe with sufficiently high resolution power.
The universality of nuclear momentum distributions discussed in the
introduction defines the framework within which we define the short-range
nucleon correlations in nuclei. The short-range correlations are baryonic
configurations in nuclei which are defined by local properties of NN
interaction rather than by the mean field nuclear properties that are
generated by the superposition of the long-range parts of the NN potentials.

The sensitivity of SRCs to the properties of the NN interaction follows from
the singularity theorem which states that if a two-body short-range
potential behaves like $V_{NN} (k) \sim k^{-n}$, with $n > 1$ at
large relative momenta, then the nuclear wave function
of the nucleon with momentum $k^2 \gg 2M|\epsilon_B|$ is expressed through
this potential in the following form:
\begin{equation}
\Psi_{A}(k_1,k_2,k_3,...k_{A-1}) \sim {V_{NN}(k_1)\over k_2^2}
f(k_3,....k_{A-1}) ~,
\label{stheorem}
\end{equation}
where $\vec k_2 \approx -\vec k_1$ and $f(k_3,....k_{A-1})$ is a smooth
function of the momenta of spectator nucleons. Such a relation can be obtained
both from Lipman-Schwinger equation~\cite{brown1976nucleon} and its
relativistic generalizations such as the Bethe-Salpeter and Weinberg
equations~\cite{Frankfurt:1981mk,Frankfurt:2008zv} using any Yukawa or
exchange-type NN interaction. Eq.~(\ref{stheorem}) defines certain key
properties of SRCs as well as the conditions of their localization:

\begin{itemize}

\item[(a)] The momentum distribution of the nucleons in the correlated 
pair is directly related to the NN potential.  It will therefore reflect the
dynamical as well as spin/isospin structure of the NN interaction at short
distances

\item[(b)] The 2N SRCs are associated with finding two high-momentum nucleons
in a nucleus with large relative momentum and a small total momentum, i.e.
$\vec{k_1} \approx -\vec{k_2}$.

\item [(c)] Isolation of 2N SRCs in nuclei will require probing nucleon
with momentum $k> k_{Fermi}$ as well as excitation energies $\sim k^2/(2M)$
which far exceed energies relevant to the nuclear shells.

\end{itemize}

In the momentum range $300 < k < 600$~MeV/c, where $V_{NN}$ is dominated
by the tensor interaction, 2N SRCs should dominantly be in an iso-singlet state
($I=0$).  Thus, one expects that the nuclear
momentum distributions at $k> k_{Fermi}$ should be nearly identical to the
deuteron ($I=0$) distribution: $n_2(k)=n_D(k)$, and the ratio of scattering
cross sections between a heavy nucleus $A$ and the deuteron should be related
to the probability of finding NN SRCs in the nucleus $A$ relative to the
deuteron.

The dominance of the deuteron-like simple picture should break down at momenta
where the central repulsive core of the NN potential dominates. This is
expected to occur for $k > k_{max} \sim 800$~MeV/c - the characteristic
momenta at which one expects that the isospin-independent repulsive core will
dominate in the NN interaction.

\begin{figure}[ht]
\begin{center}
\includegraphics[width=0.7\textwidth]{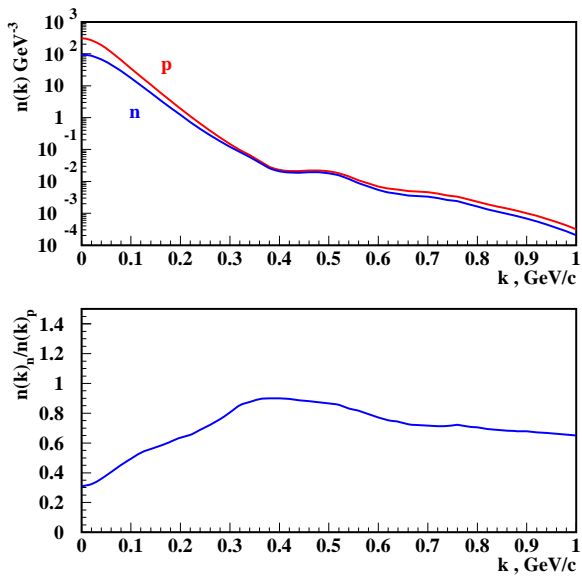}
\caption{Proton and neutron momentum distributions in $^3$He (top),
and their ratio (bottom) as a function of momentum.}
\label{he3_nk}
\end{center}
\end{figure}

The above discussed features are seen in Fig.~\ref{he3_nk}, in which momentum
distribution of the proton and neutron are compared for $^3$He nucleus.  For
$k \le 300$~MeV/c, dominated by the long range NN interaction, the average n/p
ratio is approximately equal to 0.5, which is simply the $^3$He N/Z ratio.
For $300$~MeV/c $\ltorder k \ltorder 600$~MeV/c,
the n/p ratio is close to unity, as expected if the
high-momentum distribution was generated exclusively by iso-singlet pairs.
The n/p ratio diminishes at even larger momenta as the repulsive core of the NN
central interaction begins to dominate. The calculation uses a realistic
$^3$He wave function calculated based on Argonne v18 (Av18) NN
potential~\cite{Nogga:2002qp}.  Other potentials yield nearly identical results
at low momentum but show some differences for $k > 300$~MeV/c.

Condition (b) provides a qualitative definition for NN SRCs, in that they
appear as a pair of high-momentum ($ k > k_{Fermi}$), nearly back-to-back
nucleons.  This provides an important condition for experimental
identification of 2N SRCs which should show a strong {\em angular correlation}
between the constituents of the 2N SRCs. Such correlations can be explored in
triple-coincidence experiments in which both nucleons from the SRC are
detected, as discussed in detail in section~\ref{sec:triple}.

Finally, condition (c) predicts new kind of correlation between the
excitation energy of the residual nucleus and the internal momentum of the 2N
SRCs.  As a result, nuclear excitations relevant to short-range correlations
significantly exceed the excitation energies characteristic to the nuclear
shells.

Based above discussion of the characteristics of NN SRCs the following
``definitions'' will be used throughout the text:

\begin{itemize}

\item The measurements we present use the kinematics to isolate scattering
from high-momentum nucleons based on the Plane-Wave Impulse Approximation
(PWIA).  If the value of the reconstructed initial nucleon momentum 
exceeds the characteristic Fermi momentum of the nucleus, then we infer
that the nucleon is in a short-range two-nucleon configuration.

\item The NN SRC is typically assumed to be {\em at rest}, meaning that
the two nucleons have no total momentum relative to the center-of-mass of the
nucleus.

\item The residual nucleus is considered to be ``highly excited'' when the
excitation energy of the residual (A-1) nucleus significantly exceeds the
energies characteristic to the nuclear shells.  This corresponds to the
situation where the probe interacts with a high-momentum nucleon whose
momentum is balanced by a single nucleon, as opposed to the case where the
residual system is an (A-1) nucleus where the momentum is shared more evenly
among the spectator nucleons.

\end{itemize}

While these are the basic assumptions used in high-energy studies of SRCs, we
can test or go beyond these approximation in the analysis of such experiments.
In the sections summarizing recent experimental results, we will typically
begin with these simplifying approximations and then emphasize the ways in
which we can go beyond this simplest picture of SRCs.

\section{Main Approaches of Probing Short-Range Correlations in Nuclei}
\label{sec:approaches}

This section details the desired requirements for a complete and detailed
investigation of SRCs in both light and heavy nuclei.  Some of these are
universal, applying to all hard probes, while others are reaction dependent.

\begin{itemize}

\item[I.] {\em Instantaneous interaction involving a nucleon from the SRC:}
One should have a clear way of identifying events in which the high energy
probe interacts with the correlated nucleon and instantaneously removes it
from the SRC. The requirement that the process is instantaneous is necessary
to be able to reliably infer information on the spectral functions or momentum
distributions from the measured observables.

\item[II.] {\em Struck and spectator nucleons from SRC are distinguishable:}
The ability to distinguish which nucleon is knocked out by the virtual photon
and which one is a spectator from the SRC is essential for mapping out the
dynamical mechanism of pre-existing SRCs in the ground state wave function of
the nucleus. For example, the measurement of the momentum and energy
correlations between two detected nucleons will allow for the extraction of
relative and center-of-mass momentum distribution of 2N SRCs, provided that
one clearly can identify the spectator and struck nucleons in the reaction.

\item[III.] {\em Separation of two- and three-nucleon SRCs:} SRC studies
should allow unambiguous separation of signatures for two- and three- nucleon
correlations. This will require specific reactions and kinematic conditions
which enhance either two-, three- or many-nucleon correlations.

\item[IV.] {\em Long-range two-nucleon effects are suppressed:}
Because the momentum distribution is dominated by the low-momentum components,
long-range two-step processes such as meson exchange currents (MECs) and
intermediate state resonance contributions will significantly alter the SRC
picture unless they are suppressed by kinematics or other dynamical
conditions. Typically, this means operating at higher $Q^2$ values, to
suppress final state interactions and meson exchange contributions, and at
relatively low energy transfer or missing energy, to suppress inelastic or
intermediate state resonance contributions.

\item[V.] {\em Final state interaction effects are small or well understood:}
The major problem in SRC studies is the final state interaction (FSI) effects
between particles in the final state. Hadronic FSIs do not diminish with an
increase of the transferred momentum and energy of the reaction, but do enter
into the eikonal regime. The eikonal regime (or ``regime of geometrical
optics'') has several important characteristics which follow from the high
energy and small angle scattering nature of the FSIs. One key consequence is
the approximate conservation of the $p_{-}=E-p_z$ components of the
four-momenta of scattered particles~\cite{Sargsian:2001ax}. Because of this,
one still can extract information about the high momentum component of the
nuclear wave function, even in the presence of large FSI.

Another important property of the eikonal regime of FSIs comes from the fact
that rescattering is localized in the direction almost transverse to the
direction of the fast moving particle and that the distance the fast particle
travels before rescattering decreases with increasing virtuality of the
particle. As a result, deeply bound nucleons knocked out from the SRC
propagate a short distance, comparable to the size of the SRC, before
rescattering. This implies that the bulk of the rescattering takes place 
between the nucleons in the SRC~\cite{Frankfurt:2008zv}, and thus a very large
component of the FSI is also dependent only on the structure of the two-body
system, and therefore independent of the nucleus in which the SRC appears.
This plays an critical role in preserving important kinematical correlations
between struck and spectator nucleons, even when FSI effects are sizable.

\item[VI.] {\em Improved understanding of relativistic and bound-nucleon
effects:} Nucleons in SRCs are characterized by large momenta and binding
energies. This increases significantly the effects due to the relativistic
motion of the nucleon as well as possible modification of the nucleon structure
due to the large binding effects. The comprehensive exploration of the
dynamics of SRCs requires an understanding of the reaction dynamics and
nucleon modification effects of deeply bound nucleon in the short-range
correlations. Thus the program of SRC studies should proceed in parallel with
studies of reaction dynamics and nuclear medium modification effects.

\item[VII.] {\em Detailed studies of few-body nuclei:}
Due to a certain degree of factorization between the short and long range
interactions in nuclei (similar to Eq.~(\ref{nk_src})), the above discussed
reaction dynamics, relativistic and binding effects can be studied using few
nucleon systems: $^2$H, $^3$He and $^3$H. Therefore, systematic
studies of the lightest nuclei should be an integral part of the program
to understand SRCs.

\item[VIII.] {\em Comparisons of different SRC observables and reactions:}
SRCs can be studied in electroproduction reactions with different degrees of
complexity, including inclusive A(e,e$^\prime$), semi-inclusive A(e,e$^\prime$N), and
triple-coincidence A(e,e$^\prime$NN) reactions for both polarized and unpolarized
processes. In each case, one deals with specific observables relevant to the
SRC dynamics. The assumed universality of the short-range interaction implies
that there should be clear consistency and relationship among all
observables measured in these reactions. This helps to demonstrate a reliable
separation of initial-state SRC structure and reaction-dependent corrections
related to meson exchange and resonance contributions in the scattering.

\end{itemize}

\section{SRC Kinematics}
\label{sec:kinematics}

\subsection{General kinematical considerations for SRC Studies}
\label{genkins}

The above conditions will drive the restrictions on momentum and energy
transfers in reactions aimed at probing SRCs. While the exact kinematic
requirements depend on the reaction, one can make some general estimates for
the different classes of reactions.  Condition (I), the desire to have a
nucleon from the SRC removed instantaneously, can be achieved if the energy
and momentum transfer scales are much larger than the excitation energy scale
characteristic to the nuclei in general and the SRC in particular.  This can
be achieved by requiring~\cite{Frankfurt:2008zv, Frankfurt:1981mk}
\begin{equation}
q_0 \gg V_{NN}, ~~~~ q \gg m_N/c ~,
\label{vnn}
\end{equation}
where $V_{NN}$ is the characteristic potential of the NN interaction. In this
case the residual system after the nucleon is removed from the SRC can be
considered intact at the time of the interaction and therefore a spectral
function can be introduced for the reaction which will reflect the direct
properties of SRCs.

Condition (IV), the suppression of long-range two-body interactions is
generally achieved by requiring
\begin{equation}
Q^2 \gg m^2_{meson} ~.
\label{virt}
\end{equation}
In some cases, such as for reactions close to the deuteron threshold or
in elastic scattering~\cite{VanOrden:1995eg}, exchange currents are expected to
be important even at large $Q^2$ values. In principle, they are also important
in hard break-up reactions, although they are more naturally described by
the quark-exchange diagrams~\cite{Brodsky:2003ip} due to large masses produced
in the intermediate states.  However, it is rather well established that for
the reactions we will discuss in the context of SRC studies, meson exchange
currents significantly diminish for $Q^2 \gg m_{meson}^2$.
Above $Q^2=1$~GeV$^2$, long-range meson exchange current contributions are
suppressed with respect to the direct production from SRC by an additional
factor of $Q^{-4}$~\cite{Sargsian:2001ax, Sargsian:2002wc}, and are
already a small correction in the kinematics relevant to SRC studies
at $Q^2 \ge 1$~GeV$^2$~\cite{Sargsian:2001ax, Sargsian:2002wc}.  

Contributions from intermediate state resonances are still important at
these $Q^2$ values, especially contributions from $\Delta$-isobars which, due
to the large magnetic transition form-factor, have sizable contribution in
electroproduction reactions close to the pion threshold for $Q^2 \le
4$~GeV$^2$~\cite{Stoler:1993yk,Ungaro:2006df}.  For moderate momentum
transfers (1~GeV$^2 \le Q^2 \le 4$~GeV$^2$), where the $\gamma N\rightarrow
\Delta$ transition is comparable with $\gamma N\rightarrow N$, a suppression
of these contributions can be achieved by working at low energy transfer, thus
keeping away from the inelastic threshold.  A convenient parameter for
identifying the closeness to the threshold is the Bjorken variable $x=
{Q^2\over 2 m_N q_0}$, where we use the nucleon mass in the denominator, even
though we are scattering from a heavier target. This leads to a kinematically
allowed region of $0<x<M_A/M_N$, rather than $0<x_A<1$, where $x_A$ is defined
using the mass of the nucleus. Defined this way, quasi-elastic scattering from
low-momentum nucleons corresponds to $x\approx 1$ for all nuclei, while
inelastic processes (with larger values of $q_0$) occur at smaller values of
$x$.  Choosing to probe high-momentum nucleons on the $x>1$ side of the
quasi-elastic peak yields a significant suppression of IC contributions.

\begin{figure}[htb]
\begin{center}
\includegraphics[width=0.95\textwidth,height=0.5\textwidth]{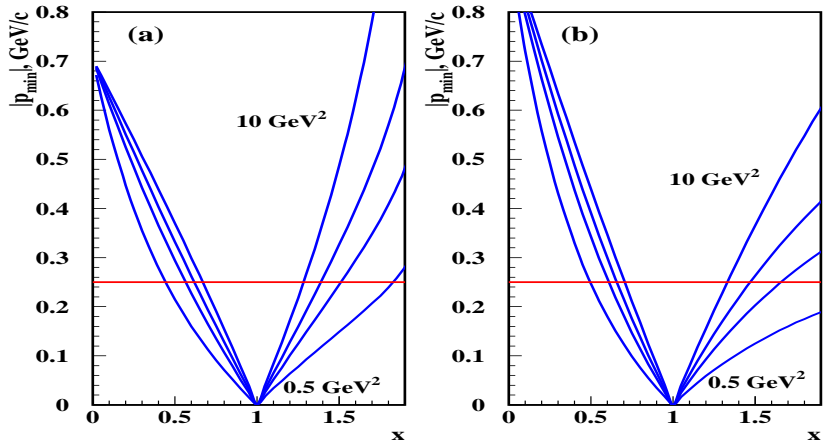}
\caption{The minimum momentum for scattering from a nucleon in deuterium (left)
and gold (right) as a function of $x$ and $Q^2$ for quasi-elastic $\gamma + 2N
\rightarrow N + N$ scattering for $Q^2$ values of 0.5, 1.5, 3, and 10 GeV$^2$.
For heavy nuclei, the minimum momentum for a given $x$ and $Q^2$ value is
somewhat smaller, as the heavier recoil system requires less kinetic energy to
balance the momentum of the struck nucleon.  This, combined with the larger
Fermi momentum for heavy nuclei, means that slightly higher $x$ or $Q^2$
values are required to fully suppress scattering from nucleons associated with
the mean-field structure.
Figure adapted from Ref.~\cite{Frankfurt:2008zv}
}
\label{fig:xpmin2nsrc}
\end{center}
\end{figure}

Figure~\ref{fig:xpmin2nsrc} demonstrates how requiring $x > 1$ and $Q^2$ above
1--2~GeV$^2$ allows one to select large internal momenta associated with
the two-nucleon correlations.  For $p_{min} > k_{Fermi}$, there is almost
no contribution from the mean-field part of the nucleon momentum distribution
in the PWIA, allowing for isolation of high-momentum nucleons associated with
SRCs.  Note that $p_{min}$ corresponds to an initial nucleon with momentum
parallel to the momentum transfer and assumes no excitation of the
residual (A-1) nucleus.  Accounting for transverse momentum and excitation of
the spectator nucleus, one expects that the true momenta probed will be
well above $p_{min}$.  The requirements that the scattering selects
high-momentum nucleons and suppresses MEC and IC contributions, together with
Eq.~(\ref{vnn}), set the necessary kinematic conditions: an energy transfer of
at least 0.5~GeV, and $Q^2$ of 1--2~GeV$^2$ or higher, while working at
$x\gtorder1$ to minimize intermediate resonance contributions.

Condition (III) requires the ability to separate two-nucleon and three-nucleon
SRCs.  This can be achieved in a rigorous way by defining the parameter
$\alpha_i$:
\begin{equation}
\alpha_i = A{E_i - p_{i,z}\over E_A - p_{A,z}} =
           A{E_i^{lab} - p_{i,z}^{lab}\over m_A} ~,
\label{alpha}
\end{equation}
where $p_A$ is the four-momentum of the nucleus $A$ and $p_i \equiv (E_i,\vec
p_i)$ is the initial four-momentum of the struck nucleon.  Note that $m_i^2
\equiv p_i^2 =  E_i^2 - \vec p_i^2 \ne m_N^2$, which reflects the
off-shellness of the nucleon. The parameter $\alpha_i$ is invariant under
boosts in the $z$ direction~\cite{Kogut:1969xa, Bjorken:1970ah,Feynman:1973xc}
and in the infinite momentum frame it describes the momentum fraction of the
nucleus carried by the interacting nucleon. The parameter $\alpha_i$ is
analogous to Bjorken-$x$ which describes the fractional momentum of the
quark constituents in deep inelastic scattering. The definition of $\alpha_i$
is such that on average, each quark carries a momentum fraction $\alpha_i
\approx 1$.  To have $\alpha_i > 1$ requires the sharing of the momentum of
at least two nucleons, while $\alpha_i > 2$ must involve at least three
nucleons.  This property can be used to separate scattering from different
SRCs, since for
\begin{equation}
j-1 < \alpha_i < j ~,
\label{SRCselec}
\end{equation}
the reaction will involve at least j nucleons, allowing one to distinguish
interactions with $j=2, 3...$ nucleons. Another important property of
$\alpha_i$ is that in the eikonal regime, discussed as part of condition (V),
this quantity is nearly unaffected by final state interactions. Therefore, the
experimentally extracted $\alpha_i$ distribution function is Lorentz boost
invariant largely insensitive to FSI.

We can further elaborate the properties of $\alpha_i$ by applying
energy-momentum conservation for the nucleon knocked out from the nucleus.
Using the kinematic relation for the produced final mass in $\gamma^*N$
interaction, $W_N^2 = (p_i+q)^2$, one obtains
\begin{equation}
\alpha_i = x\left(1 + {2p_{i,z}\over q_0 + |q|}\right) +
{W_N^2 - m_i^2\over 2m_N q_0} ~,
\label{incl_kin}
\end{equation}
where $p_{i,z}$ is the rest-frame longitudinal momentum of the initial nucleon
along the direction of the transferred momentum $q$. For quasi-elastic
scattering, $W_N = m_N$, yielding a simplified expression for $\alpha_i$. 
One can see from the above equation that for any given values of nucleon
initial momentum and $Q^2$ one can find solution for $x>1$ such that
$\alpha_i$ will satisfy the SRC selection rule of Eq.~\ref{SRCselec}.

This is very important for at least two reasons.  First, the
possibility of probing at $x>1$ kinematics will significantly suppress the 
$\Delta$-isobar contribution and secondly at sufficiently large $Q^2$ when
according to Eq.~(\ref{incl_kin}), $\alpha_i \approx x$, and so the $x>j-1$
kinematics will allow us to select $j$-nucleon SRCs in the quasi elastic
region. The latter is significant for SRC studies in inclusive reactions in
which only Bjorken-$x$ can be reconstructed (see Sec. 4.2).

The above considerations are largely independent of the specific reaction being
used to probe SRCs. We now discuss in more detail the requirements, separating
the further discussion of the ideal conditions for sensitivity to SRCs
according to the complexity of the electro-nuclear reactions used to probe the
SRCs. We discuss three main types of the reactions: {\em inclusive
A(e,e$^\prime$)X}, {\em semi-inclusive A(e,e$^\prime$N)X} and {\em
triple-coincidence A(e,e$^\prime$,NN)X} reactions. For all three cases we
consider only the quasi-elastic scattering in which case one nucleon is
knocked out from the SRC in the elastic $\gamma^*N$ interaction and the
residual nuclear system consists only of nucleons.

\subsection{Kinematics of Inclusive Reactions}
\label{kinematics:inclusive}

Because of the requirements discussed in the previous section, the cross
sections involved in quasi-elastic scattering studies of SRCs are suppressed by
both the relatively large $Q^2$ values required and the fact that one is
sampling from the high momentum tail of the nucleon momentum distributions. 
Because of this, the first attempts to isolate SRC contributions in high-$Q^2$
reactions, and the only attempts to isolate 3N SRCs to date, involved
inclusive measurements where only the scattered electron is detected. This
maximizes the cross section while still allowing for isolation of the SRC
contributions through selection of appropriate kinematics.

Note that $\alpha_i$ (Eq.~(\ref{alpha})) depends on the initial nucleon
momentum, $p_{i,z}$, which cannot be fully reconstructed in inclusive
scattering. However, it follows from Eq.~(\ref{incl_kin}) that $x \to
\alpha_i$ in the limit where $q_0$ and $|q|$ are much larger than $|p_{i,z}|$
and $W^2_N-m_i^2$. Since $x$ is defined only by the kinematics of the initial
and scattered electron, one can use $x$ in place of $\alpha_i$ to isolate
scattering from high-momentum nucleons. Because of the kinematic condition
that scattering at $x>j-1$ requires that at least $j$ nucleons be involved,
one can attempt to isolate 2N correlations by requiring $x$ to be sufficiently
larger than one, and 3N SRCs when sufficiently above $x=2$.

Extremely high $Q^2$ values are not required for this to be an effective
way of separating 2N and 3N configurations.  The kinematic limit for
scattering from a proton at rest is $x=1$ for all $Q^2$ values and scattering
at $x>1$ must involve at least {\em two}  nucleons. For $x$ slightly above
unity, the ``multi-nucleon'' contributions are dominantly quasi-elastic
scattering from a single nucleon which has a non-zero momentum, $k <
k_{Fermi}$, due to its interaction predominantly in the mean field of the
residual nucleus. The threshold in $x$ for which one is sensitive only to
nucleons with $k>k_{Fermi}$ depends on $Q^2$ (see Fig.~\ref{fig:xpmin2nsrc}).
Similarly, for $x > m_D/m_N \approx 2$, at least {\em three} nucleons must be
involved in the reaction, and configurations where the three nucleons all have
large momenta are expected to becoming dominant at some value of $x$ well
above $x=2$, the limit for contributions from an at-rest 2N SRC.

Based on Fig~\ref{fig:xpmin2nsrc}, one would expect that inclusive
A(e,e$^\prime$) reactions measured at $Q^2\gg m_N^2$ and $x \gtorder 1.4$
should allow us to probe the nucleon in 2N and higher order correlations.  At
lower $Q^2$ values, where the above approximations are not as good, 
the onset of 2N SRC dominance will require somewhat larger values of $x$.
While $\alpha_i$ cannot be reconstructed in inclusive scattering, one can
estimate $p_{i,z}$ from the inclusive kinematics under the assumption that one
is interacting with one of the nucleons in an at-rest 2N SRC. Making this
assumption, one obtains
\begin{equation}
\alpha_{2N} = 2 - \frac{q_- + 2m_N}{2m_N}
\Bigg( 1+\frac{\sqrt{W^2-m_N^2}}{W^2} \Bigg) ~,
\label{eq:alpha2n}
\end{equation}
where $W^2 = 4m_N^2 + 4m_N \nu - Q^2$. The quantity $\alpha_{2N}$, 
sometimes called $\alpha_{tn}$, is equal to $\alpha_i$ to the extent
that the two-body approximation is valid and the transverse momentum of the
initial nucleon can be neglected. This allows for a comparison of the
onset of 2N SRC dominance as a function of $x$ and $\alpha_{2N}$, where the
onset in $\alpha_{2N}$ should be $Q^2$-independent if the 2N SRC picture is
correct. For multi-nucleon SRCs, there is not a unique prescription for
determining $p_{i,z}$, as it depends on the distribution of relative momentum
of the nucleons in the SRC. A 3N SRC in a symmetric configuration, where all
three nucleons have similar momentum, will provide a different spectator
system than a linear configuration where the momentum of one nucleon is
balanced by two nucleons splitting the momentum in the opposite direction.
Thus, examination of the onset of scaling for $x>2$ using different models of
the microscopic structure of the 3N SRCs may allow for an evaluation of these
models, providing insight into the structure of these multi-nucleon
correlations.

\subsection{Final-State Interactions in Inclusive Reactions}
\label{sec:fsi}

While the cross section for hadronic interaction of the struck nucleon with 
the rest of the nucleus can be large, the inclusive scattering cross section
factorizes from later hadronic interactions, and only specific FSI
contributions are expected to impact the inclusive cross sections at $x>1$
and high $Q^2$.  Here we present a brief discussion of the current state of
understanding of FSI in this kinematical region.

It is well established that there are large FSI contributions at low $Q^2$
values~\cite{Day:1987az,CiofidegliAtti:1990rw} which provide a clear breakdown
of the scaling picture for QE scattering in the PWIA (discussed in
Sec.~\ref{sec:deuteron}). It was generally argued that for inclusive
scattering, FSI contributions would vanish when the spatial resolution of the
probe is small and the ejected nucleon has a large momentum. This was a
consequence of the factorization of the initial interaction of the probe with
the nucleon from the later rescattering of the struck nucleon, as discussed in
more detail in Refs.~\cite{Frankfurt:1988nt, CiofidegliAtti:1990rw}.
Additional studies provided more detailed predictions that FSI would be small
at sufficiently large $Q^2$~\cite{Day:1990mf, Frankfurt:1993sp}, although
there were still questions, e.g. the possible further suppression of FSI for
high-momentum nucleons because of the impact of the nucleon virtuality in the
rescattering~\cite{Frankfurt:1993sp, Uchiyama:1989vr, Benhar:1995te}. Some
more recent calculations~\cite{CiofidegliAtti:2009qc, Mezzetti:2009ch} show
FSI contributions that become very small by $Q^2$=3--4~GeV$^2$, suggesting
that the more recent inclusive data~\cite{Arrington:1998ps, Fomin:2011ng}
reach sufficient $Q^2$ values that these effects do not interfere with the
extraction of SRCs from high-$Q^2$ inclusive measurements. In addition, these
high-$Q^2$ inclusive measurements show no indication of $Q^2$ dependence, as
presented in Secs.~\ref{sec:deuteron} and~\ref{sec:inclusive}, suggesting small
contributions from FSI.

While these arguments were used as justification for analysis of inclusive
measurements in the PWIA, there are still some questions as to the
applicability of the impulse approximation. First, this relies on the
electron being insensitive to interactions of the struck nucleon once it is
``far'' from the location of the interaction (far in terms of the wavelength
of the probe and the final-state proton). However, in scattering from a
pre-existing SRC, one is selecting cases where the struck nucleon is very
close to another nucleon, and thus FSIs may not vanish, but may be localized
within the SRCs~\cite{Day:1990mf, CiofidegliAtti:1994ys, Frankfurt:2008zv}. 
In this case, the FSI within the SRC may decrease slowly with $Q^2$, making it
difficult to verify the lack of FSI contributions by examining the onset of
scaling from the $Q^2$ dependence of the data.  However, if the FSI are
contained within the SRC, and the SRCs in nuclei have the same deuteron-like
structure as the high-momentum tails in the deuteron, then the ratio of
nuclear cross sections should still yield a relative measure of the presence
of high-momentum nucleons in nuclei.

In addition, some calculations of FSI were not consistent with the prediction
that the effects should become small (or localized within the SRC) at large
$Q^2$.  Calculations by Gurvitz and Rinat~\cite{Gurvitz:1986dn, Gurvitz:1988ft,
Gurvitz:2001qm} and Correlated Glauber Approximation (CGA) calculations by
Benhar and collaborators~\cite{Benhar:1991af, Benhar:1993ja, Benhar:1995xa,
Benhar:1995ph} indicated significant FSI even at relatively large $Q^2$
values.  For some of the earlier calculations, questions were raised about the
consistency of the real and imaginary potentials used~\cite{Day:1990mf} and
the sensitivity of the calculations to the inclusion of the N--N correlation
function\footnote{The correlation function encodes the fact that the
short-range repulsive core of the NN potential significantly suppresses the
probability for another nucleon to be found very close to the struck
nucleon~\cite{Benhar:1993ja}}.  The calculations of Benhar, \textit{et al.},
include significant color transparency effects which reduce the FSI
contributions at large $Q^2$, but such large color transparency effects appear
to be ruled out in A(e,e$^\prime$p) experiments~\cite{Makins:1994mm,
O'Neill:1994mg, Abbott:1997bc, Garrow:2001di} at the range of $Q^2$ considered
in Ref.~\cite{Benhar:1991af}. It would be beneficial to have updated versions
of the CGA calculations, including both the A and $Q^2$ dependence
of FSI effects. This would allow for evaluation of the results against
existing measurements of absolute cross sections and ratios for a variety of
nuclei and a significant range of $x$ and $Q^2$ values~\cite{Day:1993md,
Arrington:1998ps, Fomin:2011ng}.

It has been argued~\cite{Frankfurt:2008zv, Frankfurt:2009vv} that the CGA
calculations yield an overestimate of the inclusive final-state interactions
because they do not include inelastic diagrams needed to restore unitarity. 
Details of this argument are presented in Sec.~\ref{sec:fsi2}. However, because
the quantitative effect of these diagrams has not been fully evaluated, it is
not yet clear whether the issue of unitarity provides an explanation for the
large FSI results in the CGA calculations.  Therefore, it is important to
remember that the inclusive measurements are typically analyzed under the
assumption that FSI are small or at least cancel in the ratio, as discussed in
Sec.~\ref{sec:inclusive}, and this remains a critical assumption in the
examination of the inclusive results.

\subsubsection{Inelastic Processes and their Impact on Final-State Interactions}
\label{sec:fsi2}

\begin{figure}[htb]
\begin{center}
\includegraphics[width=0.7\textwidth,height=0.45\textwidth]{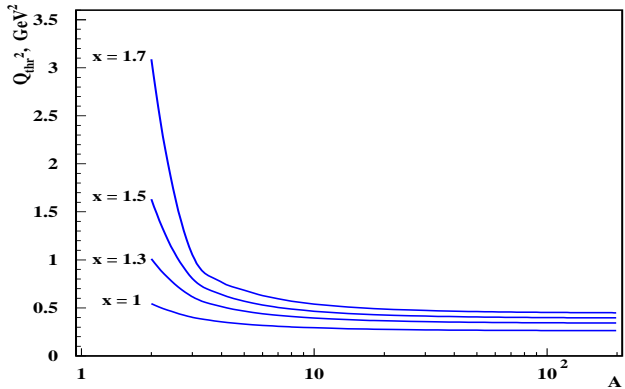}
\caption{The dependence of the threshold value of $Q^2$ for inelastic
processes in inclusive $A(e,e')X$ reactions to the mass number of nuclei,
${A}$ at different values of Bjorken $x$.}
\label{q2_thr}
\end{center}
\end{figure}

In inclusive scattering, where only the scattered electron is detected, one
cannot kinematically ensure pure quasi-elastic scattering. Indeed the
threshold of $Q^2$ at which inelastic processes can happen is defined from the
condition
\begin{equation}
(q + M_A)^2 \ge (M_A+ m_\pi)^2 ~,
\label{th1}
\end{equation}
where $q$ is the four-momentum of the virtual photon $M_A$ and $m_\pi$ are
masses of the target nucleus and pion. From Eq.~(\ref{th1}) one obtains the
threshold value of $Q^2$ for inelastic processes as:
\begin{equation}
Q^2 \ge Q^2_{thr} = {2M_Am_\pi + m_\pi^2\over {M_A\over m_N x} - 1}
 \label{thr2}
\end{equation}
The inelastic threshold is shown in Figure~\ref{q2_thr}.  For $Q^2 \gtrsim
2$~GeV$^2$, the inelastic processes is relevant for all nuclei in the
range $1.3<x<1.5$ where one expects to be able to study contributions from SRCs.
 
\begin{figure}[htb]
\begin{center}
\includegraphics[width=0.95\textwidth]{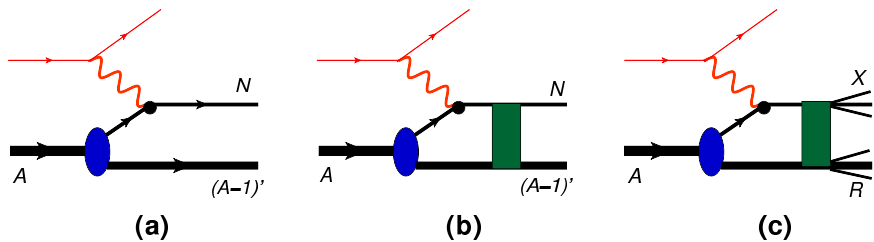}
\caption{Diagrams contributing to inclusive $A(e,e')X$ reactions: (a) plane
wave impulse approximation, (b) elastic and (c) inelastic final state
interaction contributions.}
\label{pwia_fsi}
\end{center}
\end{figure}

In this case, one must include the inelastic scattering diagrams in the form
of Fig.~\ref{pwia_fsi}(c)\footnote{Note that in this case inelastic
contributions from the PWIA diagrams are strongly suppressed due to large
virtuality involved in the inelastic $\gamma N$ vertex at $x>1$.} in addition
to the quasi-elastic scattering diagrams (Fig.~\ref{pwia_fsi}(a) and
\ref{pwia_fsi}(b)). Models in which only quasi elastic
(Fig.~\ref{pwia_fsi}(b)) channels are included in the FSI diagram may
become increasingly inaccurate at large $Q^2$.

The direct calculation of the diagram of Fig.~\ref{pwia_fsi}({c}) is
practically impossible, due to an increasing number of inelastic channels
involved in the scattering.  The high energy approach to this problem is based
on the fact that at sufficiently high energies, in which eikonal approximation
is valid, the FSI can be expressed through the sequential diffractive elastic
and inelastic rescatterings of nucleons in the nuclei, for which one can apply
the optical theorem in the form $ Im\{f^{el}_{NN}(t=0)\} = \sigma^{tot}_{NN}$
with $f^{el}_{NN}(t) = \sigma^{tot}(i + \alpha)e^{{B\over 2} t}$. Such an
approach within Regge theory of diffractive scattering for inclusive processes
was discussed in Ref.~\cite{Abramovsky:1973fm}, which resulted to well known
Abramovsky-Kanchelly-Gribov~(AGK) cutting rules. Similar rules have been
discussed within eikonal approximation by Bertocchi and
Treleani~\cite{Bertocchi:1976bq} for inclusive hadron-nucleus scattering. They
demonstrated that including only elastic rescattering amplitudes (Glauber
theory) in the inclusive scattering violates the unitarity condition for the
nuclear scattering amplitude, which is restored with the inclusion of the
inelastic rescatterings. The main essence of these cutting rules is that,
because of the unitarity relations between inelastic and elastic NN
scattering amplitudes, it is necessary to account for the cancellations
between rescattering amplitudes.

\begin{figure}[htb]
\begin{center}
\includegraphics[width=0.95\textwidth]{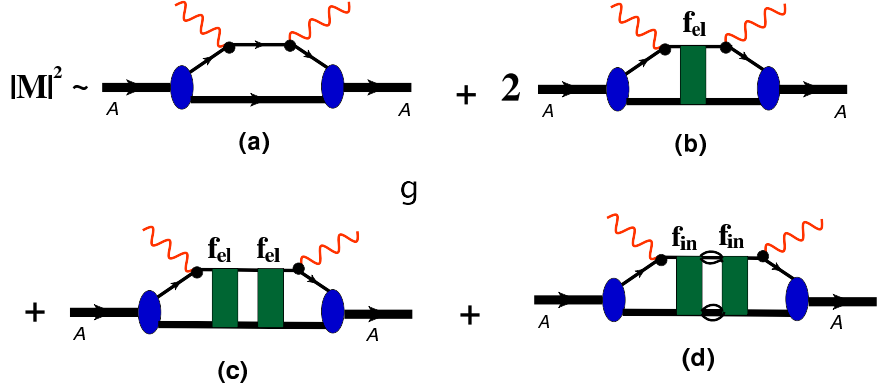}
\caption{The main terms contributing to the cross section of inclusive
$A(e,e')X$ cross section in quasi-elastic kinematics.}
\label{pwia_fsi_square}
\end{center}
\end{figure}

\begin{figure}[htb]
\begin{center}
\includegraphics[width=0.95\textwidth]{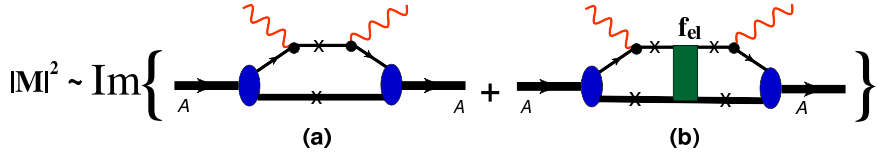}
\caption{The imaginary part of the forward nuclear Compton scattering
amplitude.}
\label{fcompton_im}
\end{center}
\end{figure}

The qualitative aspects of the application of AGK type cutting rules can be
seen from Fig.~\ref{pwia_fsi_square}, in which the inclusive cross section is
defined by the sum of the four terms, where Fig.~\ref{pwia_fsi_square}(b)
represents the interference between PWIA and elastic rescattering amplitude
and in the eikonal regime is predominantly destructive. The main effect of the
application of AGK rules in the sum of the terms in Fig.~\ref{pwia_fsi_square}
is that the sum of the squares of elastic (c) and inelastic (d) terms cancels
half of the interference term (b) resulting to the net contribution of
diagrams as given in Fig.~\ref{fcompton_im}. The latter is related to the
imaginary part of the forward nuclear Compton scattering~\cite{CS}.

The above discussion illustrates the potential importance of maintaining
unitarity through the inclusion of inelastic rescattering diagrams and also
shows that inclusion of only the elastic rescatterings in the FSI (Glauber
theory) can overestimate the final state interaction contribution. This is
seen in Ref.~\cite{Benhar:1991af} where authors used the correlated Glauber
approximation to calculate the cross section of inclusive A(e,e$^\prime$)X
scattering. While they obtained reasonably good description of the data at
small $Q^2$, their FSI calculation significantly overestimated the data at
large $Q^2$ (where according to Fig.~\ref{q2_thr} inelastic processes become
important). In Ref.~\cite{Benhar:1991af} the agreement with the data is
achieved only after the inclusion of a large color transparency effect.

As mentioned above, the FSI in the inclusive A(e,e')X reaction can be
calculated through the imaginary part of the forward nuclear virtual Compton
scattering amplitude (Fig.~\ref{fcompton_im}). Ref.~\cite{Frankfurt:2008zv}
used these amplitudes to analyze the space time properties of the final state
interaction at $x>1$. This analysis (Sec.~3.4 of Ref.~\cite{Frankfurt:2008zv})
demonstrated that the distance the knocked-out nucleon propagates before the
rescattering in the amplitude of Fig.~\ref{fcompton_im}(b) is inversely
proportional to the initial longitudinal component of the nucleon $|p_{iz}|
\sim |p_{min}|$, where the dependence of $p_{min}$ on $x$ and $Q^2$ is given in
Fig.~\ref{fig:xpmin2nsrc}. This suggests that already at $300$~MeV/c, the
reinteraction distances are $\lesssim$1~fm, indicating that at least the
first rescattering will take place within the 2N SRC.  So while FSI between
nucleons in the SRC may survive at high $Q^2$, these very short-distance
interactions within the SRC should be identical for scattering from a deuteron
or from a deuteron-like pair within a heavy nucleus. This
situation naturally explains the persistence of the scaling in the ratios of
inclusive cross sections measured for nuclei A and the deuteron or $^3$He.
However, a full, quantitative evaluation of FSI has not yet been performed,
although numerical estimates are in progress~\cite{CS}.

\subsection{Kinematics of semi-inclusive A(e,e$^\prime$N) reactions}
\label{kinematics:coincidence}

We now examine the semi-inclusive A(e,e$^\prime$N) reaction with kinematics
chosen such that the struck nucleon, with momentum $\vec{p}_N$, is
detected in coincidence with the scattered electron.  The goal is to isolate
the single nucleon knock-out channel where only the (A-1) spectator nucleons
are undetected and the initial kinematics of the struck nucleon can be
reconstructed in the PWIA.  Clean identification of the nucleon as being struck
requires a large momentum transfer such that that the final nucleon momentum
is much larger than the momentum of any of the spectator nucleons in the
residual nucleus.  This is achieved by requiring
\begin{equation}
\vec{p}_N = \vec{p}_i + \vec{q} \gg k_{Fermi} ~.
\label{eep_kin0}
\end{equation}
where $\vec{p}_i$ will represent the initial momentum of the struck nucleon
within the PWIA picture of the scattering.

For further discussion we start with the "standard"  definition of the
kinematics of knock-out $A(e,e'N)$ reactions(e.g. \cite{Kelly:1996hd,
Lapikas:1003zz,Frullani:1984nn}) by introducing missing momentum $p_m$ and
missing energy $E_m$ of the reaction as follows:
\begin{equation}
\vec p_m = \vec q - \vec p_N~,~~~~ E_m = E_R - T_{A-1} ~,
\label{eep_kin1}
\end{equation}
where $T_{A-1} = {p_m^2\over 2 m_{A-1}}$ is the center of mass kinetic energy
of the residual $A-1$ nucleus. The above definition was justified from the
point of view of the shell-model studies. Indeed applying the energy and
momentum conservation for the reaction $\gamma^* + A \rightarrow N + (A-1)^*$
one obtains
\begin{equation}
E_m = E_{A-1}-T_{A-1} - m_{A-1} + E_\alpha ~,
\end{equation}
with $E_{A-1}$, and $m_{A-1}$ being the total energy and ground state mass of
the recoil (A-1) nucleus, while $E_\alpha$ is the removal energy of the
knocked-out nucleon. It follows from the above equation that for the situation
in which the (A-1) nucleus is in its ground state the missing energy, $E_m$,
equals to the removal energy of the nucleon from the particular shell-$\alpha$.

However, for semi-inclusive reactions aimed at studies of SRC structure of the
nucleus the above definition of the missing energy, $E_m$, is somewhat
inconvenient.  The major reason is the fact that the  structure of the
residual $A-1$ nucleus is more complex and strongly depends on  whether 2N or
3N SRCs are probed\footnote{For example in the case of the  knockout of the
nucleon from 2N SRC the residual (A-1) system consists predominantly  of slow
(A-2) nucleus and fast nucleon that was in correlation with the struck
nucleon.}. For this reason much more relevant quantity is the recoil nuclear
energy, $E_R$~\cite{Frankfurt:2008zv} which is related to $E_m$ through:
\begin{equation}
E_R =  E_m +  T_{A-1} = q_0 - (\sqrt{m_N^2 + p_N^2} - m_N),
\label{eep_kin2}
\end{equation}
and represents more natural variable describing SRCs, as shown below.

In the PWIA, it is straightforward to relate the four momentum of the initial
nucleon $p_i \equiv (E_i, \vec p_i)$ to missing momentum and residual nuclear
energy as follows:
\begin{equation}
\vec p_i = - \vec p_{m} ~~~~ \mbox{and} ~~~~ E_i = m_N - E_R ~.
\label{pwia_kin}
\end{equation}
Using these variables, we define the {\em necessary} condition for probing
SRCs as
\begin{equation}
|p_m| = |q-p_N| > k_{Fermi} ~,
\label{nec}
\end{equation}
as this corresponds to scattering from a nucleon with $p_i > k_{Fermi}$ in the
PWIA. This condition is not {\em sufficient}, since it is derived within PWIA
and final state interactions could significantly alter the actual relation
between $p_m$ and $p_i$.

Another observation that can be drawn from the PWIA picture is the relation
between $p_m$ and $E_R$ (or $E_m$)\footnote{The missing momentum $p_m$ and
energy $E_m$ (or $E_R$) are in general independent kinematical variables for
A(e,e$^\prime$p) reactions.}. If one assumes that the interaction took place
with 2N SRCs which are well factorized from the mean field of (A-2) nucleus,
one obtains for the quasi-elastic kinematics the relation
\begin{equation}
E_R \approx \sqrt{m_N^2 + p_m^2} - m_N ~,
\label{empmcor}
\end{equation}
i.e. the kinetic energy associated with the recoil of the correlated nucleon.
Thus $p_m$ and $E_R$ (or $E_m$) will be correlated quantities if the SRC is
probed in the interaction. Note, however, that this correlation arises from a
purely kinematical condition, so any two-body current (such as meson exchange
currents or $\Delta$-isobar contributions) may induce similar correlations. To
enhance the role of the 2N SRCs in the observation of $p_m-E_R$ correlations,
the additional conditions on $Q^2$ and $x$ discussed in Sec.~\ref{genkins}
should be imposed.

\subsection{Kinematics of triple-coincidence A(e,e$^\prime$NN)X reactions}
\label{kinematics:triple}

Triple-coincidence measurements, while adding additional complexity, also have
the potential to provide more comprehensive information about the 
structure of SRCs. In such experiments, one couples the semi-inclusive
single-nucleon knock-out measurement of a high-momentum nucleon with detection
of the spectator nucleon from the initial state SRC.  We start with
same kinematic requirements as for the case of semi-inclusive A(e,e$^\prime$N)
scattering from a nucleon in an SRC (Eqs.~\ref{eep_kin0} and~\ref{nec}).
In addition, we detect another nucleon with a recoil momentum $p_r \approx
-p_i$, which we identify as the recoil nucleon from the 2N SRC.

This is convenient experimentally, because while one is potentially
interested in a large region of initial nucleon momenta, these should all
be found in a relatively narrow cone around the momentum transfer for
sufficiently large $|q|$. Once these kinematic conditions are satisfied,
several signatures can be explored to study SRCs.

The first step is to identify events consistent with pre-existing SRCs. In the
PWIA, there should be a strong correlation between initial momentum of the
struck nucleon ($p_i$) and recoil momentum (initial and final momentum $p_r$
in the PWIA) if both were in the SRC prior to the scattering:
\begin{equation}
\vec p_m \approx - \vec p_i \approx \vec p_r ~,
\label{momentum_corr}
\end{equation}
where these momenta are equal for an at-rest SRC, where the entire missing
momentum is carried by the spectator nucleon from the SRC.  If the SRC has
a non-zero net momentum, one can reconstruct the initial momentum of both
nucleons in the PWIA, in principle allowing for a measure of the momentum
distribution of the pair in the nucleus.  Similarly, the SRC picture predicts
a correlation between missing energy of the reaction and the kinetic energy of
the recoil nucleon. This correlation is a generalization of
Eq.~(\ref{empmcor}) which yields:
\begin{equation}
E_R \equiv q_0-T_p \approx T_r ~,
\label{Tcorr}
\end{equation}
where $T_p$, $T_r$ are the kinetic energies of struck and recoil nucleons.

The above two correlations also yield a relation between light-cone momentum
fractions of initial struck nucleon $\alpha_i$ and recoil nucleon $\alpha_r$:
\begin{equation}
\alpha_i + \alpha_r \approx 2 ~,
\label{alphacorr}
\end{equation}
which is a statement of the condition that 2N SRC carries a fraction
of approximately $2/A$ of
the light cone momentum of the nucleus $A$, as expected for a two-nucleon
system whose total momentum is small. One important advantage of
Eq.~(\ref{alphacorr}) as compared to Eqs.(\ref{momentum_corr},\ref{Tcorr}) is
that while latter relations are strictly speaking correct only within the
PWIA, the former survives the FSI in the eikonal regime.

Perhaps the most interesting advantage of the triple-coincidence experiments is
the possibility to study the isospin-dependence of the SRCs by detecting
different possible final states: pp, pn, nn. Generalizing these reactions to
the case in which the recoiling system represents a baryonic resonance, such as
$\Delta$-isobar, one can use these reactions to gain access to the
non-nucleonic components of SRCs~\cite{Allasia:1986kg, Frankfurt:2009vv}.

Many of the detailed studies possible in a triple-coincidence reaction
rely on small final-state interactions, although the approximate conservation
of $\alpha_i$ in the eikonal regime leave some information intact even in
the presence of large FSIs.  These issues will be discussed in more detail in
Sec.~\ref{sec:triple}.

\section{Studies of the Deuteron}
\label{sec:deuteron}

The deuteron plays a special role in SRC studies. The measurement of the
momentum distribution is necessary for identification of the iso-singlet
component of 2N SRCs in nuclei, which is expected to be a universal feature in
the high-momentum tail of all nuclei (Eq.~(\ref{nk_src})). As the simplest
nucleus, the deuteron also provides an ideal testing ground for many issues
related to details of the reaction mechanism: meson exchange contributions
(MECs), isobar contributions (ICs), final state interactions (FSIs), as well
as modification of the properties of nucleons due to their off-shellness.

\begin{figure}[htbp]
\begin{center}
\includegraphics[width=0.6\textwidth,height=0.65\textwidth]{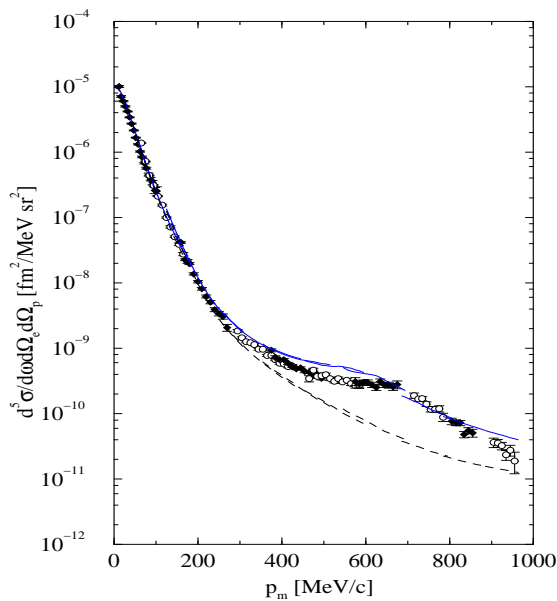}
\caption{The $p_m$ dependence of the D(e,e$^\prime$p)n differential cross
section~\cite{Blomqvist:1998fr} at $Q^2=0.13-0.33$~GeV$^2$. Dashed line is
PWIA contribution, solid line is PWIA+MEC+IC~\cite{Arenhovel:1995be}.
Figure reprinted with permission from Ref.~\cite{Blomqvist:1998fr}
}
\label{fig:d_lowQ2}
\end{center}
\end{figure}

Extensive measurements of single nucleon knockout, both inclusive and
coincidence, have been made to probe the low-momentum region and high-momentum
tails of the deuteron momentum distribution.  For nucleon momenta above
$\sim$250~MeV/c, calculations of the cross sections showed growing sensitivity
to the choice of nucleon-nucleon potential in exactly the region where
correlations are expected to dominate the momentum distribution. It was thus
expected that measurements at large values of missing momentum would allow one
to study the details of the short-range part of the nucleon--nucleon potential.

In 1981, Saclay measured the D(e,e$^\prime$p)n reaction with a 500~MeV electron
beam~\cite{Bussiere:1981mv} and extracted a momentum distribution up to
$\approx$300~MeV/c. The sub-GeV beam energies that were available meant that
subsequent experiments at higher $p_m$ were forced to make measurements in the
$x_B < 1$ region, near the $\Delta$ resonance. So while these experiments were
able to measure missing momenta out to 950~MeV/c, the cross section for $p_m$
above 400~MeV/c was dominated by meson exchange and isobar
currents~\cite{Blomqvist:1998fr} and it was not possible to extract information
about the underlying distribution (see Fig.~\ref{fig:d_lowQ2}).

\begin{figure}[htb]
\begin{center}
\includegraphics[width=0.8\textwidth]{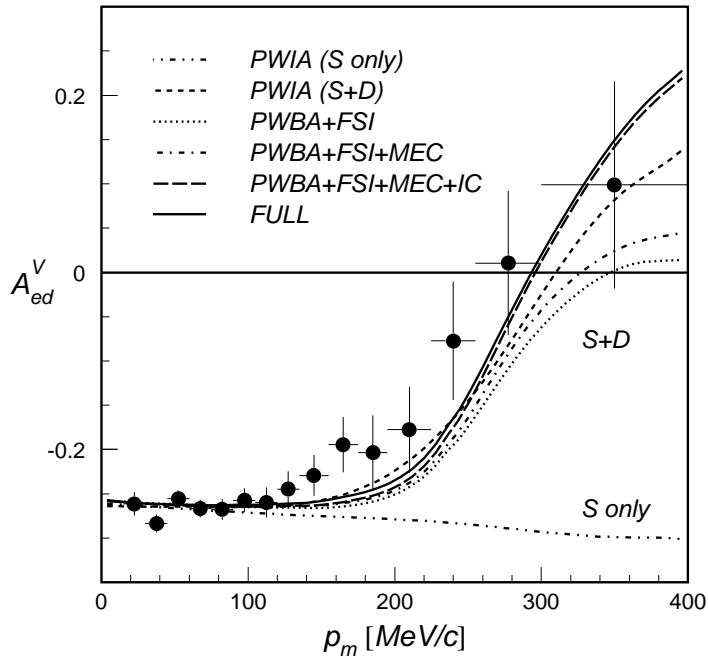}
\caption{
Spin correlation parameter $A^V_{ed}$ as function of missing
momentum for the $\vec{\rm D}(\vec e,e^\prime p)n$ reaction at $Q^2 =
0.21~({\rm GeV}/c)^2$~\cite{Passchier:2001uc}. The short-dashed and dot-dot-dashed curves are PWIA
predictions with and without inclusion of the $D$-wave, respectively. The
other curves are calculations~\cite{Ritz:1997cq, Leidemann:1992fp} including
additional contributions, folded over the detector acceptance.
Figure reprinted with permission from Ref.~\cite{Passchier:2001uc}
}
\label{Fig:d_pol_lowQ2}
\end{center}
\end{figure}

The situation is somewhat improved when exploring polarization observables in
the deuteron electrodisintegration reaction. For example, the NIKHEF polarized
D(e,e$^\prime$p)n measurements~\cite{Passchier:2001uc} shown in
Fig.~\ref{Fig:d_pol_lowQ2} provide information on the structure at high $p_m$,
with reduced sensitivity to FSI and MEC.  While these play a role, the change
of sign provided a clear demonstration of the importance of including both the
S and D wave components of the deuteron. However, a detailed description of
the data still requires contributions due to isobar and meson exchange
currents.

The first high-energy studies involving the deuteron were the studies of
inclusive high-$Q^2$ reactions at SLAC in late 1970s, 1980s and early
90s~\cite{Schutz:1976he, Rock:1982gf, Lung:1992bu}. These experiments for the
first time probed $x>1$ kinematics. The most comprehensive high
$Q^2$ inclusive electron-deuteron scattering measurements for $x>1$ were
performed in Hall C at Jefferson Lab.  Cross sections for the inclusive 
D(e,e$^\prime$) reaction were taken over a range of $x$ and $Q^2$ using a
4~GeV electron beam in 1996~\cite{Arrington:1998hz, Arrington:1998ps,
Arrington:2001ni} and the kinematics significantly extended in a subsequent
measurements at 6~GeV~\cite{Fomin:2008iq,Fomin:2010ei,Fomin:2011ng}. These
inclusive measurements extract the deuteron momentum distribution by means of
$y$-scaling analysis~\cite{West:1974ua, Sick:1980ey, Pace:1982xi, Day:1990mf,
Benhar:2006wy, Fomin:2011ng}. This analysis assumes the dominance of
quasi-elastic scattering and involves reconstructing the minimum initial
momentum of the struck nucleon ($y$) consistent with the kinematics of the
scattering.  The minimum initial momentum can be obtained from energy
conservation assuming that the momentum is entirely along the
direction of the transferred momentum $q$, and that the residual (A-1)
nucleus is left in an unexcited state:
\begin{equation}
q_0 + m_d = \sqrt{m^2_N+(q+y)^2} + \sqrt{m^2_N+y^2} ~.
\label{y_par}
\end{equation}
We obtain the $y$-scaling function, $F(y,Q^2)$, by dividing out the
elastic e-N cross section:
\begin{equation}
F^{exp}(y,Q^2) = {{d\sigma\over dE^\prime_e,d\Omega} \over \sigma_{ep} +
\sigma_{en}} ~,
\label{Fexp}
\end{equation}
where $\sigma_{eN}$ is the elastic scattering cross section off the bound
nucleon.

\begin{figure}[htb]
\begin{center}
\includegraphics[width=0.95\textwidth]{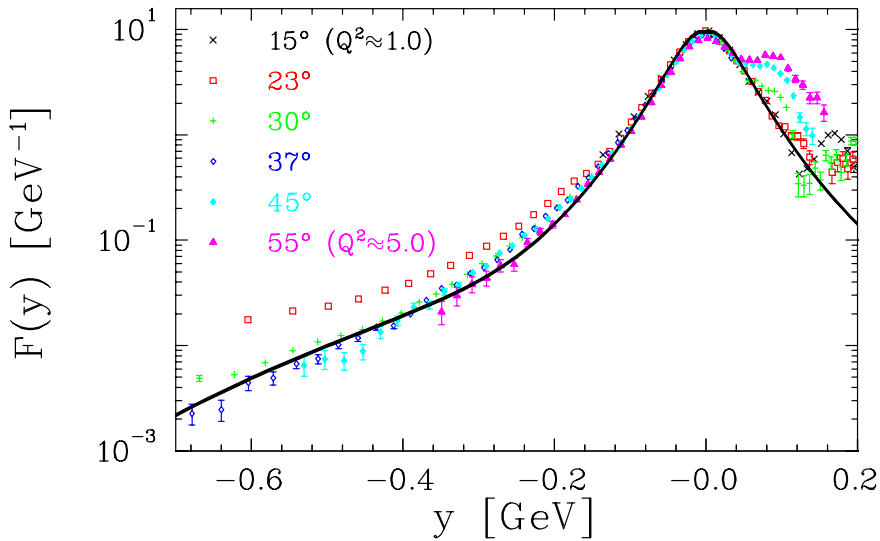}
\caption{Scaling function $F(y$) for deuteron from
Ref.~\cite{Arrington:1998hz,Arrington:2003tw} for scattering at E=4.045 GeV
and 15$\deg \le \theta \le 55\deg$. The solid line represents the expected $F(y)$
based on the calculation of the deuteron momentum distribution using the Av14
NN potential. Note that the inelastic contribution for $y \ge 0$ has been
subtracted using a model. Figure adapted from Ref.~\cite{Arrington:2003tw}}
\label{Fig:d}
\end{center}
\end{figure}

In the scaling limit, $F(y,Q^2) \to F(y)$ if the assumptions of
the scaling analysis are valid: quasielastic scattering, PWIA (no final-state
interactions), and low excitation of the final spectator system.  Inelastic
scattering is suppressed by going to low energy transfer, corresponding to
$x>1$, while for the deuteron, the final spectator system is a single nucleon
where there are no options for low energy excitations. The observation of
scaling in the data, shown in Fig.~\ref{Fig:d}, supports the assumptions of
the $y$-scaling analysis suggesting that FSI, which are expected to fall
rapidly with $Q^2$, are small.

\begin{figure}[htb]
\begin{center}
\includegraphics[width=0.65\textwidth,angle=270]{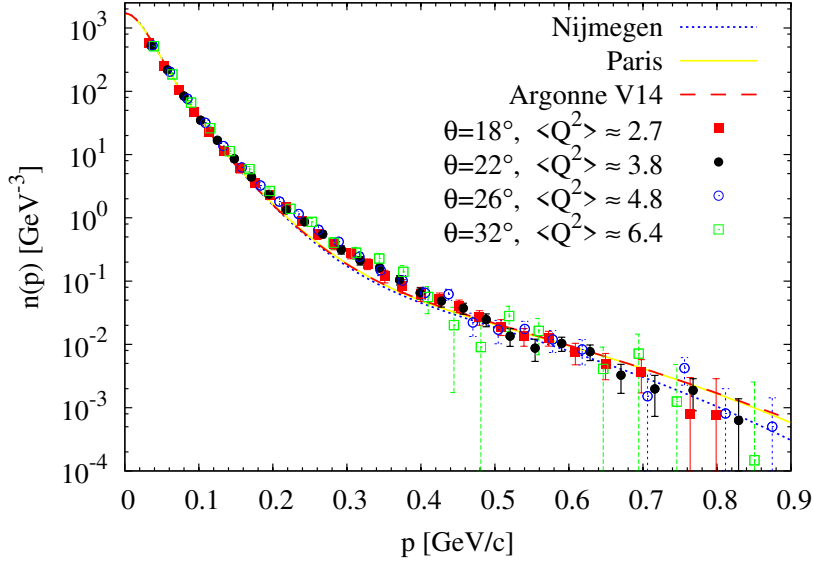}
\caption{Deuteron momentum distribution, $n(p)$, as extracted from inclusive
scattering at $x>1$ from JLab E02-019~\cite{Fomin:2011ng}.  The different
symbols show the data sets corresponding to different scattering angles,
and thus different $Q^2$ values, and the curves are calculations based on
three different NN potentials.  Figure adapted from Ref.~\cite{Fomin:2011ng}.}
\label{Fig:ndk}
\end{center}
\end{figure}

In the PWIA picture of quasi-elastic scattering, the above extracted scaling
function can be directly related to the nucleon momentum distribution,
as the cross section is simply a convolution of the e-N elastic cross
section and the momentum distribution of the nucleons in the nucleus:
\begin{equation}
F(y) \approx 2\pi \int\limits_{|y|}^{\infty} n(p) p dp ~~,~~
n(p) \approx \frac{-1}{2 \pi p} \frac{dF(p)}{dp} ~.
\label{Fteor}
\end{equation}
The relation between the scaling function and the momentum distribution can be
understood in a simple picture.  The kinematics of the scattering set the
minimum initial nucleon momentum that can contribute to the cross section
($y$). While the inclusive cross section is an integral that involves all
momenta above $y$, it is dominated by momenta close to $y$, due to the rapid
falloff of the nucleon momentum distribution.  So the main difference between
the cross section at kinematics corresponding to $y_1$ and $y_2=y_1+\delta y$
is the loss of the scattering from nucleons in the narrow momentum region
$\delta y$ between $y_2$ and $y_1$.  Thus, the momentum distribution is
connected to the derivative of the scaling function.

Figure~\ref{Fig:ndk} shows the deuteron momentum distribution, $n(p)$ as
extracted from Eq.~(\ref{Fteor}) out to 900~MeV/c from the more recent
measurement at higher energy~\cite{Fomin:2011ng}. While the agreement with the
deuteron momentum distribution calculated from the NN potential again suggests
that final state interactions are not too large in this region, it is
difficult to set precise limits, as other modern NN potential yield
predictions that vary by 10--20\% for these nucleon momenta above 400~MeV/c.
Thus, further studies of FSI contributions for scattering from high-momentum
nucleons are still necessary. For inclusive scattering, the FSI contributions
are less of
a concern than in exclusive reactions, due to the closure approximation which
is increasingly well satisfied as $Q^2$ increases, leading to FSI
contributions that largely cancel.  When the struck nucleon is extremely close
to another nucleon, as in the case of scattering from one nucleon in an SRC,
the FSIs are confined within the correlated nucleons. Thus, while the
observation of a $Q^2$-independent scaling function supports the $y$-scaling
picture in general, the high momentum tails of the distribution may still have
contributions from a nearly $Q^2$-independent FSI contribution due to
rescattering of the struck nucleon from the correlated nucleon in the SRC.

\begin{figure}[t]
\begin{center}
\includegraphics[width=0.9\textwidth]{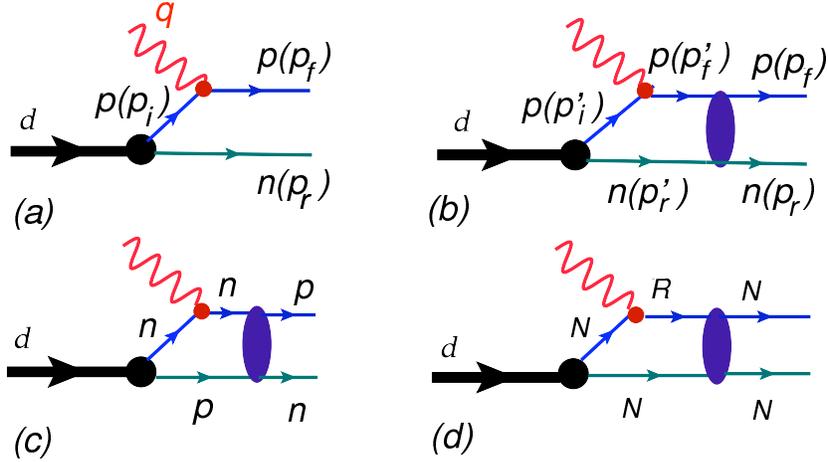}
\caption{Impulse approximation (a), diagonal (b) and change-exchange (c) final
state interactions and intermediate resonance contributions (d) to the
D(e,e$^\prime$p)n reaction.
Figure reprinted with permission from Ref.~\cite{Sargsian:2009hf}
}
\label{Fig:edepn_diagrams}
\end{center}
\end{figure}

Exclusive D(e,e$^\prime$p)n measurements can provide additional information, as the
initial kinematics of the struck proton can be fully reconstructed within the PWIA.
However, while these reactions have the potential to provide a more
complete understanding of the high momentum components of the nucleon
distribution, they are also more sensitive to the details of the reaction
mechanism.  The deuteron, then, provides the best testing ground for both
the high momentum wave function and the competing effects including final
state interactions, isobar contributions, and reaction dynamics.  This is
one of the main reasons why these reactions have been so intensively studied
by several groups~\cite{Jeschonnek:2000nh, CiofidegliAtti:2000xj,
Laget:2004sm, Jeschonnek:2008zg, Laget:2004sm, Jeschonnek:2009tq,
Jeschonnek:2009ds, Frankfurt:1994kt, Frankfurt:1996xx, Sargsian:2009hf}.

One important aspect of exclusive D(e,e$^\prime$N)N reactions is that they provide
the possibility to map out the role of intermediate inelastic transitions
such as IC contributions and to study the onset of the eikonal regime in the
final state interaction of two outgoing nucleons. This is an extremely
important issue because, unlike MEC contributions, IC and FSI contributions do
not decrease rapidly with increasing $Q^2$ values\footnote{For $Q^2 <
4$~GeV$^2$, the $\gamma^*N\rightarrow \Delta$ transition amplitude is
comparable with the elastic $\gamma^*N\rightarrow N$ form factor and the FSI
contribution is essentially $Q^2$ independent, at least up to $Q^2 \approx
8$~(GeV/c)$^2$~\cite{Garrow:2001di,Sargsian:2002wc}.}. As a result, the most
relevant processes contributing to the exclusive D(e,e$^\prime$p)n reactions at
large $Q^2$ are the PWIA single nucleon knock-out, final state interactions,
and intermediate resonance contributions (mainly the $\Delta$-isobar) of
Fig.~\ref{Fig:edepn_diagrams}.

\begin{figure}[th]
\centering\includegraphics[scale=0.44]{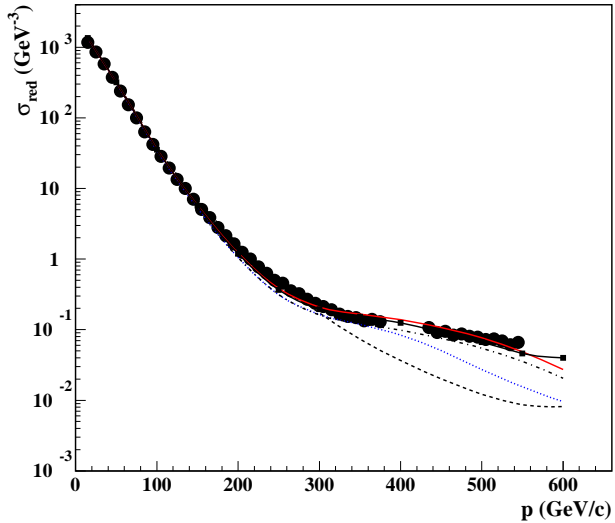}
\caption{The ratio of the D(e,e$^\prime$p)n cross section to the elastic ep
cross section as a function of $p_m$, with the ep kinematics taken such that
it reduces to the deuteron momentum distribution in the PWIA. The data are
from Ref.~\cite{Ulmer:2002jn}. Dashed line -- PWIA calculation; dotted line --
PWIA + FSI term with only on-shell pn $\rightarrow$ pn rescattering included;
dash-dotted line -- PWIA + FSI with on- and off-shell pn $\rightarrow$ pn
rescattering; solid line -- PWIA + full FSI (pn $\rightarrow$ pn and
charge-exchange pn $\rightarrow$ np rescattering); and solid line with squares
-- PWIA + full FSI + contribution from the mechanism in which the proton is a
spectator and the neutron was struck by the virtual photon. Figure reprinted
with permission from Ref.~\cite{Sargsian:2009hf}.}
\label{Fig:ulmer_exp}
\end{figure}

The first experiments on D(e,e$^\prime$p)n with the momentum transfer close to
the GeV/c region were performed at SLAC~\cite{Bulten95} and Jefferson
Lab~\cite{e94004}. The SLAC measurement covered $1.2 < Q^2 < 6.8$~GeV$^2$,
but was limited to $p_m$ values below 300~MeV/c. Even so, the cross section,
and in particular the asymmetry between positive and negative $p_m$ values,
showed the importance of having detailed relativistic
calculations~\cite{Tjon92, Beck92}.
The Jefferson Lab measurement was performed at $Q^2 \approx 0.7$~(GeV/c)$^2$,
where data were taken for $x \approx 1$ (on top of the quasi-elastic
peak)~\cite{e94004}.  The extracted momentum distribution follows the early
Saclay results up to $\approx 300$~MeV/c, but above this, final state
interactions dominate the results~\cite{Ulmer:2002jn}. Studies of
this experiment~\cite{Jeschonnek:2008zg, Sargsian:2009hf} demonstrated that
FSIs, calculated within the eikonal approximation, can reasonably well
describe the absolute cross section of the reaction (see
Fig.~\ref{Fig:ulmer_exp}), even when FSI yield 80\% of the total cross
section.  This provides a significant test of the validity of the eikonal
approximation, suggesting that the corrections can be reliably applied in
kinematics where the contributions can be further suppressed. Another result of
these measurements is the observation that kinematical suppression of IC for
the low energy transfer region ($x \ge 1$) is effective at these kinematics,
with no substantial IC contribution observed in these measurements.

\begin{figure}[htb]
\begin{center}
\includegraphics[width=0.8\textwidth]{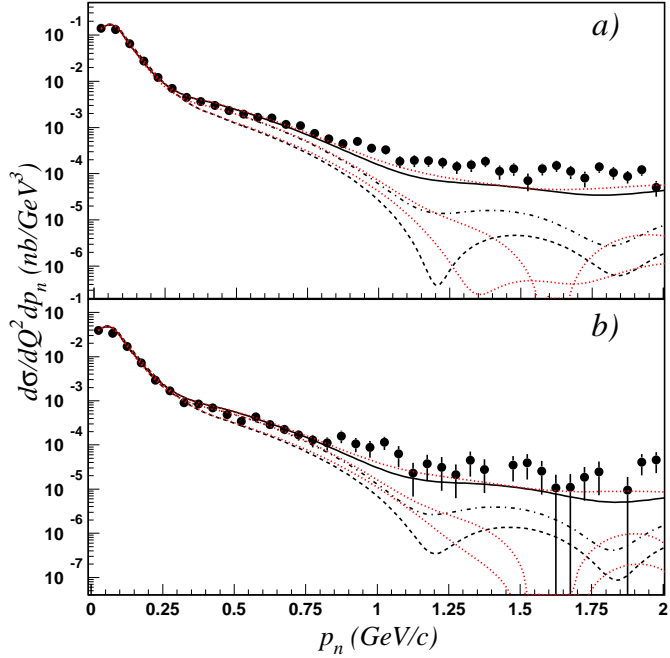}
\caption{The recoil neutron momentum distribution for (a)
$Q^2=4\pm0.5$~GeV$^2$ and (b) $Q^2=5\pm 0.5$~GeV$^2$. The dashed, dash-dotted
and solid curves are calculated~\cite{Laget:2004sm} using
the Paris potential~\cite{Lacombe:1980dr} and averaged over the CLAS
acceptance for PWIA, PWIA+FSI and PWIA+FSI+MEC+N$\Delta$, respectively. The
red dotted curves are calculated using Av18 potential~\cite{Wiringa:1994wb}.
Figure reprinted with permission from Ref.~\cite{Egiyan:2007qj}
}
\label{fig:egiyan_deep}
\end{center}
\end{figure}

These measurements were followed by a measurement at high-$p_m$ and $Q^2$ up
to 5~(GeV/c)$^2$~\cite{Egiyan:2007qj} where for the first time, the momentum
and energy transfer in the reactions exceeded the nucleon mass.
The momentum distribution shows a substantial contribution
from FSI at large missing momenta ($p_m>300$~MeV/c) when integrated over all
recoil neutron angles, as shown in Fig.~\ref{fig:egiyan_deep}.  The main
advantage of this experiment was that it provided access to the recoil-nucleon
angular dependence of the cross section at large $p_m$.  This is critical for
distinguishing the onset of the eikonal regime of FSI which is characterized
by a very specific angular distribution, with nuclear screening minima and
incoherent rescattering maxima~\cite{Frankfurt:1994kt, Frankfurt:1996xx}.

\begin{figure}[htbp]
\centering\includegraphics[scale=0.46]{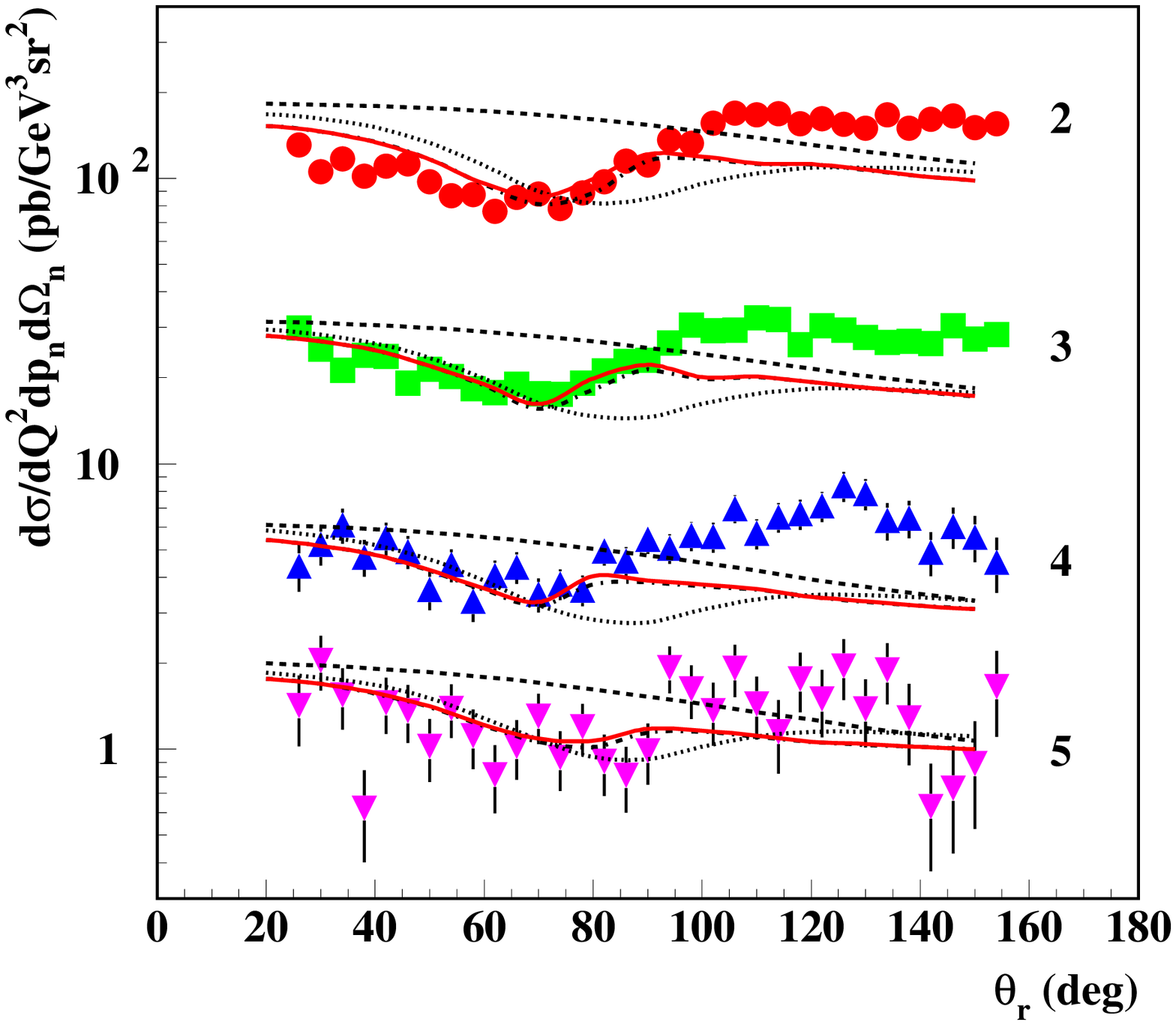} \\
\centering\includegraphics[scale=0.46]{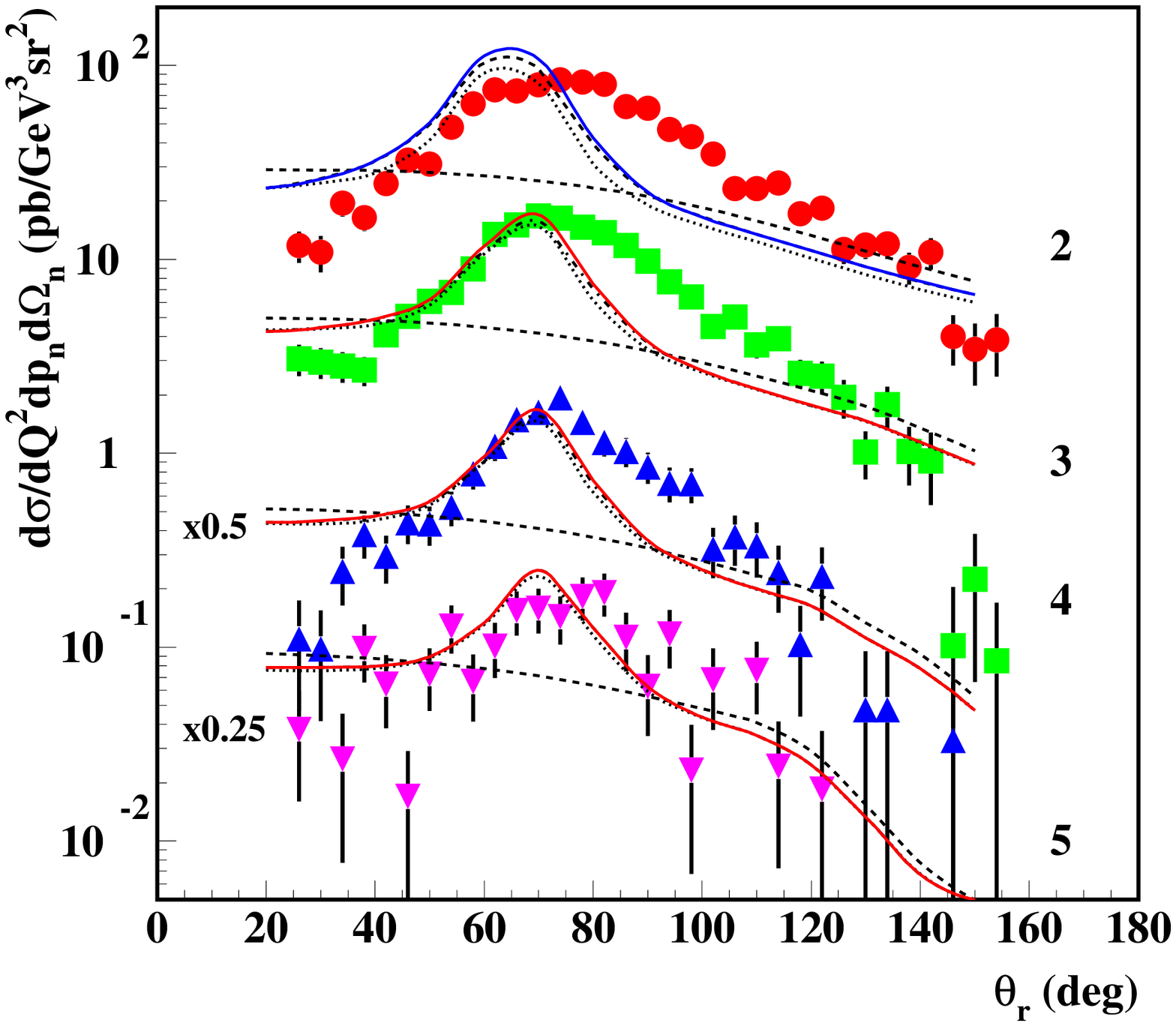}
\caption{\textbf{Top:} Dependence of the differential  cross section on the direction of
the recoil neutron momentum. The data are from Ref.~\cite{Egiyan:2007qj}.
Dashed line - PWIA calculation; dotted line - PWIA + FSI term that contains
only on-shell part of the pn $\rightarrow$ pn rescattering; dash-dotted line  -
 PWIA+ FSI term that contains both, off- and on- shell parts of the
pn $\rightarrow$ pn rescattering; solid line - PWIA + full FSI containing
pn $\rightarrow$ pn  and charge exchange pn $\rightarrow$ np rescatterings. The
momentum of the recoil neutron is restricted to $200 < p_r < 300$~MeV/c.  The
labels 2, 3, 4 and 5 correspond $Q^2$ bins of $Q^2 = 2\pm0.25, 3\pm0.5,
4\pm0.5,$ and $5\pm0.5$~GeV$^2$, respectively. \textit{Bottom:} Same as the
top figure, but with the momentum of the recoil neutron is restricted to $400
< p_r < 600$~MeV/c.  The results for
bins ``4'' and ``5'' are scaled by factors of 0.5 and 0.25 respectively.
Figure reprinted with permission from Ref.~\cite{Sargsian:2009hf}
}
\label{Fig:kim_250_500}
\end{figure}

Figure~\ref{Fig:kim_250_500} show the comparison of the
generalized eikonal approximation calculations~\cite{Sargsian:2009hf}
and the data from Ref.~\cite{Egiyan:2007qj}. The calculation in the comparison
is evaluated over identical kinematic integrations as the experiment. Despite
these integrations, which lessen the sensitivity of the cross section to
various dynamical details of the process, several important observations can
be made:

\begin{itemize}

\item First, the angular distribution exhibits the expected eikonal features,
with a clear minimum or maximum in the top and bottom plots of
Fig.~\ref{Fig:kim_250_500}, respectively, at nearly transverse kinematics due
to the final state interaction.  The maximum of the FSI contributions is at
recoil angles of $70\deg$, in agreement with the generalized eikonal
approximation prediction of Refs.~\cite{Frankfurt:1994kt, Frankfurt:1996xx},
as opposed to the Glauber theory calculations (not shown) which predict peak
FSI contributions at $90\deg$.

\item The calculation disagrees with the data for $\theta_r > 70\deg$, which
corresponds to the $x<1$ kinematic range close to the inelastic threshold.
This could be due to intermediate isobar contributions. The comparisons also
suggest that the relative strength of the $\Delta$-isobar contribution may
diminish with increasing $Q^2$ and for neutron production angles $\theta_r
\rightarrow 180\deg$.  If this is confirmed in more precise measurements it
will have an important implication for SRCs in deep-inelastic reactions, since
in this case fast backward recoil nucleons will not be affected by the final
state interaction with the products of DIS scattering.

\item The forward direction of the recoil nucleon momentum, being far from the
$\Delta$-isobar threshold, exhibits a relatively small contribution from
FSI.  This indicates that the forward recoil angle region is best suited for
studies of PWIA properties of the reaction such as the deuteron wave function
and off-shell electromagnetic current.

\end{itemize}

\begin{figure}[htb]
\begin{center}
\includegraphics[width=1.0\textwidth]{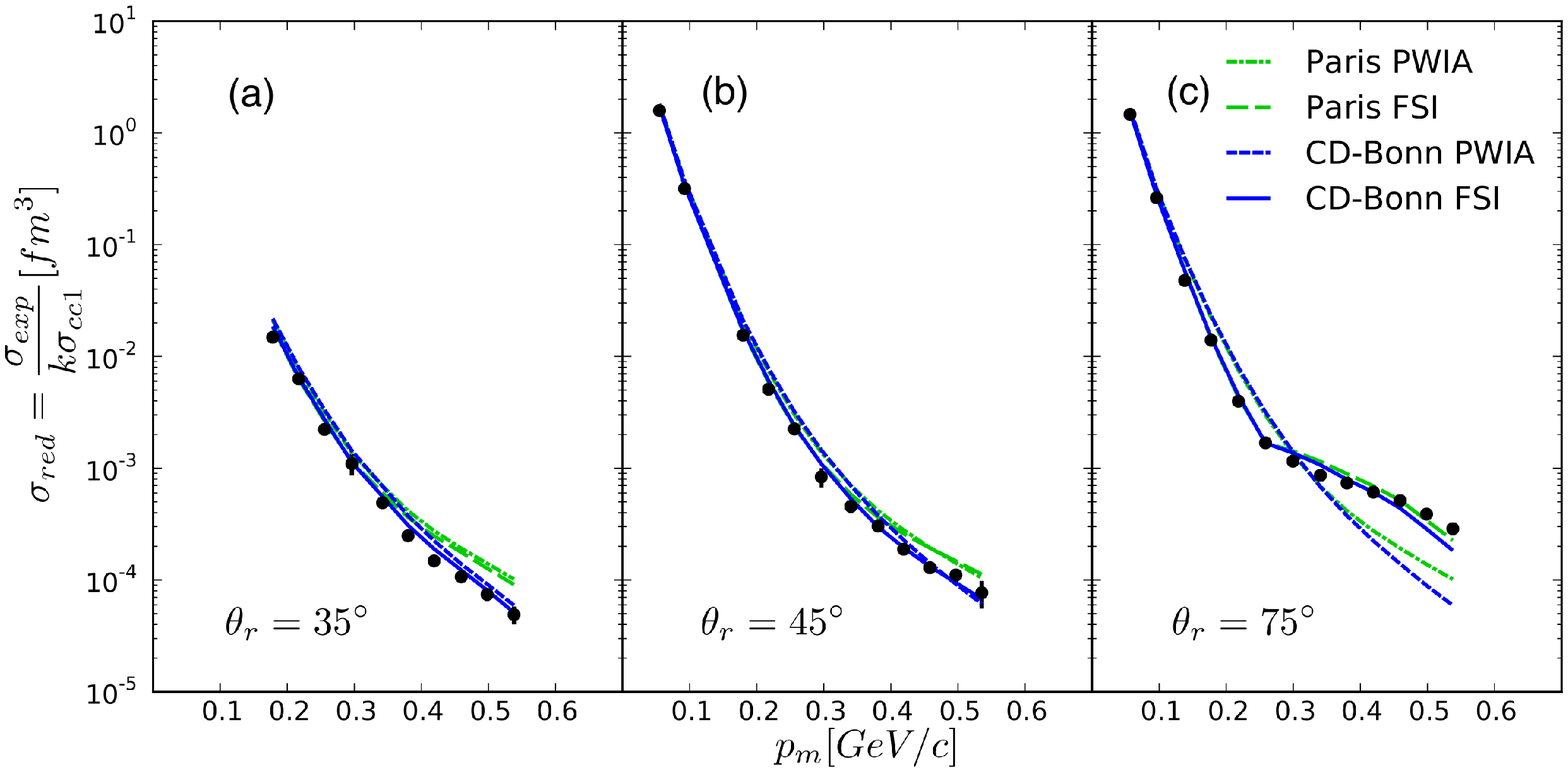}
\caption{
The reduced cross section as function of missing momentum is shown in panels
a, b and c for $\theta$ = 35; 45 and 75$^\circ$, respectively.  CD-Bonn
potential: dashed (blue) lines PWIA , solid (blue) lines FSI.  Paris
potential: dash-dot (green) lines PWIA, long-dashed (green) lines FSI. The
PWIA results are for all angles identical.  Figure reprinted with permission
from Ref.~\cite{Boeglin:2011mt}.}
\label{fig:boeglin}
\end{center}
\end{figure}

The results of another related experiment~\cite{e01020, Luminita} have
recently been submitted for publication~\cite{Boeglin:2011mt} and are shown
in Fig.~\ref{fig:boeglin}.  In this
experiment, the reaction is measured at forward recoil angles and for $Q^2$ up
to 3.5~GeV$^2$, with extremely high precision.  Their data extend to
$p_m = 550$~MeV/c and the cross sections are examined as a function of
$\theta_r$ from 20$\deg$ to 90$\deg$.  The results are compared to three
models of scattering from the deuteron, with and without the inclusion of FSI,
MEC, and IC, which all yield a reasonable description of the data.  
These data will allow for more detailed evaluations of the models of the
FSI contributions over a range of $\theta_r$ and $p_m$ values.  In addition,
they show that for $\theta_r \approx 40\deg$, FSI contributions are relatively
small and $\theta_r$-independent over the full $p_m$ range of the experiment,
suggesting that these kinematics may provide the most direct access to the
high-momentum tail of the deuteron momentum distribution.

These high-$Q^2$ experiments on deuteron electrodisintegration have already
shown great potential to resolve several issues
related to the exploration of nucleon with high initial momentum and missing
energy.  This will allow for a clean extraction of the high-momentum components
of the deuteron momentum distribution in future measurements focused on
kinematic regions where FSI, MEC, and IC are suppressed.  Future experiments
can also provide further data where these effects are large to provide a
more quantitative validation of calculations of these effects in a well
understood nucleus.  This will be critical in moving to similar detailed
studies for heavier nuclei.

\clearpage
\section{Inclusive Reactions Beyond the Deuteron}
\label{sec:inclusive}

The first high-$Q^2$ experiments that probed the high-momentum component of
the ground state wave functions of more complex nuclei were
inclusive measurements performed at SLAC in the 80s~\cite{Rock:1981aa,
Bosted:1982gd} and 90s~\cite{Meziani:1992xr, Filippone:1992iz, Bosted:1992fy,
Day:1993md, Arrington:1995hs}, followed by series of measurements at Jefferson
Lab~\cite{Arrington:1998hz, Arrington:1998ps, Egiyan:2003vg, Egiyan:2005hs,
Fomin:2008iq, Fomin:2010ei}. A compilation of the data from the many of
these experiments as well as a detailed discussion of quasielastic 
scattering (with a focus on probing the mean-field structure of the nucleus)
can be found in Ref.~\cite{Benhar:2006er}.

To probe SRCs in these experiments, kinematic conditions similar to those
discussed for inclusive scattering from the deuteron should be considered.
Thus, they require $Q^2 > 1$~GeV$^2$ measurements at $x>1$, such that the
scattering is far from the inelastic pion production threshold and
dominated by quasi-elastic scattering from the bound nucleon.

\begin{figure}[htb]
\begin{center}
\includegraphics[width=0.9\textwidth]{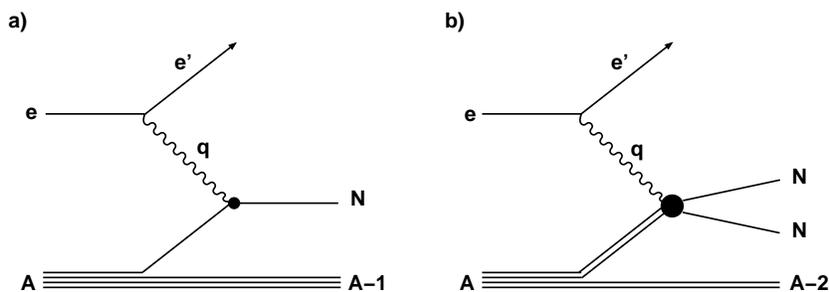}
\caption{\label{src:diagrams}Shown are the dominate diagrams of the
A(e,e$^\prime$) reaction for x$>$1 and $Q^2 > 1$(GeV/c)$^2$. Figure a) shows
single nucleon scattering while figure b) shows scattering from a correlated
initial-state pair.}
\label{spec_A-2}
\end{center}
\end{figure}

One of the first approaches in probing the momentum distribution of nucleons
in heavy nuclei was the framework of $y$ scaling~\cite{Day:1990mf} where the
$y$ parameter was defined as a solution of the following equation:
\begin{equation}
q_0 + m_A  =  \sqrt{m^2_N+(q+y)^2} + \sqrt{m^2_{A-1}+y^2} ~.
\label{y_par_A}
\end{equation}
This is a generalization of the case for the deuteron (Eq.~(\ref{y_par})),
with the assumption that the final state is a two-body system consisting of
the struck nucleon and a spectator (A-1) system.  Note that the value used
for $m^2_{A-1}$ is related to the excitation of the residual system, and so
for the case of an unexcited (A-1) spectator, corresponding to to panel a) of
Fig.~\ref{spec_A-2}, this is the mass of the ground state (A-1) spectator
nucleus.  There can also be small excitation of the (A-1) system, yielding
a larger effective mass for the A-1 system, while the case of
Fig.~\ref{spec_A-2}b corresponds to a significantly larger value for
$m^2_{A-1}$, due to the large kinetic energy of the spectator nucleon from
the initial-state SRC.

nucleus in its ground state,
corresponding to to panel a) of Fig.~\ref{spec_A-2}.  

\begin{figure}[tb]
\begin{center}
\includegraphics[width=0.8\textwidth]{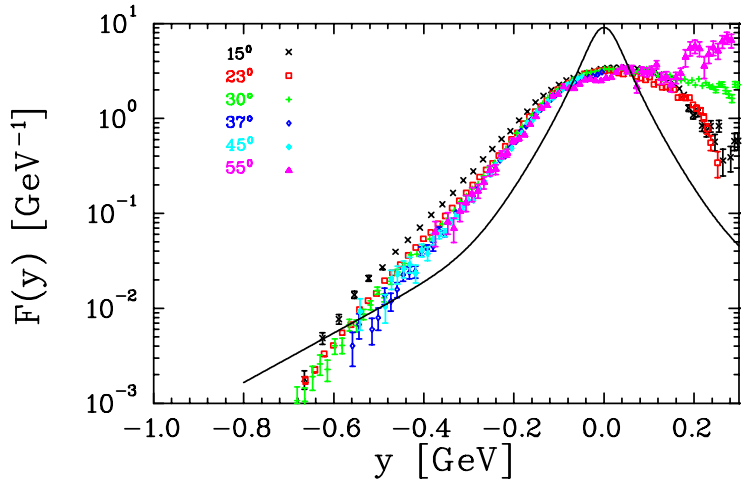}
\caption{Extracted $y$-scaling function $F(y)$ for
Fe~\cite{Arrington:1998ps, Arrington:2006pn}, compared to a fit to the
deuterium data (solid line).  Note that a model of the inelastic contributions
which are significant for $y>0$ has been subtracted to isolate the
quasielastic contributions.  Figure adapted from Ref.~\cite{Arrington:2006pn}}
\label{Fig:A_y}
\end{center}
\end{figure}

The requirement that measurements be made for $x>1$ corresponds to negative
values of $y$. While scattering at large negative $y$ is sensitive to the
tails of the momentum distribution, there is an additional complication that
arises in applying the $y$-scaling approach for complex nuclei. The extraction
of $y$ from Eq.~(\ref{y_par_A}) assumes an unexcited (A-1) spectator nucleus
in the final state. This is a reasonable approximation for low values of $y$,
where one is removing a nucleon from one of the nuclear shells, but not for
high momentum nucleons which are part of an SRC.  In this case,
one expects that in addition to the struck nucleon, there will be one high
momentum spectator nucleon (the 2nd nucleon from the SRC) along with an
unexcited (A-2) spectator, as shown in panel b) of Fig.~\ref{spec_A-2}.  For
large values of $y$, there is a significant difference in the kinetic energy
of the spectator system in these two pictures. This is not an issue for the
deuteron, where the two-body final state is well justified if one avoids
resonance excitations, but can be very important for heavier nuclei.
Figure~\ref{Fig:A_y} shows $F(y)$ extracted from the JLab Hall C measurements
on Fe~\cite{Arrington:1998ps, Arrington:2006pn} at the same kinematics as the
deuterium measurement from Fig.~\ref{Fig:d}, and the definition of $y$ from
Eq.~(\ref{y_par_A}). The result shows two unexpected features.  First, the
peak is not symmetric about $y=0$. While the model-dependent subtraction
of the inelastic contributions is significant for $y>0$ and large $Q^2$, 
the agreement among the low $Q^2$ data sets suggests that the subtraction is
not the source of the asymmetry.  Second, the tail of the distribution falls
off significantly more rapidly for Fe than it does for deuterium, suggesting
that Fe has fewer very high momentum nucleons.  One would expect both the
mean-field and SRC contributions to yield more high-momentum nucleons in Fe
than in $^2$H. Both of these unexpected features are consequences of treating
the spectator system as an unexcited (A-1) nucleus.

This issue can be addressed by adjusting the assumptions that go into defining
Eq.~(\ref{y_par_A}) or by explicitly calculating corrections to $F(y)$ to
account for the modified final state. The first approach was to replace $y$
with $y_2$~\cite{Ji:1989nr, CiofidegliAtti:1997km}:
\begin{equation}
q_0 + m_A  =  \sqrt{m^2_N+(q+y_2)^2} + \sqrt{m^2_N+y_2^2} + m_{A-2} ~,
\label{y2_par_A}
\end{equation}
which follows from the assumption that scattering from a high momentum nucleon
follows the scenario of Fig.~\ref{spec_A-2}b in which 2N SRC is factorized
from the low momentum residual (A-2) nucleus, assumed to be at rest.  This
resolves the problem in the high-momentum tail, yielding a falloff similar to
that observed in the deuteron, but alters the extracted scaling function
$F(y)$ near $y=0$, where the standard definition of $y$ gives a reasonable
approximation.

Later approaches attempted to improve the definition of $y$ by incorporating
the impact of SRCs at large values of $|y|$ while maintaining appropriate
behavior at small $|y|$ values.  One such attempt was the introduction of
$y_{CW}$ by Ciofi degli Atti and West~\cite{CiofidegliAtti:1999is}:
\begin{equation}
q_0 + m_A  =  \sqrt{m^2_N+(q+y_{CW})^2} + E^*_{A-1} ~,
\label{ycw_par_A}
\end{equation}
where $E^*_{A-1}$, the average energy of the residual (A-1) system is taken
from a model where the two-nucleon SRC dominates for large values of $|y|$.
The model also includes a contribution for the average center-of-mass motion
of the SRC and corrections such that at low $|y|$, the excitation energy of
the residual nucleus approaches the expected value for the (A-1) spectator
nucleus.  Note that the calculation of $E^*_{A-1}$ was refined over time (most
recently in Ref.~\cite{CiofidegliAtti:2009qc}) yielding somewhat different
definitions of $y_{CW}$ in different works.

\begin{figure}[tb]
\begin{center}
\includegraphics[width=0.8\textwidth]{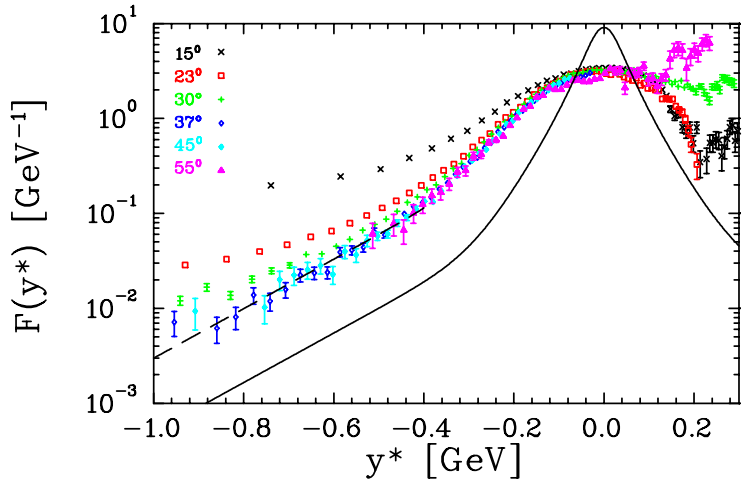}
\caption{Extracted $y^*$-scaling function $F(y^*)$ for
Fe~\cite{Arrington:2006pn}, compared to a fit to the high-$Q^2$
deuterium data (solid) from Fig.~\ref{Fig:d} and a scaled up version of the
high-momentum tail of the deuterium fit (dashed).  Note that a model of the
inelastic contributions which are significant for $y>0$ has been subtracted to
try and isolate the quasielastic contributions. Figure adapted from
Ref.~\cite{Arrington:2006pn}.}
\label{Fig:A_ystar}
\end{center}
\end{figure}

Another approach uses a simple convolution of the SRC distribution (taken to
be identical to the deuteron momentum distribution) with an estimated
center-of-mass motion of the SRC.  From this, one can determine the average
contribution from both the C.M. motion of the pair and the relative momentum
of the nucleons in the SRC to determine the excitation energy of the residual
system, neglecting transverse momentum, allowing for the extraction of the
initial nucleon momentum, $y^*$~\cite{Arrington:2003tw}:
\begin{equation}
q_0 + m_A  =  \sqrt{m^2_N+(q+y^*)^2} + \sqrt{m^2_N+(K_{CM}/2-k_{rel})^2} +
\sqrt{(m_{A-2}+K_{CM})^2} ~,
\label{ystar_par_A}
\end{equation}
where the initial nucleon momentum, $y^*$, is broken down into the
contributions from the C.M. motion of the SRC and the relative motion within
the pair: $y^*=K_{CM}/2+k_{rel}$. The underlying picture is the same as the
$y_{CW}$ approach, but rather than calculating the expected excitation, the
motion of the SRC is adjusted so that the convolution reproduces the observed
distribution, thus yielding an estimate of average contribution from the C.M.
motion as a function of $y^*$.  Figure~\ref{Fig:A_ystar} shows the result of
the analysis~\cite{Arrington:2003tw, Arrington:2006pn} using the modified
$y^*$ scaling variable. The shape of the $F(y^*)$ distribution for heavy
nuclei at large negative $y^*$ values is nearly identical to that of the
deuteron in the scaling (large $Q^2$) limit, and also provides a more
symmetric peak about $y^*=0$. This situation confirms the expectation that
high momentum tail of the nuclear momentum distribution is generated
predominantly by NN short-range correlations.  Note that use of the $y_{CW}$
variable yields very similar results, and that for the deuteron, both $y^*$
and $y_{CW}$ reduce to the standard definition of $y$.

While these modified scaling variables produce results that are consistent
with the SRC expectation, they require a model of the excitation of the
residual (A-1) system.  This is determined in a picture that is consistent
with the initial state assumed in the SRC picture, but in both cases the
correction is model dependent.  This makes it difficult to use this approach
to make precise, quantitative extractions. Similarly, there have been attempts
to apply corrections to the scaling function, rather the scaling variable, i.e.
$F(y,Q^2) = f(y,Q^2) - B(y,Q^2)$, where $F(y,Q^2)$ is the measured scaling
function, $B(y,Q^2)$ is the calculated ``binding correction'' for the
target, and $f(y,Q^2)$ is the ``corrected'' scaling function which can be
related to the nucleon momentum distribution in the
nucleus~\cite{CiofidegliAtti:1990rw,CiofidegliAtti:1999is}.  Note that for the
deuteron, $B(y,Q^2)=0$ and $F(y,Q^2)=f(y,Q^2)$.

In spite of the fact that these analyses require a model-dependent correction
to either the scaling variable or scaling function, they all yield results
consistent with the expectations of Eq.~(\ref{nk_src}), supporting the idea that
the high-momentum components have a universal behavior. If this behavior is
universal, then the \textit{ratio} of cross sections in the region dominated
by 2N SRCs should allow for an extraction of the relative strength of SRCs in
heavy nuclei and the deuteron~\cite{Frankfurt:1981mk, Frankfurt:1988nt,
CiofidegliAtti:2009qc}. For a perfectly defined $y$-scaling variable, the
ratio $F^{A}(y)/F^{d}(y)$ for $|y|>k_{Fermi}$ should be constant and yield a
measure of the relative contribution of SRCs in nucleus A relative to
deuterium, $a_2(A,Z)$ of Eq.~(\ref{nk_src}). However, the fact that a
model-dependent correction in the definition of $y$ is required for heavy
nuclei may yield a modified value in the extracted ratio or introduce an
artificial $y$ dependence. Thus, we would like to compare the high-momentum
tails in nuclei in a way that does not require such model-dependent
corrections.

To compare the relative strength of the high-momentum tail in various
nuclei, it is useful to examine the ratio of cross sections from different
nuclei as a function of $\alpha_i$ (Eq.~(\ref{incl_kin})) or, for inclusive
scattering, Bjorken $x$ (since $\alpha_i \to x$ at high $Q^2$).
To isolate two-nucleon SRCs, one
wants to be well above $x=1$, where single nucleons with $k \ltorder
k_{Fermi}$ dominate, but below $x=2$, above which the contributions of 3N SRCs
may become significant.  In this region, if the underlying SRC distributions
are identical, one expects to see the same $x$ dependence in the cross section,
with the ratio providing an indication of the relative contribution related to
SRCs.  In this case one can define the per-nucleon cross section ratio and
the ``isoscalar'' ratio:
\begin{eqnarray}
r(\hbox{A}, ^2\hbox{H}) = \frac{\sigma_{eA}/A}{\sigma_{ed}/2} ~,~
r_{iso}(\hbox{A}, ^2\hbox{H}) = \frac{\sigma_{eA}/(Z\sigma_{ep} + N\sigma_{en})}
{\sigma_{ed}/(\sigma_{ep}+\sigma_{en})} ~,
\label{a2}
\end{eqnarray}
where $r_{iso}$(A,$^2$H) accounts for the difference between the $e-p$ and
$e-n$ elastic scattering cross sections, assuming that the relative
contribution of protons and neutrons in the high-momentum tail scales like the
total number of protons and neutrons.  This was the assumption used in most
of the analyses that used the inclusive cross section ratios to extract the
relative contribution of the SRCs between heavy nuclei and the deuteron.
More recently, two-nucleon knock-out measurements (discussed in
Sec.~\ref{sec:triple}), suggested that np SRCs dominate the high-momentum tail,
as one expects if the high-momentum tails are dominated by the
isospin-dependent tensor interaction.  A dominance of iso-singlet np pairs
would yield identical proton and neutron contributions to the high-momentum
tail, such that the isoscalar correction does not need to be applied, and
the relative cross section per nucleon in the tails, $r$(A,$^2$H), yields
a direct measure of the relative strength of the high momentum tails in 
the nuclei. This will be discussed more at the end of this section where we
present results from these extractions.

The simple SRC model predicts that $r$(A,$^2$H)
will be independent of $x$ and $Q^2$ for $Q^2 \gtrsim 1.5$~GeV$^2$ and
$1.5< x < 2$.  The lower limit of $x$ depends on $Q^2$, and illustrated in
Fig.~\ref{fig:xpmin2nsrc}, while the upper limit on $x$
is to ensure the dominance of the 2N SRCs, which would be zero above $x=2$ for
stationary SRCs and die out rapidly above $x=2$ after accounting for motion of
the pair. In the region where scaling is observed, that is to say the ratio is
independent of $x$ and $Q^2$, Eq.~(\ref{nk_src}) allows us to relate the
magnitude of $r$(A,$^2$H) to $a_2(A)$, the relative (to the deuteron)
probability of a nucleon being part of a 2N SRC.

Examination of the ratio of high-$x$ scattering between two nuclei has an
additional advantage.  As discussed in Sec.~\ref{sec:deuteron}, if the
scattering occurs from one of a pair of nucleons that is very close together,
then the final state interactions between the nucleons need not fall 
rapidly with $Q^2$ and the cross section may be sensitive to these final
state interactions between nucleons in the SRC.  This could lead to
a modification of the cross section for both the deuteron and heavier nuclei.
However, because these FSIs occur \textit{within} the two-nucleon correlation,
it is expected that they will be identical for scattering from a 2N SRC in 
a heavy nucleus or for scattering from a high-momentum component in the
deuteron and should thus cancel in the A/D ratio. This is supported by more
detailed evaluations of the space-time properties of final state
interaction~\cite{Frankfurt:1993sp, Frankfurt:2008zv}, which demonstrate that
the distance the struck nucleon propagates before rescattering shrinks with
increasing $Q^2$, and is comparable to the size of NN correlations for
$Q^2>1$~GeV$^2$ and $x>1.5$, as discussed in Sec.~\ref{sec:fsi}.

In the PWIA, the momentum distribution in the high-momentum tail of a nucleus
relative to the deuteron is $a_2 = n_A(k)/n_2(k) = (2/A) \sigma_A/\sigma_2$
for large enough $x$ values.  As discussed in Sec.~\ref{sec:introduction},
this reduces to $r$(A,$^2$H)$=(2/A)\sigma_A/\sigma_{^2H}$ if only isosinglet
np pairs contribute, as one would expect if the tensor interaction dominates. 
This is under the assumption that the deuteron-like pairs in the nucleus
are at rest, and thus the high-momentum tails are unaffected by motion of
the 2N SRC in the nucleus.  The impact of the total SRC momentum will be
discussed later.

\begin{figure}[htb]
\begin{center}
\includegraphics[width=0.77\textwidth]{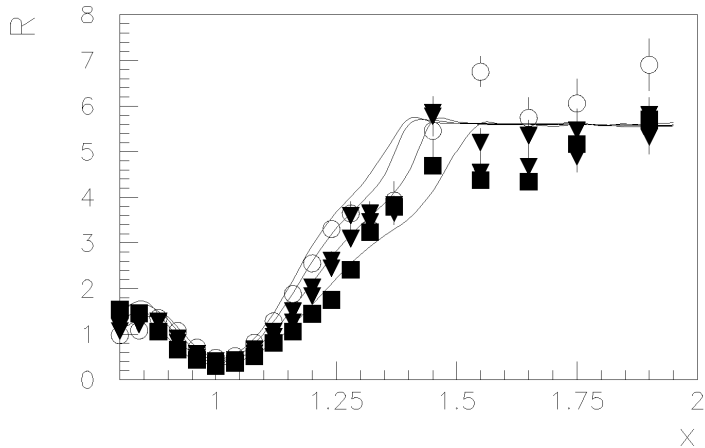}
\caption{The $x$ dependence of the ratio $r$(A,$^2$H) for Fe nucleus for
different values of $Q^2=1.2$--2.9~GeV$^2$.  The ratios and calculations
are taken from Ref.~\cite{Frankfurt:1993sp}.}
\label{fig:A_d}
\end{center}
\end{figure}

\begin{figure}[htb]
\begin{center}
\includegraphics[width=0.48\textwidth,angle=0]{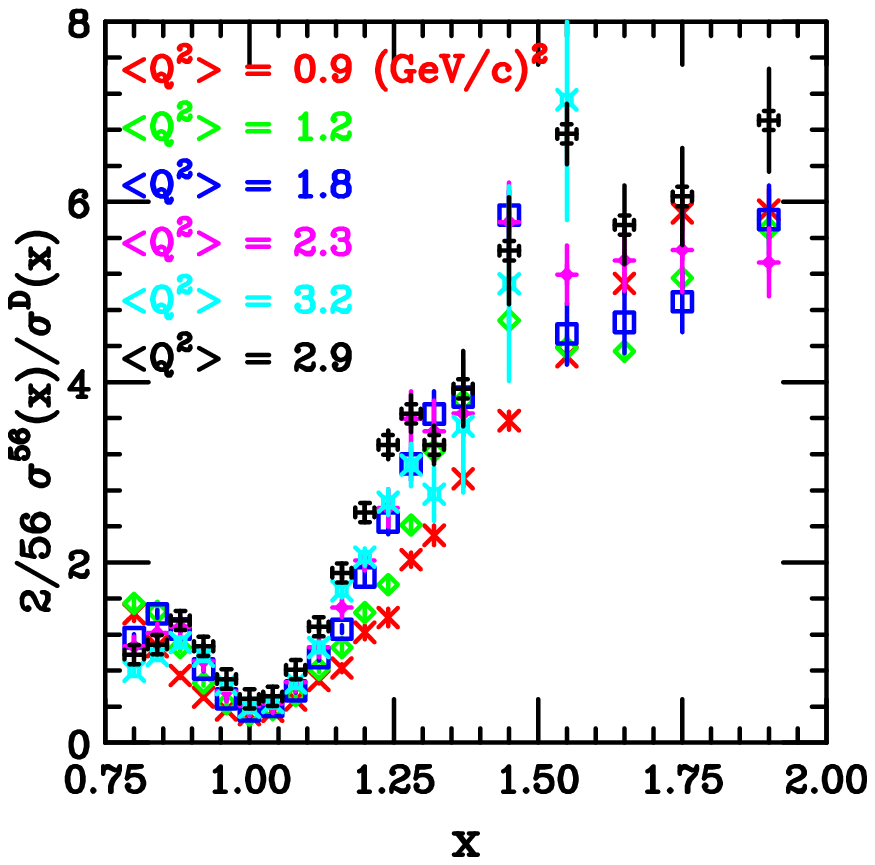}
\includegraphics[width=0.45\textwidth,angle=0]{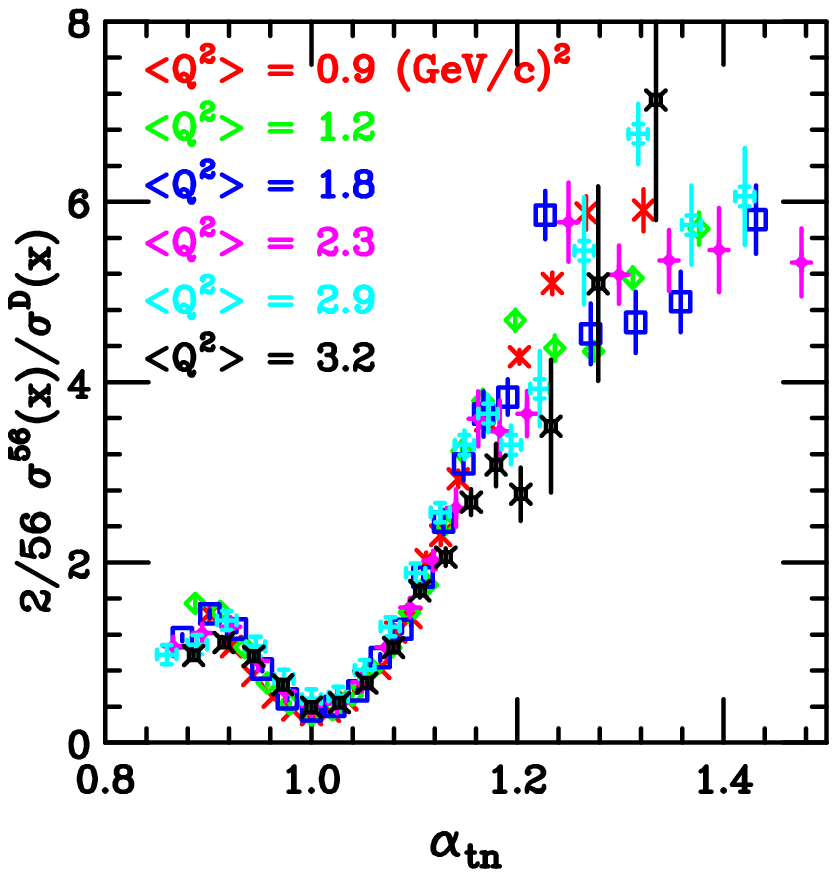}
\caption{Ratio of Fe to $^2$H for different $Q^2$ values, plotted as a
function of $x$ (left panel) and $\alpha$ (right
panel)~\cite{Frankfurt:1993sp}.}
\label{fig:x_vs_alpha}
\end{center}
\end{figure}

One of the first extractions of $r$(A,$^2$H)~\cite{Frankfurt:1993sp} combined
cross sections from multiple SLAC data sets to construct the ratios of
Eq.~(\ref{a2}). The statistics were rather limited
and the deuteron data were measured at different kinematics, so comparisons to
the other nuclei required nontrivial extrapolations. The result of these
analysis and a comparison with the theoretical prediction based on the 2N SRC
model is given in Fig.~\ref{fig:A_d}. Note that in their extraction of the
ratio, they applied an isoscalar correction to extract $r_{iso}$(A,$^2$H)
for $^3$He, where the correction is large, but neglected the small correction
for iron, for which they examine the ratio $r$(A,$^2$H).

As discussed in Sec.~\ref{kinematics:inclusive}, one expects reduced $Q^2$
dependence if the ratio is taken as a function of $\alpha_{tn}$, which 
is an approximation to the light cone momentum fraction of the struck nucleon
based on the assumption of interaction with a nucleon from a 2N SRC. The
data are shown as functions of $x$ and $\alpha_{tn}$ in
Fig.~\ref{fig:x_vs_alpha}.  There is a clear $Q^2$ dependence when taken
as a function of $x$ in the transition region between the QE peak ($x \approx 1$)
and the SRC-dominated region, while this is significantly smaller when
taken as a function of $\alpha_{tn}$.  Because $\alpha_{tn}$ is an approximation
to the real light cone momentum $\alpha_i$ based on the assumption of a pair
of nucleons with large relative momentum and zero total momentum, this again
supports the picture of scattering from a nearly at-rest SRC.

When the SLAC measurements were published, this comparison provided the
strongest indication of the existence of scaling in the ratios $r$(A,$^2$H).
Subsequent experiments at Jefferson Lab~\cite{Arrington:1998ps,
Arrington:1998hz, Fomin:2008iq, Fomin:2011ng} provided data for deuterium and
a variety of nuclei, extending to higher $Q^2$ with improved statistics, and
confirmed the observation of scaling in the target ratios. These data allowed
for further examination of the correlation between $Q^2$ and the $x_{min}$
value at which scaling of the ratio is observed (Fig.~\ref{fig:xpmin2nsrc}).

\begin{figure}[htb]
\begin{center}
\includegraphics[width=0.75\textwidth,angle=270]{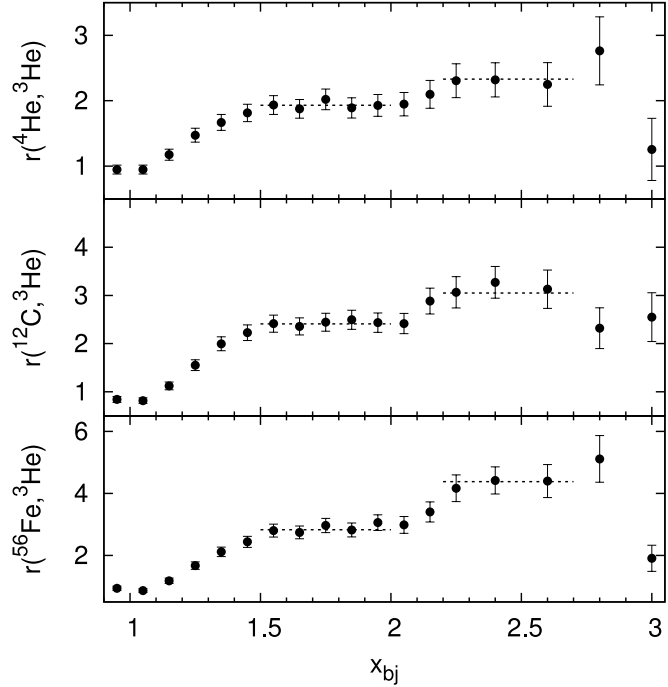}
\caption{Cross section ratios, $r_{iso}$ (Eq.~\ref{a2})) for (a) $^4$He/$^3$He
(b) $^{12}$C/$^3$He and (c) Fe/$^3$He from Ref.~\cite{Egiyan:2005hs} as a
function of $x_B$ for $Q^2>1.4$~GeV$^2$.  The horizontal dashed lines indicate
the two-nucleon and three-nucleon scaling regions used to calculate the
per-nucleon probabilities for two- and three-nucleon short-range correlations
in nucleus $A$ relative to $^3$He.}
\label{src:inclusive-ratios}
\end{center}
\end{figure}

Because the structure in the 2N SRC scaling region is expected to be
universal, one can make similar comparisons between any two nuclei.
Measurements using the CEBAF Large Acceptance Spectrometer (CLAS) in Hall B at
Jefferson Lab~\cite{Egiyan:2003vg, Egiyan:2005hs} looked at the ratio
$r_{iso}$(A,$^3$He) (Fig.~\ref{src:inclusive-ratios}).  The $Q^2$ coverage of
the data made it possible to examine the onset of scaling in the 2N SRC regime,
where a plateau was visible at $x>1.4$ for $Q^2$ values above 1.4~GeV$^2$.

\begin{figure}[htb]
\begin{center}
\includegraphics[width=0.55\textwidth,angle=270]{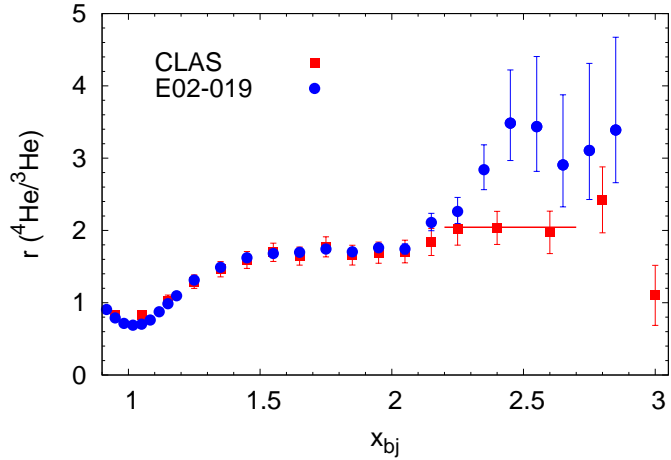}
\caption{Comparison of the $^4$He/$^3$He ratios from CLAS~\cite{Egiyan:2005hs}
and E02-019~\cite{Fomin:2011ng}.  Note that the isoscalar correction has
been removed from the CLAS ratios, and both data sets represent the direct
ratio of the measured cross section per nucleon.}
\label{src:clas_e02019}
\end{center}
\end{figure}

Examining the ratios of heavy nuclei to $^3$He also allows for an extension of
these measurements out beyond $x=2$, which may allow for the identification
of high-momentum configurations involving three nucleons which share
significant momentum (3N SRCs).  While data for $^3$He had been taken in
previous SLAC measurements, the rapid fall of the cross sections at large
$x$ meant that the statistics for $x>2$ were insufficient to provide a clear
extraction of such ratios, although an early extraction based on preliminary
cross section results was shown in Figure 8.3 of Ref.~\cite{Frankfurt:1988nt}.
The higher statistics of the second CLAS measurement~\cite{Egiyan:2005hs}
provided an improved look at scaling in the $x_B > 2$ region for
$Q^2>1.4$~GeV$^2$.  The data above $x=2.25$ are consistent with a constant
value, as shown in Fig.~\ref{src:inclusive-ratios}, which was interpreted
as a dominance of 3N SRCs.  Results from the recent measurement from Hall C at
JLab (E02-019)~\cite{Fomin:2011ng} are compared to the CLAS data in
Fig.~\ref{src:clas_e02019}.  The new data are in excellent agreement with the
CLAS measurement in the 2N SRC region, and differ slightly near the QE peak
due to the lower momentum resolution of the CLAS measurement.  Both
extractions have limited statistics, but the Hall C data show a
significantly larger ratio for $x>2$. The Hall C data in this region are taken
at $Q^2 \approx 2.9$~GeV$^2$, while the CLAS data cover a wide range in $Q^2$
with an average $Q^2$ value near 1.6~GeV$^2$, suggesting that the CLAS data
may be too low in $Q^2$ to cleanly isolate the signature of 3N SRCs. However,
the preliminary results from the early SLAC measurements (Figure 8.3 of
Ref.~\cite{Frankfurt:1988nt}) show no indication of a $Q^2$ dependence from 1
to 2.4~GeV$^2$, and yield a ratio in between the two JLab measurements.

Unfortunately, none of these measurements have data with sufficient statistics
and $Q^2$ range to precisely determine the kinematics at which scaling
sets in.  The difference could be related to a true $Q^2$ dependence in
the ratio or an inconsistency between the measurements.  While
$Q^2=1.4$~GeV$^2$ is sufficient to observe scaling in the 2N SRC region,
it is not as clear what $x$ and $Q^2$ values are required to isolate 3N SRC
contributions.  From Fig.~\ref{fig:xpmin2nsrc}, we observe that for $Q^2=1.5$
and $x>1.5$, the minimum initial nucleon momentum is above 250~MeV/c, and
thus one expects the single-particle contribution to the momentum distribution,
dominant at $k < k_{Fermi}$, to be strongly suppressed.  So it is not
surprising that 2N SRC contributions dominate in this region.  However, it is
not clear what $p_{min}$ value is required for the 2N SRC contributions to be
strongly suppressed, and it may be that a larger $Q^2$ value is required to
isolate 3N SRC contributions. An experiment was recently
completed~\cite{e08014} that focused on the $x>2$ regime with the goal of
mapping out the $x$ and $Q^2$ dependence with sufficient precision to confirm
the onset of scaling and determine the scaling threshold in the 3N SRC regime.

As mentioned above, one can use the cross section ratio in the 2N SRC regime
to determine the contribution of two-nucleon correlations in heavy nuclei
relative to the contribution in the deuteron. In Ref.~\cite{Egiyan:2005hs},
the probability of a nucleon-nucleon correlation in the deuteron was taken to
be the fraction of the deuteron momentum distribution above $k = 275$~MeV/c. 
These are momenta that have been shown in $\vec{D}(\vec{e}, e^{\prime}p)$
asymmetry measurements to be dominated by highly correlated
nucleons~\cite{Passchier:2001uc}.  Based on a calculation of the deuteron
structure~\cite{CiofidegliAtti:1995qe}, they find that $(4.1\pm0.8)$\%
of the time, the nucleons in deuterium had $k>275$~MeV/c, and were thus
assumed to be in a short-range, high-momentum configuration. Taking this as a
measure of the contribution of 2N SRCs in deuterium, and using the cross
section ratios in the plateau region as a measure of the relative contribution
of 2N SRCs in heavier nuclei, one can extract the absolute contribution for
the nuclei where $r$(A,$^2$H) measurements exist.  One can also combine the
CLAS $r_{iso}$(A,$^3$He) measurements with a measured or calculated value for
$r$($^3$He,$^2$H).  The latter was taken to be $r$($^3$He,$^2$H)=1.97$\pm$0.1
based on the average of the SLAC extraction~\cite{Frankfurt:1993sp}
(1.7$\pm$0.3), and a calculated value~\cite{Forest:1996kp} (2.0$\pm$0.1).

A similar extraction of the contribution of 3N SRCs was also performed,
yielding a contribution of less than 1\% for carbon or
iron~\cite{Egiyan:2005hs}.  However, as noted above, it is not clear at
this point if this measurement was able to isolate the 3N SRC contributions,
and the more recent data~\cite{Fomin:2011ng} suggest a larger contribution
from 3N SRCs.

In extracting relative contributions of SRCs from the cross section ratios,
it is assumed that final state interactions do not modify the ratios in the
scaling region.  This is expected to be valid for sufficiently high $Q^2$, as
interactions between the struck nucleon and spectators at large distance
scales will fall rapidly with $Q^2$, and interactions with nucleons that are
very close (i.e. the other nucleon(s) in the SRC) should be identical for SRCs
in light or heavy nuclei, and thus cancel in the ratio.  While we expect that
FSI contributions should not impact these extractions, a more quantitative
evaluation, especially for the data at modest $Q^2$ values, would be
worthwhile.

At this point, we reiterate that, except for the recent extraction from the
Hall C measurement at 6 GeV~\cite{Fomin:2011ng}, these extractions are based on
the simple SRC model, where the scattering is assumed to have occurred from a
nucleon in a stationary 2N (or 3N) SRC, where the 2N SRCs are assumed to be
isospin-independent and thus of equal importance for pp, np, and nn pairs.
In reality, the SRCs have small but non-zero total momentum and the NN
potential that generates the bulk of the high-momentum pairs is strongly
isospin-dependent and accounting for this will modify the results for the
extracted 2N and 3N SRC probabilities.  These effects require corrections
to the standard extraction of the probability of nucleons being found in SRCs
in nuclei.

First, these values were obtained from
measurements of $r_{iso}$(A,$^3$He), as defined in Eq.~(\ref{a2}), which
assumes that the contribution of high-momentum protons and neutrons scales
with $Z$ and $N$. For the case of isoscalar dominance, the tensor interaction
is the primary source of high momentum nucleons, and so only np SRCs are
formed.  In this case, the neutron and proton have equal high-momentum
distributions and this correction should not be applied.
Note that using the raw $r$(A,$^3$He) ratio means assuming total dominance
of np pairs, while calculations and the triple-coincidence measurements 
yield small but non-zero contributions from iso-triplet pairs.  Thus, while
using the uncorrected ratio is a much better approximation than using
$r_{iso}$(A,$^2$H), it is probably a small underestimate of the more true
value.

Table~\ref{src:inclusive-table} includes values for $r$(A,$^3$He) as published
(with the isoscalar correction), and with the isoscalar correction removed. 
The raw cross section ratios are 15--18\% below the values with the isoscalar
correction applied. If the conversion from A/$^3$He ratios to
A/$^2$H ratios had been done using the measured $^3$He/$^2$H ratio from SLAC,
which applied the same isoscalar correction to $^3$He, then only the small
factor applied to the non-isoscalar Fe target would have to be removed.
However, in the analysis of Ref.~\cite{Egiyan:2005hs}, the value used is
$r$($^3$He/$^2$H)=1.97, which is mainly based on a calculated
ratio~\cite{Forest:1996kp} and which does not include the isoscalar correction
factor applied in the A/$^3$He measurements.  Note that the SLAC results
quoted for $^3$He, $r_{iso}$($^3$He/$^2$H)=1.7$\pm$0.3 correspond to
an uncorrected ratio of $r$($^3$He/$^2$H)=2.0$\pm$0.3, in good agreement with
the calculation of Ref.~\cite{Forest:1996kp}.

\begin{table}[htb]
\begin{center}
\caption{\label{src:inclusive-table}
Shown are the average ratios, $r$(A/$^3$He) for the scaling regions
$1.5<x_B<2$ from the CLAS data as published~\cite{Egiyan:2005hs}, and after
removing the isoscalar correction that was applied (``raw''). For the CLAS
data, the A/$^3$He cross section ratio is converted to A/$^2$H, taking
$r$($^3$He/$^2$H)=2.0 as extracted from the raw $^3$He/$^2$ data from
SLAC~\cite{Frankfurt:1993sp} (no isoscalar correction) or the calculation of
Ref.~\cite{Forest:1996kp}. $R_{2N}(A)$ is extracted from the $r$(A,$^2$H) by
applying a correction for the CM motion of the pair as estimated in
Ref.~\cite{Fomin:2011ng}, which yields a relative 2N SRC contribution
10--20\% lower than the raw cross section ratio.}
\begin{tabular}{lcc|ccc}
          &$r_{iso}$(A/$^3$He)  & $r$(A/$^3$He) & $R_{2N}$(A) & $R_{2N}$(A) 	& $R_{2N}$(A) \\
	  & (published)	  & (raw)	  & 	CLAS	  & SLAC	& E02-019 \\ \hline
$^3$He    & -             & -		  & -		  & 1.8$\pm$0.3	& 1.93$\pm$0.10 \\
$^4$He    & 1.93$\pm$0.17 & 1.69$\pm$0.15 & 2.85$\pm$0.29 & 2.8$\pm$0.4	& 3.02$\pm$0.17	\\
$^9$Be	  & -		  & -		  & -		  & 		& 3.37$\pm$0.17 \\
$^{12}$C  & 2.49$\pm$0.15 & 2.11$\pm$0.18 & 3.55$\pm$0.35 & 4.2$\pm$0.5	& 4.00$\pm$0.24 \\
Fe(Cu) 	  & 2.98$\pm$0.18 & 2.38$\pm$0.19 & 3.96$\pm$0.38 & 4.3$\pm$0.8	& (4.33$\pm$0.28) \\
Au	  & -		  & -		  & -		  & 4.0$\pm$0.6	& 4.26$\pm$0.29 \\
\end{tabular}
\end{center}
\end{table}

Once we have removed the isoscalar correction, the results in
Table~\ref{src:inclusive-table} represent the raw A/$^3$He per-nucleon cross
section ratio, which should correspond to the relative strength in the
high-momentum tails.  If the 2N SRC is at rest, then the relative strength
in the high-momentum tails corresponds to the relative probability for a
nucleon to be in a 2N SRC.  If we relax this constraint, and allow for the
pair to have a non-zero total momentum, then the direct connection between
the high-momentum contribution and the relative strength of SRCs is broken.
The high-momentum tail of the deuteron-like pair can be ``smeared'' by the
pair motion, yielding an overall increase in the high-momentum tails.  Note
that this is a real enhancement of the high-momentum tails in the nucleus, so
the ratio still provides a direct measure of the tail of the momentum
distribution.  This correction simply separates the enhancement of the
high-momentum tail into a relative probability of finding SRCs in the nucleus
and an enhancement of the SRC high-momentum tail due to motion of the pair.

The smearing effect was expected to be
$\sim$20\% for heavy nuclei~\cite{CiofidegliAtti:1995qe}.  While this effect is
discussed in Ref.~\cite{Egiyan:2005hs}, no correction or uncertainty is
applied when extracting the relative 2N SRC contribution from the ratios.
In the analysis of the E02-019 data~\cite{Fomin:2011ng}, a correction for
the impact of the CM motion is applied when extracting the relative
contribution of 2N SRCs.  Table~\ref{src:inclusive-table} shows the 
SLAC, CLAS, and E02-019 data extractions of $R_{2N}(A)$ which is the
ratio $r$(A,$^2$H) with the CM correction ((20$\pm$6)\% for iron) removed,
and is related to the relative number of SRCs, as opposed to their relative
contribution to the high-momentum tails.
The same correction is applied to all experiments, and the CLAS ratios
are converted from $r$(A,$^3$He) to $r$(A,$^2$H) taking $r$($^3$He/$^2$H)=2.0.

The question of the isospin structure of the 2N SRCs has been examined in
recent triple-coincidence reactions, discussed in Sec.~\ref{sec:triple}, and
will be further studied in inclusive scattering from nuclei of similar mass
but with different N/Z ratios, e.g. recent measurements of $^{40}$Ca and
$^{48}$Ca at Jefferson Lab~\cite{e08014} and future measurements planned for
$^3$H and $^3$He.  It is less clear what impact this would have in the $x>2$
regime, where one is attempting to isolate 3N SRCs.

While there are still some issues to be resolved, the significant body of
inclusive data already taken demonstrates that it is feasible to kinematically
isolate scattering from 2N SRCs.  We have mapped out the strength of the 2N
SRCs as a function of A for several nuclei.  The inclusion of $^9$Be in the
most recent measurement~\cite{Fomin:2011ng} demonstrates an anomalous behavior in the A dependence
of the strength of the 2N SRC contributions.  For non-interacting nucleons,
one would expect that the probability for two nucleons to be close enough to
interact via the short-range tensor or repulsive core of the NN interaction
would be proportional to the nuclear density.  However, while $^9$Be is a very
low density nucleus, comparable to $^3$He, the extracted value of $R_{2N}(A)$
is much larger, on par with the significantly denser $^4$He and $^{12}$C.  This
is similar to the behavior observed in measurements of the EMC effect in these
same light nuclei~\cite{Seely:2009gt}, which was interpreted as an effect of the
strong cluster structure in $^9$Be.  One would expect a similar effect here,
where clustering among the nucleons in the nucleus will yield an increase in
the fraction of the time that nucleons are close together, even though the
average density can be very low if the clusters are spread apart.  Thus, this
observation reinforces the interpretation of the A dependence of the EMC effects.

While the results from the initial extension of inclusive measurements to the
three-nucleon correlation region have yet to demonstrate a clear sensitivity
to 3N SRC, a recently completed experiment~\cite{e08014} will address this
question in detail.  Future inclusive measurements will map out the strength
of 2N and 3N SRCs in a large set of light and medium nuclei~\cite{e1206105}.
These data may also be able to provide information on the isospin structure
of three-nucleon configurations by comparing $^3$H to $^3$He~\cite{e1211112}.
While the isospin structure of the 3N SRC is well defined for these nuclei,
if the momentum sharing is not identical for the singly- and doubly-occurring
nuclei, one will see a modified ratio in the 3N SRC region.

However, while we have learned a great deal from these inclusive measurements,
more complex reactions, in which one or more nucleons from the SRC are
detected, are important for providing more detail on the structure of the
hadrons and additional information on questions such as the center-of-mass
motion and isospin structure of the SRCs.  These issues will be discussed in
the next sections.

\section{Semi-Inclusive A(e,e$^\prime$N) Knock-out Reactions beyond the Deuteron}
\label{sec:knockout}

Semi-inclusive A(e,e$^\prime$N) reactions on nuclei, in which the struck nucleon is
detected in coincidence with the scattered electron, provides the next highest
level of complexity along with new details of the SRC structure.  The
conditions required for isolation of the structure of SRCs are discussed in
Sec.~\ref{kinematics:coincidence}. While inclusive reactions within the PWIA
framework can provide some information about the momentum distribution of the
nucleon, semi-inclusive reactions can, in principle, fully probe the nuclear
spectral function. In the PWIA, the spectral function is related to the
measured cross section as follows~\cite{Benhar:2006wy}:
\begin{equation}
{d\sigma\over d\Omega_{e^\prime}dE_{e^\prime} d^3p_N d E_m }=
{F_N\over F_A}\sigma_{eN}\cdot S_A(p_m,E_m) ~,
\label{eepr-ttt}
\end{equation}
where $F_A$ is the nuclear flux factor and $F_N$ is the flux factor calculated
for moving bound nucleon with momentum $\vec p_i= - \vec p_m$,
and $\sigma_{eN}$ is the cross section for scattering from the bound nucleon.
Within the PWIA, the nuclear spectral function, $S(p_m,E_m)$, can be related to
the ground state ($\psi_A$) and residual ($\phi_{A-1}$) nuclear wave functions
as follows:
\begin{equation}
S_A(p_m, E_m) = \mid \langle\psi_{A-1}|\delta(H_{A-1}-E_m)a(p_i)|\psi_{A}
\rangle \mid ^2 ~,
\label{spectral}
\end{equation}
where $a(k)$ is an annihilation operator for the nucleon with momentum $k$.
Several remarkable properties of the spectral function define the specific
utility of the semi-inclusive reactions:

\begin{itemize}

\item The cross section is sensitive to the composition of the (A-1) residual
state through the reconstruction of the missing energy, $E_m$.  This provides
a measure of the excitation of the residual system, allowing us to separate
a one-body spectator system consisting of an (A-1) nucleus (ground state or
excited state) or a multi-body final state.

\item For the case of the coherent (A-1) state, the parameter $E_m$
represents the nucleon removal energy of the particular nuclear shell.

\item At a fixed value of $E_m$ associated with a specific nuclear shell,
the $p_m$ dependence of the spectral function is related to the momentum
distribution of the nucleon in the given nuclear shell.

\item The occupation number of the given shell (labeled $\alpha$) can be
extracted:
\begin{equation}
n_\alpha = \int\limits_{\Delta E_m}\int S(p_m,E_m) d^3p_m dE_m ~,
\label{nal}
\end{equation}
where integration over the $E_m$ covers the missing energy width of the given
nuclear shell.

\end{itemize}

Historically the measurements of the occupation number in
Eq.~(\ref{nal})~\cite{Kelly:1996hd, Lapikas:1003zz} gave the first glimpse of
the role that correlations play in the redistribution of probabilities in the
ground state nuclear wave functions. Namely the experimental measurements
revealed a substantial missing strength in the values of $n_\alpha$ for all
shells over a range of nuclei, which was attributed in part to the short-range
interaction of the nucleon at the given shell.

Probing SRCs directly in high-$Q^2$ A(e,e$^\prime$N) reactions requires of taking
the spectral function to large values of missing momentum ($p_m> k_{Fermi}$)
and energy ($E_m> 100$~MeV).  In this case, $E_m$ characterizes the nucleon
removal energy in the continuum (rather than particular nuclear shell).  From
the practical point of view, the more relevant quantity will be the nuclear
recoil energy $E_R$ defined in Sec.~\ref{kinematics:coincidence}. The
advantage of considering $E_R$ is that the most prominent and new signature of
SRCs in coincidence reactions (as compared to the inclusive scattering) is the
correlation relation between $E_R$ and missing momentum $p_m$ according to
Eq.~(\ref{empmcor}):
\begin{equation}
E_R \approx \sqrt{m_N^2 + p_m^2} - m_N ~.
\label{empmcor2}
\end{equation}
Note however that this is merely a kinematical correlation and its observation
only indicates the dominance of the two-body currents, but not necessarily 2N
SRCs.  Such a correlation was observed for example in $^4$He(e,e$^\prime$p) data at
NIKHEF~\cite{vanLeeuwe:2001uc, vanLeeuwe:1997wq} at $Q^2=0.34$~GeV$^2$.
However, the theoretical analysis demonstrated that the strength of the cross
section was dominated by the long range two-body currents such as MEC and IC.

To go beyond the correlation relation of Eq.~(\ref{empmcor2}) and probe the
part of the spectral function dominated by SRC, the factorized relation
between the electron-nucleon cross section and the spectral function must be
valid. For A(e,e$^\prime$p) scattering, this factorization was studied in
Ref.~\cite{Sargsian:2005ru, Sargsian:2004tz} and it was demonstrated that in
the limit of $Q^2 \gg p_i^2$, Eq.~(\ref{eepr-ttt}) is valid, even when
including final state interactions within the eikonal approximation. 
The FSI do not become negligible, even at very large $Q^2$, and so detailed
experimental and theoretical studies, such as the deuteron measurements
discussed in Sec.~\ref{sec:deuteron} are critical.  While significant FSI
always remain, their implementation is simplified at high energy. In this
case the spectral function is defined within the distorted wave impulse
approximation~(DWIA) ($S^{DWIA}_A(p_m,E_m,q)$) and depends strongly on the
relative angle of $\vec p_m$ and $\vec q$ (see e.g.
Ref.~\cite{Sargsian:2005ru} and Figs.~\ref{Fig:kim_250_500}
and~\ref{fig:boeglin}). The advantage of the DWIA is that if its
application is justified, then one can use the knowledge of $\sigma_{eN}$
obtained from the studies of exclusive D(e,e$^\prime$N)N reaction
(Sec.~\ref{sec:deuteron}) at given virtuality of the bound nucleon to extract
$S^{DWIA}_A(p_m,E_m,q)$ from Eq.~(\ref{eepr-ttt}).

\begin{figure}[htbp]
\begin{center}
\includegraphics[width=0.65\textwidth]{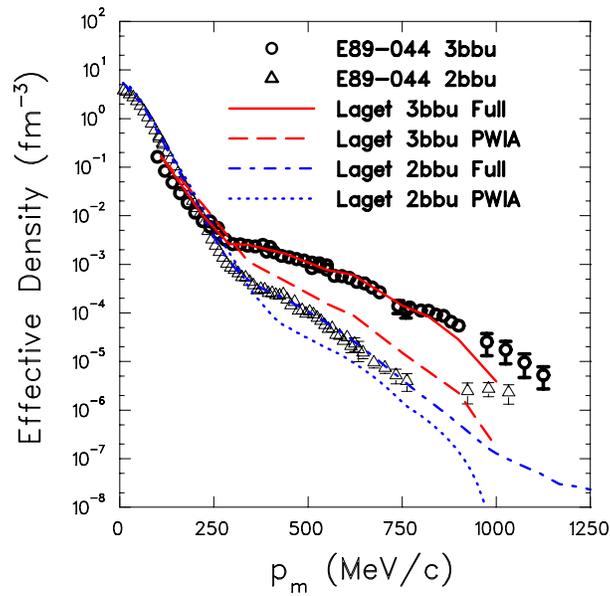}
\caption{\label{src:he3density} Proton effective momentum density distributions
extracted from $^3$He(e,e$^\prime$p)pn three-body break-up (3bbu - open
black circles) and the $^3$He(e,e$^\prime$p)D two-body break-up (2bbu - open
black triangles)~\cite{Benmokhtar:2004fs}.  The 3bbu integration covers $E_M$
from threshold to 140~MeV.  The curves are calculations from
J.-M.~Laget~\cite{Laget:2004sm} which show dominance of the continuum cross
section at large missing momentum, along with strong FSI contributions. Figure
reprinted with permission from Ref.~\cite{Benmokhtar:2004fs}}
\end{center}
\end{figure}

The first measurement of high-$Q^2$ semi-inclusive reactions at $p_m$ values
corresponding to probing SRCs was performed at
Jefferson Lab using a $^3\mbox{He}$ target. Measurements of the
$^3$He(e,e$^\prime$p)D and $^3$He(e,e$^\prime$p)pn reactions were taken in
Hall A with a beam energy of 4.8~GeV, for $Q^2 = 1.5$~(GeV/c)$^2$ and $x_B
\approx 1$. When this experiment was proposed, it was expected that these
kinematics would cleanly isolate short-range correlations at missing momenta
greater than 300~MeV/c.  What was observed was a much greater strength in the
high missing momentum region then expected~\cite{Benmokhtar:2004fs,
Rvachev:2004yr} as shown in Fig.~\ref{src:he3density}. In this figure, the
three-body break-up (3bbu) has been integrated over missing energy so that it
could be plotted together with the two-body break-up (2bbu) and the strengths
of the two reactions compared. The measurements clearly demonstrated the
dominance of the three-body break-up channel at high $p_m$,
as expected in the SRC picture.

One important result of this measurement was that it demonstrated the validity
of the eikonal approximation in calculating the final state interaction, which
provides the dominant contribution at high $p_m$. The enhanced cross section
was explained as an interference between correlations in the initial state and
final-state interaction~\cite{Laget:2004sm}, where neither effect alone could
explain the observed cross section.  Subsequently, calculations within the
generalized eikonal approximation (GEA)~\cite{degliAtti:2005dh,
CiofidegliAtti:2005qt, CiofidelgiAtti:2007qu, Alvioli:2009zy} which used the
realistic spectral function of Ref.~\cite{Wiringa:1994wb}, in which
correlations are included in a self-consistent fashion, achieved a
parameter-free description of the data.

\begin{figure}[htp]
\begin{center}
\includegraphics[width=0.65\textwidth]{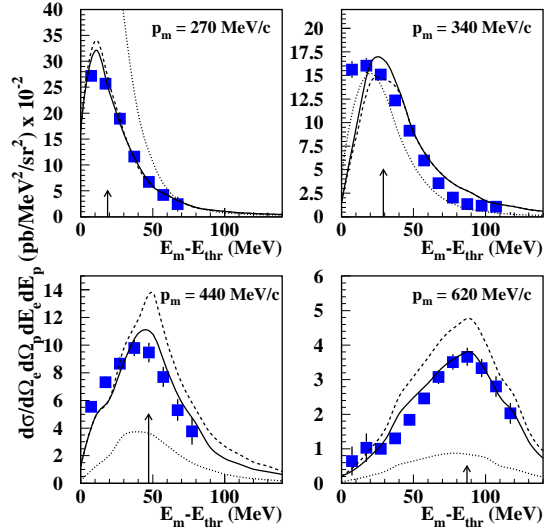}
\caption{The dependence of the differential cross section on the missing
energy for $^3\mbox{He}$ three-body break-up reactions at different values of
initial nucleon momenta ($p_i$ from Eq.~(\ref{pwia_kin})). Dotted, dashed and
solid curves corresponds to PWIA, PWIA + single rescattering and PWIA + single
+ double rescatterings~\cite{Sargsian:2004tz, Sargsian:2005ru}. Data are from
Ref.~\cite{Benmokhtar:2004fs}. Arrows define the correlation according to
Eq.~(\ref{empmcor2}). Similar description of the data is achieved in
Refs.~\cite{CiofidegliAtti:2005qt,Laget:2004sm}.
Figure reprinted with permission from Ref.~\cite{Frankfurt:2008zv}}
\label{3bbu}
\end{center}
\end{figure}

Another important result of the above mentioned measurements is that in the
three-body break-up channel, the clear correlation between $E_R$ and $p_m$ of
Eq.~(\ref{empmcor2}) is observed, as seen in Fig.~\ref{3bbu}. The comparison of
theoretical calculations based on generalized eikonal approximation within
DWIA~\cite{Sargsian:2004tz, Sargsian:2005ru} with the data of
Ref.~\cite{Benmokhtar:2004fs} demonstrated that the large contribution from
final state re-interaction not only preserves the pattern of the correlation
of Eq.~(\ref{empmcor2}) but actually reinforces it. This indicates a rather
new phenomenon, that in high energy kinematics sensitive to SRC, FSI is
dominated by single rescattering of the struck nucleon with a spectator
nucleon from the SRC. As a result, this rescattering does not destroy the
correlation structure of the spectral function, as it is related to 2-body
interactions contained within the SRC.  The validity of this assertion
is crucial for understanding the role of the FSI in the $x$-scaling of the
inclusive cross section ratios at $x>1$ (Sec.~\ref{sec:inclusive}). If FSI is
confined within the SRC, then it will cancel out in the ratios of
Eq.~(\ref{a2}) without distorting the factorization of the 2N SRC from
the (A-2) residual nuclear system.

The fact that the eikonal approximation provides a satisfactory description
of the data for $Q^2$ as low as 1.5~GeV$^2$ demonstrates the feasibility of
working in a well understood kinematic region with experimentally feasible
kinematic restrictions. To suppress the absolute magnitude of FSI in the
semi-inclusive cross section, one requires similar kinematic restrictions to
those discussed in Sec.~\ref{kinematics:coincidence} (see also
Refs.~\cite{Rohe:2005vc, Barbieri:2004up}). The study of the
issue of the suppression in the final state interaction at $x>1$ kinematics
within generalized eikonal approximation~\cite{Frankfurt:1996xx,
Sargsian:2001ax} demonstrated that an additional restriction needs to be
imposed on measured magnitudes of missing momentum and recoil energy $E_R$
such that:
\begin{equation}
|p_m|-{q_0\over |q|}E_R > k_{Fermi} ~.
\label{nures}
\end{equation}
Thus it will be very important to extend the $^3$H(e,e$^\prime$p) experiments
to the $x>1$ domain, which will allow us to verify the prediction of FSI
suppression. Equation~(\ref{nures}) is a consequence of the eikonal regime of
FSI, and no such condition exists for FSI at low energy. In the high-$Q^2$
region where the eikonal regime is established, A(e,e$^\prime$p) experiments will be
able to verify that this condition is sufficient.  If so, then it should
be possible to extract the spectral function in a region where FSI are small
and can be reliably calculated.

In Jefferson Lab's Hall C, A(e,e$^\prime$p) measurements done in parallel kinematics
determined the strength of the high momentum region instead of simply
inferring it from the absence of the strength seen in the low missing momentum
region.  This was the first proton knock-out measurement aimed at extracting
the high-momentum component at kinematics where the corrections to the PWIA
appear to be small. However, because of the $x<1$ kinematics probed in this
reaction, the large missing energy range was restricted due to inelastic
contributions (as seen in Fig. 3 of Ref.~\cite{Rohe:2004dz}).  Nonetheless,
the observed strength at large $p_m$ in the region that was free of large
contributions from other reactions was found to be in reasonable agreement
with the theoretical expectation~\cite{Benhar:1989aw, Muther:1995bk}

Comprehensive studies of semi-inclusive reactions on light nuclei at $x>1$
kinematics are crucial for extending such measurements reliably to medium to
heavy nuclei.  At present, there are two approaches being used to further
these measurements.  Some experiments have been performed in kinematics which
minimize the reaction mechanism effects in order to focus on probing the
structure of the SRCs, while others are aimed at better understanding the FSI,
MEC, and IC, so that models of these contributions can be constrained well
enough to reliably correct for such effects in kinematics optimized to reduce
their impact.  To date, the studies focused on more fully understanding the
reaction mechanism have focused on the deuteron~\cite{Ulmer:2002jn, 
Egiyan:2007qj, Boeglin:2011mt}, with some additional measurements on
$^3$He~\cite{Benmokhtar:2004fs, Rvachev:2004yr}.

\section{Triple-coincidence Reactions}
\label{sec:triple}

A newer approach to resolving the detailed structure of short-range correlations
involves making nucleon-knockout measurements in which two outgoing nucleons
are detected, either in photo-nuclear reactions~\cite{Hehl:1995iq} or in
coincidence with the scattered electron in electroproduction
reactions~\cite{Kester:1995zz, Blomqvist:1998gq, Onderwater:1998zz,
Groep:2000cy}. For most of these initial measurements, two protons were
detected so as to minimize meson exchange effects, since only neutral mesons
can be exchanged between two protons.  However, since the energy and momentum
transfer were only few hundred MeV, it was practically impossible to
distinguish the struck and recoil nucleons.  Theoretical studies of these
reactions demonstrated that in fact the dominant contribution comes from
a two-step processes where there is an electromagnetic interaction with one
proton, with the second proton knock-out coming from IC or FSI
effects~\cite{Ryckebusch:1993tf, Ryckebusch:1994zz, Giusti:2007fn}. Even
though these reactions demonstrated sensitivity to high momentum component of
the nuclear wave function~\cite{Ryckebusch:1995ze}, the SRC evidence was
indirect and the dominant mechanisms were final state interactions and
long-range nucleon correlations~\cite{Ryckebusch:1989nn}. It is interesting to
note that two-nucleon knock-out reactions in which the outgoing nucleons carry
approximately same momenta are produced mainly due to the final state
interactions, even in the large momentum transfer
region~\cite{Frankfurt:1999ik, Sargsian:2008zm, Pomerantz:2009sb}.

Thus, the kinematics which can maximize the contributions from SRC should
correspond to the ``asymmetric'' situation in which one nucleon (final
momentum $p_N$) can be identified as a struck nucleon with momentum close to
the large (few GeV/c) momentum transfer and the another (final momentum $p_r$)
identified as the recoil (spectator) nucleon from within the SRC, with
momentum exceeding characteristic Fermi momentum of the nucleus, but well
below the scale of the momentum transfer, as discussed in
Sec.~\ref{kinematics:triple}.

In this case, one can characterize the A(e,e$^\prime$$N_N$$N_r$) cross section in the
PWIA as follows:
\begin{equation}
{d\sigma\over d\Omega_{e^\prime}dE_{e^\prime} d^3p_N d d^3p_r} =
{F_N\over F_A}\sigma_{eN}\cdot D_A(p_i,p_r,E_r) ~,
\label{eepr}
\end{equation}
where the decay function $D_A(p_i,p_r,E_r)$~\cite{Frankfurt:1988nt,
Sargsian:2005ru, Frankfurt:2008zv} :
\begin{equation}
D_A(p_i,p_r, E_R) = \mid \langle \psi_{A-1}
a^\dagger(p_r)|\delta(H_{A-1}-(E_R-T_{A-1})a(p_i)|\psi_{A}\rangle
\mid ^2 ~,
\label{decay}
\end{equation}
describes the probability that after a nucleon with momentum $p_i$ is
instantaneously removed from the nucleus, the residual (A-1) nucleon system
will have excitation energy $E_m = E_R-T_{A-1}$ and contain a nucleon with
momentum $p_r$. The integration of this function by $d^3p_r$ represents the
part of the nuclear spectral function (Eq.~(\ref{spectral})) corresponding to
the case of the break-up of (A-1) residual nucleus with at least one nucleon
in the continuum.  Note that, as in the case of A(e,e$^\prime$N) reactions, if
factorization of the nucleon electromagnetic current is justified then the
decay function can be generalized within the DWIA to account for the final
state interaction effects.

If a 2N SRC is probed, one of the prominent signatures will be that the decay
function will exhibit strong correlation between $p_i$ and $p_r$, peaking at
\begin{equation}
\vec p_i \approx - \vec p_r ~,
\label{pi_pr}
\end{equation}
as discussed in Sec.~\ref{kinematics:triple}. The first attempt to probe such
angular correlation was made in the two-proton knockout experiment of
Ref.~\cite{Blomqvist:1998gq} in which the $^{12}C(e,pp)X$ cross section was
measured as a function of relative angle of two outgoing protons for parallel
kinematics. Even though theoretical calculations showed sensitivity of the
angular function to the SRCs, no clear correlation signature was observed due
to impossibility of separating struck and spectator proton in low momentum
transfer kinematics.

\begin{figure}[hbt]
\begin{center}
\includegraphics[height=0.40\textheight,width=0.7\textwidth]{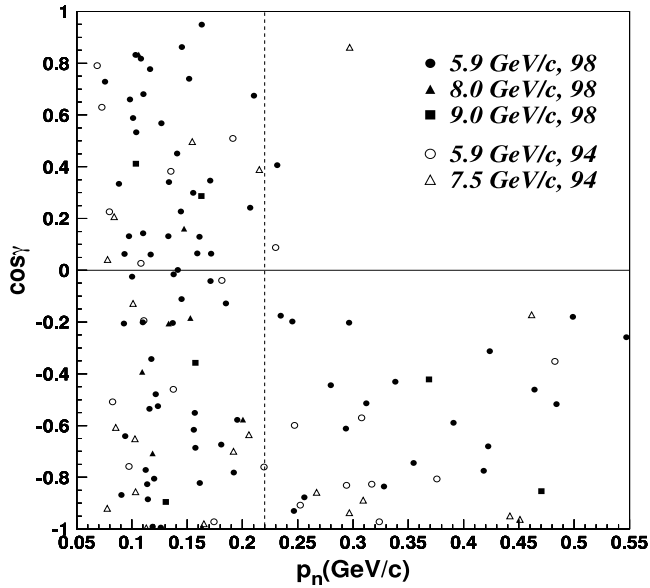}
\caption{The correlation between the magnitude of the recoil neutron momentum
$p_n$ and its direction $\gamma$ relative to $\vec p_i$. The legend gives
the initial proton beam momentum and year of the run, with hollow points
(labeled 94) from Ref.~\cite{Aclander:1999fd} and solid points (labeled 98)
from Ref.~\cite{Tang:2002ww}. In all cases, no angular correlation is observed
for neutron momenta less than $k_{Fermi}$, while neutrons with larger momentum
are produced preferably in the backward hemisphere of the reaction.}
\label{p2pn}
\end{center}
\end{figure}

The first time such an angular correlation was observed unambiguously in the
high momentum transfer regime was the $p+^{12}\mbox{C} \rightarrow p + p + n +
X$ measurement of E850 at Brookhaven National Lab~\cite{Aclander:1999fd,
Tang:2002ww}. The experiment looked for recoil neutrons produced in
coincidence with single proton knockout, and the results are shown in
Figure~\ref{p2pn}. Neutrons with momenta $p_n$ below the Fermi momentum were
uniformly distributed in $\cos{\gamma}$, where $\gamma$ is the angle between
the neutron momentum and the reconstructed initial momentum of the struck
proton, $p_i$. Neutrons above the Fermi momentum were almost exclusively
observed for $\cos{\gamma}<0$, i.e. opposite to the direction of $\vec p_i$.
This is consistent with the assumption that high-momentum nucleons are the
results of SRCs, in which the two nucleons have large relative momenta but a
small total momentum, and thus have initial momenta that are large and
back-to-back. Using the data from Ref.~\cite{Tang:2002ww}, they extracted
the probability that a neutron with $p_n > 220$~MeV/c $\approx k_{Fermi}$ is
found in association with a proton with $p_i > 220$~MeV/c, after correcting for
acceptance and efficiency. They determined that ($49\pm13$)\% of events with a
fast proton had an associated backwards-going fast neutron.

This is essentially the probability that the fast recoil neutron is observed in
coincidence with the fast proton.  However, a pn pair with large relative
momenta may not yield a high-momentum neutron in the final state; the neutron
could be reabsorbed as it passes through the nucleus or the pn pair might be
moving in the mean field of the (A-2) nucleus such that the final neutron
momentum in the rest frame of the nucleus is below $k_{Fermi}$. A more detailed
analysis of the above experiment was made in Ref.~\cite{Piasetzky:2006ai},
which was based on the modeling of the spectral and decay functions of the
reaction within the light-cone approximation~\cite{Yaron:2002nv}. This
analysis allowed for an extraction of the quantity $P_{pn/pX}$, which
represents the underlying probability of finding a pn correlation in the
''$pX$'' configuration that is defined by the presence of at least one proton
with $p_i> k_{Fermi}$.  The analysis was similar to that of
Ref~\cite{Tang:2002ww}, except that only events with nucleons above 275~MeV/c
were examined, to be more clearly in the region where SRC are expected to
dominate, and corrections were applied to account for the possibility that the
high-momentum nucleons might not be observed in the final state, yielding
\begin{equation}
P_{pn/pX} = 0.92^{+0.28}_{-0.18} ~,
\label{P_pn_exp}
\end{equation}
This result indicates that the removal of a proton from the nucleus with
initial momentum $275-550$~MeV/c is accompanied by the emission of a
correlated neutron that carries momentum roughly equal and opposite to the
initial proton momentum $92^{+8}_{-18}\%$ of the time (where the upper limit
is simply truncated to ensure that the maximum value for this quantity is
100\%).

Even though no recoil protons were measured in the experiments of
Ref.~\cite{Aclander:1999fd, Tang:2002ww}, it was possible to estimate the
upper limit for the ratio of probabilities of pp and pn
SRCs~\cite{Piasetzky:2006ai}, using the relationship $2P_{pp} + P_{pn} \le
P_{pX}$, where $P_{pp}$ has an additional factor of two in the weight as it
will yield a ``$pX$'' configuration with a high momentum proton if either of
the protons is initially at the appropriate kinematics to be detected
when struck by the beam proton.  This leads to:
\begin{equation}
P_{pp/pX} \le {1\over 2} (1-P_{pn/pX})= 0.04_{-0.04}^{+0.09} ~,
\label{pp_pn_BNL}
\end{equation}
where the upper limit occurs if all configurations with a detected
high-momentum proton are associated with a high-momentum spectator nucleon
and the lower limit corresponds to no pp pairs\footnote{In
Ref.~\cite{Piasetzky:2006ai}, this expression was assigned to $P_{pp}/P_{pn}$
rather than $P_{pp/pX}$, but these quantities differ only by a factor of
$P_{pn/pX}$, which is close to unity in this case.}.  Note that $P_{pp/pX}$
does \textit{not} correspond to the probability that an observed
high-momentum proton would have an associated high-momentum recoil proton
because it applies the factor of two weighting to pp pairs based on the 
assumption that for a finite acceptance experiment, you will ``double count''
the pp SRCs since either proton can be the high-$p_m$ proton.  Therefore,
$P_{pp/pX}$ is half of what would be observed in a direct measurement, and
has an upper limit of $P_{pp/pX}=0.5$ if only pp SRCs are present.

By making triple-coincidence measurements where both recoil proton and
neutrons are detected, one can more directly probe probability that a
high-momentum nucleon is associated with an SRC and extract the isospin
structure of SRCs in the nucleus.  Such measurements were carried out for the
first time in at Jefferson Lab~\cite{Shneor:2007tu,Subedi:2008zz} 
where scattered electrons and knocked-out protons were detected in
coincidence, with a large acceptance detector used to look for an associated
recoil proton or neutron in $^{12}$C(e,e$^\prime$pN) reactions.

The Hall A triple-coincidence experiment used an incident electron beam of
4.627~GeV and the two Hall A high-resolution spectrometers
(HRS)~\cite{Alcorn:2004sb} to identify the $^{12}$C(e,e$^\prime$p) reaction.
The $^{12}$C(e,e$^\prime$p) kinematics were chosen such that the measurement was at
$Q^2 \approx 2$~GeV$^2$ and $x \approx 1.2$, with missing momenta from
300--600~MeV/c.  The BigBite~\cite{deLange:1998au} large-acceptance
spectrometer was used to detect recoil protons with $\vec p_r \approx - \vec
p_m$ from $^{12}$C(e,e$^\prime$pp) events, while a neutron detector placed
behind BigBite was simultaneously used to detect recoil neutrons from
$^{12}$C(e,e$^\prime$pn) reactions.

Reaching sufficiently large values of $Q^2$ and Bjorken-$x$ was required to
reduce MEC and IC contributions as well as moderating the effect of the FSI
(see discussion in Sec.~\ref{sec:knockout}), while the $p_m$ range was
chosen to map out the region of 2N SRC dominance.  A cut was applied to the
missing energy reconstructed from the $^{12}$C(e,e$^\prime$p) kinematics which significantly
suppressed the contribution associated with the excitation of an intermediate
$\Delta$ resonance~\cite{Shneor:2007tu}.

\begin{figure}[htb]
\begin{center}
\includegraphics[width=0.8\linewidth]{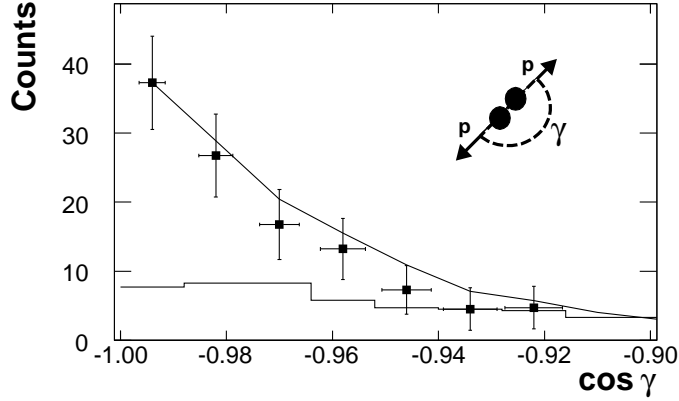}
\caption{\label{src:bigbitecosine} For the $^{12}$C(e,e$^\prime$pp) reaction
at $Q^2 > 1$~(GeV/c)$^2$, the distribution of the cosine of the opening angle
between the $\vec p_{miss}$ and $\vec p_{rec}$ for the $p_m=0.55$~GeV/c
setting. The curve is a simulation of the scattering off a moving pair with a
width of 0.136~GeV/c for the pair center-of-mass momentum and the histogram
shows the distribution of random events.
Figure reprinted with permission from Ref.~\cite{Shneor:2007tu}
}
\end{center}
\end{figure}

The first signature of SRCs was the correlation between
strength of the cross section and the relative angle ($\gamma$) between
initial momentum of the struck nucleon $p_i$ and momentum of the recoil
nucleon $p_r$.  Figure~\ref{src:bigbitecosine} shows the distribution of
events in $\cos{\gamma}$ for the highest $p_m$ setting of
550~MeV/c~\cite{Shneor:2007tu}, which is strongly peaked near
$\cos{\gamma}=-1$, corresponding to the back-to-back initial momenta of the
struck and recoil protons.  The solid curve is the simulated distribution for
scattering from a moving pair, with the pair center-of-mass momentum taken to
be a gaussian distribution with a width of 0.136~GeV/c. That width was varied
to best reproduce the data, but is also consistent with the width deduced from
the $(p, ppn)$ experiment at Brookhaven National Lab~\cite{Tang:2002ww} as well
as a theoretical calculation based on the convolution of two independent
single particle momentum distributions~\cite{CiofidegliAtti:1995qe}. Also shown
in Fig.~\ref{src:bigbitecosine} is the angular correlation for the random
background as defined by a time window offset from the coincidence peak, which
shows the effect of the acceptance of the spectator proton detector.

\begin{figure}[htbp]
\begin{center}
\includegraphics[width=1.0\linewidth]{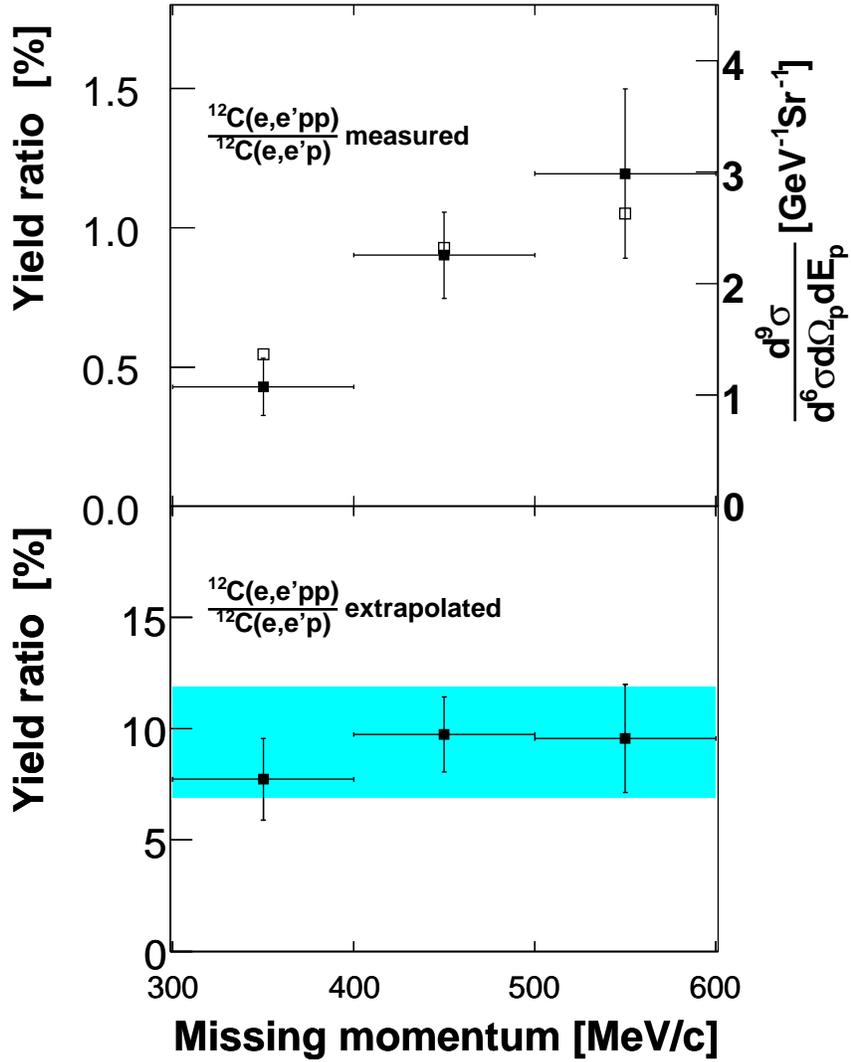}
\caption{\label{src:yieldratios}The measured and extrapolated ratios of yields
for the $^{12}$C(e,e$^\prime$pp) and the $^{12}$C(e,e$^\prime$p)
reactions.  The solid squares indicate the raw yield ratio, while the
hollow squares show the ratio of the extracted cross section.
The shaded area represents the range of results associated with varying the
gaussian width of the pair momentum distribution over the range
0.136$\pm$0.040~GeV/c (corresponding to the 2$\sigma$ range in the extracted
value of this width).
Figure reprinted with permission from Ref.~\cite{Shneor:2007tu}
}
\end{center}
\end{figure}

The other main quantity of interest for the two-proton knockout case is the
fraction of high-$p_m$ events in which there is a high-momentum,
backward-angle correlated proton, i.e. the ratio of $^{12}$C(e,e$^\prime$pp)
to $^{12}$C(e,e$^\prime$p) events.  The measured ratio must be corrected to
account for the finite acceptance of BigBite.  This is done using a PWIA
simulation of events for scattering from a pp SRC with a total pair momentum
distribution obtained from the measured angular correlation.  The raw and
acceptance-corrected $^{12}$C(e,e$^\prime$pp) / $^{12}$C(e,e$^\prime$p)
ratios are shown in Fig.~\ref{src:yieldratios}.  It was observed that for $300
< p_m < 600$~MeV/c, approximately 9\% of the $^{12}$C(e,e$^\prime$p) events
had a high-momentum recoil proton with $\vec p_r \approx - \vec p_m$,
independent of the value of $p_m$ within this region. The shaded region
indicates the range of results obtained by varying the width of the pair
momentum distribution within the $\pm 2\sigma$ range on the extracted width.
Averaged over the entire $p_m$ range and accounting for the experimental 
uncertainties and impact of the pair motion yields
$^{12}$C(e,e$^\prime$pp)/$^{12}$C(e,e$^\prime$p)=(9.5$\pm$2)\%~\cite{Shneor:2007tu}.
Note that absorption of the second proton as it exits the nucleus was
estimated to decrease this ratio by 15--20\%, while the effect of
charge-exchange FSI were estimated to yield a similar increase due to the
more common pn SRCs yielding a pair of protons in the final state.  Because
these effects were estimated to be similar in magnitude but of opposite sign,
no correction (or uncertainty) was applied for these effects.

The experiment also extracted the ratio of A(e,e$^\prime$pn) to
A(e,e$^\prime$p) yields using a similar procedure~\cite{Subedi:2008zz}.
Taking into account the finite acceptance of the neutron detector and the
$\sim$40\% neutron detection efficiency, it was found that $96 \pm 22$\% of the
A(e,e$^\prime$p) events with a missing momentum above 300~MeV/c had a
recoiling neutron.  This is in agreement with the
$(p, 2pn)/(p, 2p)$ ratio~\cite{Piasetzky:2006ai} (Eq.~(\ref{P_pn_exp}))
extracted from the Brookhaven proton beam measurement~\cite{Tang:2002ww}.

\begin{figure}[htb]
\begin{center}
\vspace{0.4cm}
\includegraphics[angle=-90, width=0.7\textwidth]{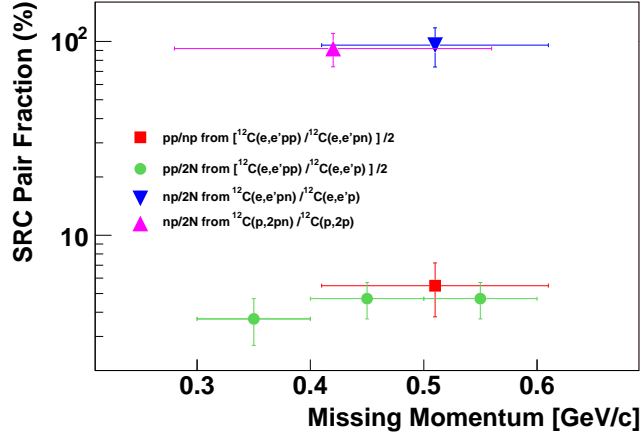}
\caption{\label{src:pairfraction} The fraction of correlated pair combinations
in carbon as obtained from the A(e,e$^\prime$pp) to A(e,e$^\prime$pn)
reactions~\cite{Subedi:2008zz}, as well as from previous $(p, 2pn)$
data~\cite{Piasetzky:2006ai}.  Note that the pp/np ratio has been corrected
for charge-exchange FSI, but the other ratios have not.
Figure reprinted with permission from Ref.~\cite{Subedi:2008zz}
}
\end{center}
\end{figure}

The largest advantage of the Jefferson Lab triple-coincidence experiment was
that both pp and pn pairs were measured, whereas for the Brookhaven
result, only an upper limit on the rate of pp correlations could be deduced.
Since the $^{12}$C(e,e$^\prime$pp) and $^{12}$C(e,e$^\prime$pn) data were
collected simultaneously with detectors covering nearly identical
solid angles, the $^{12}$C(e,e$^\prime$pn) / $^{12}$C(e,e$^\prime$pp) ratio
could be determined with reduced systematic and theoretical uncertainties,
e.g. the acceptance correction associated with the total momentum of the pair.
Correcting only for detector efficiencies, a ratio of $8.1\pm 2.2$ was
obtained.  The acceptance and attenuation of the recoiling protons and
neutrons was assumed to be equal, and the only correction related to
final-state interactions of the measured $^{12}$C(e,e$^\prime$pn) to
$^{12}$C(e,e$^\prime$pp) ratio is due to single charge exchange interactions.
Because the (e,e$^\prime$pn) cross section is much larger than the
(e,e$^\prime$pp) cross section, the main impact of the charge exchange FSI
will be the leakage of (e,e$^\prime$pn) events into the observed
(e,e$^\prime$pp) sample.  A Glauber approximation
calculation~\cite{Mardor:1992sb} estimated this yields an 11\% decrease in the
$^{12}$C(e,e$^\prime$pn)/$^{12}$C(e,e$^\prime$pp) ratio, thus implying an
initial $pn/pp$ ratio of $9.0\pm 2.5$.

Since the experiment triggered only on forward $^{12}$C(e,e$^\prime$p) events,
the probability of detecting pp pairs was twice that of pn pairs; thus, we
conclude that the ratio of pn/pp pairs in the $^{12}$C ground state is
$18 \pm 5$. This result translates to the following ratio of the pp to pn
two-nucleons SRCs:
\begin{equation}
{P_{pp}\over P_{pn}} = 0.056^{+0.021}_{-0.012} ~,
\label{pp_pn_JLab}
\end{equation}
in agreement with Eq.~(\ref{pp_pn_BNL}). Since these two experiments yield
consistent results at very different kinematics, it furthers the
interpretation of the process as being due to scattering off a correlated pair
of nucleons~\cite{Higinbotham:2009zz}.

Note that the $pn/pp$ ratio quoted in~\cite{Subedi:2008zz} is not calculated
using the quoted value for the $pp/pX$ ratio from the analysis
of~\cite{Shneor:2007tu}.  The $pp/pX$ and $pn/pX$ extractions apply different
cuts and corrections, so cannot be used to extract the $pn/pp$ ratio.
The original extraction of $pp/pX$ used the combined result from all $p_m$
settings, while the $pn/pX$ ratio was extracted using only the high-$p_m$ data,
and thus required different corrections for acceptance, efficiency, etc.... 
Therefore, a new extraction of $pp/pX$ was performed~\cite{RajSubedi:2007zz}
using only the largest $p_m$ setting to match the kinematics of the $pn/pX$
measurement. No correction was applied for nuclear transparency or
charge-exchange FSI, as the transparency was assumed to cancel in the pn/pp
ratio and the charge-exchange FSI is applied to the ratio as a separate
correction.  This analysis yielded $pp/pX$=(11.8$\pm$1.8)\% for the high-$p_m$
data set, larger than the (9.5$\pm$2)\% quoted in Ref.~\cite{Shneor:2007tu}
and with a smaller uncertainty, as the uncertainty due to the total pair
momentum is neglected because it is assumed to cancel in the $pn/pp$ ratio.

In both the BNL and Jefferson Lab experiments, SRCs are identified
kinematically, by looking for two nucleons which, in the PWIA, reconstruct to
a pair of nucleons with a large relative but a small total momenta. 
As with the $p_m$-$E_r$ correlation (Fig.~\ref{3bbu}), the kinematics only
tell us that the reaction involves a two-body process with low total momentum;
one can obtain the same final state in a two-step process where one
low-momentum nucleon is struck and a secondary interaction leads to a second
high-momentum nucleon in the final state.  Thus, one must consider the
possibility of FSI in both cases.  In the BNL measurement, the hard elastic $p p
\to p p$ cross section falls as $s^{-10}$.  Therefore, there is a strong
enhancement in the cross section for events at lower $s$, i.e. where the
struck proton has a large momentum in the direction of the proton beam, which
amplifies the signal from pre-existing SRCs relative to scattering from a
low-momentum nucleon with a subsequent reinteraction~\cite{Yaron:2002nv,
Farrar:1988mf}.  For the Jefferson Lab experiment, the electron kinematics are
chosen to suppress the contribution of FSI by taking data at $x=1.2$ and
$Q^2=2$~GeV$^2$ which, in the PWIA, corresponds to a minimal longitudinal
momentum of $\sim 200$~MeV/c.  The hadron rescattering cross section does not
go away at high $Q^2$, but a rescattering of the struck proton will not modify
$x$, reconstructed from only the electron kinematics, unless the rescattering
takes place within a short distance of the initial interaction, $\sim$1~fm for
this experiment.  Thus, the measurement is expected to be sensitive only to
scattering from short-range configurations.

While we normally associate such short-range configurations with high-momentum
components, the triple-coincidence reactions demonstrate that only short-range
iso-singlet np configurations yield high momentum. While the short timescale
relevant for FSI at $x>1$ limits the effects to rescattering between nucleons
that are already very close together, it does not require that these nucleons
be in what we call a short-range correlation, i.e. they do not have to be in a
high relative momentum state.  Iso-singlet np pairs feel a strong tensor
interaction when are close together, generating high-momentum states, while
nn, pp, and iso-triplet np pairs can be close together without interacting
strongly, at least until they are close enough to interact via the short-range
repulsive core. There are also pp pairs at short distance, but which are not
in a high relative momentum configuration because they do not have a strong
tensor interaction at short distances. Scattering from such a low-momentum
nucleon which rescatters from another close-by low-momentum nucleon would be
indistinguishable from PWIA scattering from an initial-state high-momentum
pair, and thus the measurement will potentially be sensitive to a combination
scattering from SRCs and scattering from low-momentum pairs which are close
enough together that they can undergo final-state interactions.

This raises some interesting questions which may make the triple-coincidence
results even more significant.  The analysis of the $^{12}$C(e,e$^\prime$p)
events at high $p_m$~\cite{Monaghan:2008zz} suggests that the FSI contributions
are small when the final state is reconstructed to be a $^{11}$B nucleus,
but dominate the cross section for more inelastic scattering.  Because
the experiment reconstructs that nearly 100\% of these high-$p_m$
$^{12}$C(e,e$^\prime$p) events are associated with a high-momentum recoil nucleon
corresponding to an initial SRC (in the PWIA), the majority of the
triple-coincidence events must also involve a final-state interaction.
Had the measurements observed np pairs and pp pairs in equal numbers,
then it could mean that both iso-singlet and iso-triplet pairs occur
frequently or else it could mean that there is a significant contribution
coming from scattering from low relative momentum iso-triplet pairs with a
significant final state interaction generating two high-momentum nucleons
which reconstruct to a small total momentum and a large relative momentum.

The fact that FSI contributions dominate the triple-coincidence
channel combined with the observed dominance of np pairs demonstrates that
FSI between low-momentum pp pairs with small relative separation does not
contribute significantly.  This would appear to imply one of the following
options: few short-distance pp pairs in the initial state, some suppression
of FSI when scattering from such pp pairs, or an isospin-dependent FSI where
low momentum pairs (pp, np, or nn) have rescattering which mimics an initial
state SRC at these kinematics only for np pairs.
This opens up the possibility that
a comparison of the isospin structure in the triple-coincidence measurement,
which has large FSI contributions, and the inclusive measurements on nuclei
with different N/Z ratios could provide a way to determine the role
of short-distance nucleon pairs that do not generate high momentum states.
However, a more detailed examination of the role of final-state interactions
in the triple-coincidence measurements is needed to more fully understand the
implications of these measurements.

\begin{figure}[htb]
\begin{center}
\includegraphics[width=0.75\textwidth,height=0.57\textwidth]{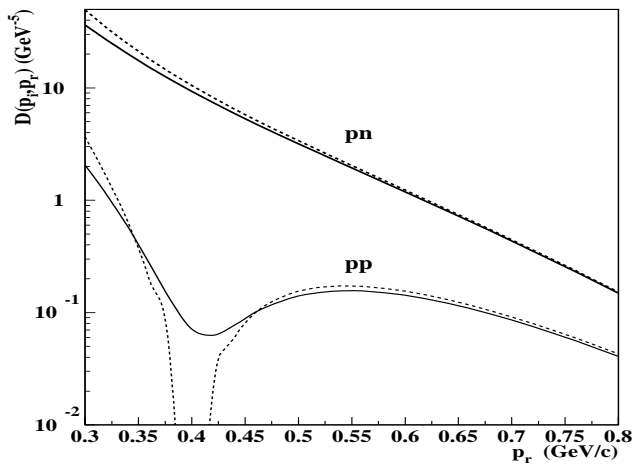}
\caption{\label{pp_to_pn}
Recoil nucleon momentum dependence of the decay function calculated for
$^3$He(e,e$^\prime$$N_N$$N_r$)N reaction for 2N SRC kinematics in which
initial and recoil nucleons are correlated back-to-back, i.e. for NN pairs with
zero total momentum. Dashed line PWIA, solid line DWIA calculations of
Ref.~\cite{Sargsian:2005ru}. Labels ``pp'' and ``pn'' mark the results for
pp and pn 2N SRCs respectively.}
\end{center}
\end{figure}

The startlingly small ratio of proton-proton to neutron-proton SRCs was
explained by several groups~\cite{Sargsian:2005ru, Schiavilla:2006xx,
Alvioli:2007zz} to be a consequence of the dominance of the tensor interaction
in this $p_m$ range. In Ref.~\cite{Sargsian:2005ru}, the triple-coincidence
reaction was studied for $^3$He(e,e$^\prime$NN)N reactions at high $Q^2$.
These reactions are the most simple from the point of view of
triple-coincidence measurements and allow the calculation of the above defined
decay function through the realistic wave function of $^3\mbox{He}$ nucleus.
The decay function was calculated in the distorted wave impulse approximation
with FSI implemented within the generalized eikonal approximation. The
result of the calculations for $^3$He(e,e$^\prime$pp)n and
$^3$He(e,e$^\prime$pn)p reactions in the kinematics that maximize the
contribution from 2N SRCs are presented in Fig.~\ref{pp_to_pn}.  One can see
that for recoil momenta of 300--600~MeV/c, the pp correlations are
suppressed by an order of magnitude or more compared to pn correlations.

\begin{figure}[htb]
\begin{center}
\includegraphics[angle=-90, width=0.8\textwidth]{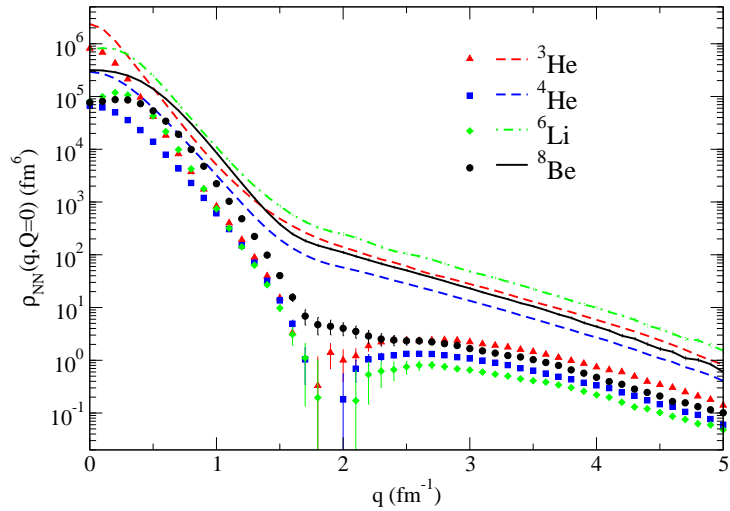}
\caption{The momentum distribution for neutron-proton pairs (lines) and
proton-proton pairs (symbols) are shown for various nuclei. The calculations
assume pairs at rest as a function of the relative momentum of the nucleons in
the pair. The dip in the proton-proton pair momentum distribution around 2
fm$^{-1}$ (~400 MeV/c) is due to nucleon-nucleon tensor correlations.
Figure reprinted with permission from Ref.~\cite{Schiavilla:2006xx}
}
\label{theory-ratios}
\end{center}
\end{figure}

Very similar results are obtained also for other
nuclei~\cite{Schiavilla:2006xx, Alvioli:2011aa} in calculations of the
two-nucleon relative momentum distributions (shown in
Fig.~\ref{theory-ratios}) for the ground states of light nuclei (A$<$8) using
variational-Monte-Carlo wave functions derived from a realistic Hamiltonian
with two- and three-nucleon potentials. Their calculation for the relative
nucleon momentum distribution for at-rest nucleon pairs shows the same large proton-neutron
to proton-proton ratio in the relative momentum range of 300--600 MeV/c. Note
that within PWIA, the two-nucleon relative momentum distribution with the pair
at rest in the center-of-mass is related to the decay functions that enter
into the cross section of triple-coincidence reactions.

The universality of the above two results is based on the fact that in the
momentum range of 300--600~GeV/c, the NN interaction is dominated by the
tensor force which means that the iso-triplet pp, pn, and nn channels
are strongly suppressed relative to the iso-singlet pn channel. These
theoretical conclusions are in agreement with earlier studies of nuclear
spectral functions for large momenta that also indicated the dominance of
tensor correlations~\cite{Roth:2004ua, CiofidegliAtti:1995qe}.  Note that
in the inclusive measurements, the cross section is sensitive to a range
of momenta with $p > p_{min}$ (Fig.~\ref{fig:xpmin2nsrc}).  While this
includes very large momenta, where iso-triplet pairs are not as strongly
suppressed, the contribution to the inclusive cross section from this region
is small because there are so few of these extremely high momentum nucleons.
So the cross section is dominated by scattering from nucleons in the
300--500~MeV/c range, where there is a strong suppression of iso-triplet
SRCs.

Finally, studies of two-nucleon knockout in $^3$He are a special case,
as the measurement of the scattered electron in coincidence with two of
the nucleons provides for a complete kinematic reconstruction of the final
three-body system, provided that particle production can be suppressed.
This affords the possibility to study a two-nucleon
SRC in a unique way, namely by interacting with the third nucleon which
is not part of the SRC.  After the fast removal of the spectator nucleon,
one expects the residual (A-1) system, i.e. the 2N SRC, to decay into two
nucleons, allowing one to directly measure the momenta of both correlated
nucleons in the absence of FSI.

\begin{figure}[htb]
\begin{center}
\includegraphics[width=0.75\textwidth]{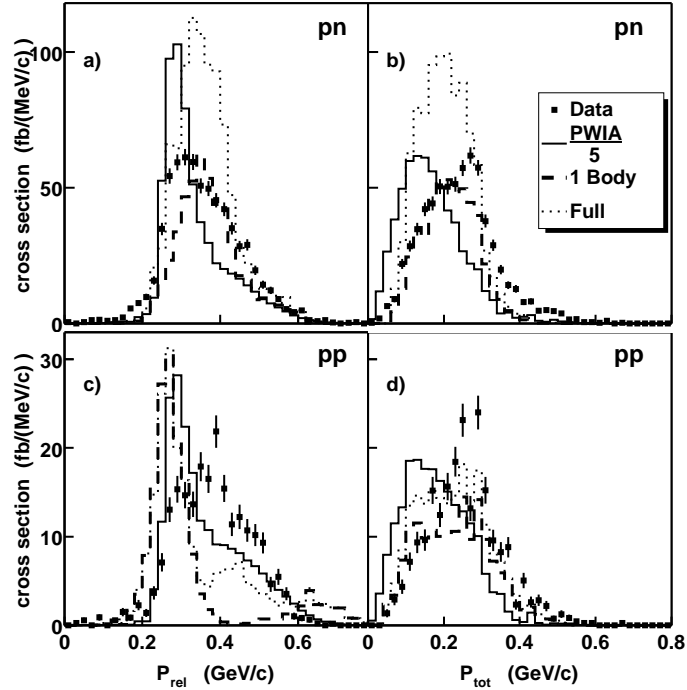}
\caption{\label{src:spectator}
Figure a) shows lab frame cross section versus pn pair relative momentum.
Points show the data, solid histogram shows the PWIA calculation multiplied by
a factor 0.2, dashed histogram shows Laget's one-body
calculation~\cite{Laget:1987jr, Audit:1996tq, Laget:1987zz}, dotted histogram
shows Laget's full calculation; b) the same for total momentum of the pair; c), d) the
same for pp pairs.
Figure reprinted with permission from Ref.~\cite{Niyazov:2003zr}
}
\end{center}
\end{figure}

Such a strategy was realized in the Jefferson Lab experiment of
Ref.~\cite{Niyazov:2003zr} in which the $^3$He(e,e$^\prime$pp)n reaction was studied
using 2.2 and 4.4~GeV electron beams and detecting the scattered electron and
ejected protons in CLAS over a wide kinematic range. In the PWIA, interacting
with a nucleon inside of an SRC would yield a high-momentum struck nucleon, a
recoil nucleon from the SRC with $k>k_{Fermi}$, and a low momentum spectator
with $k<k_{Fermi}$.  However, if the nucleon that is not in the SRC is struck,
there will be three high-momentum nucleons in the final state.  The struck
nucleon will have a momentum close to $q$, while the two nucleons in the SRC
will come out with large, back-to-back momentum of $k>k_{Fermi}$.  The
experiment observed that when all three final state nucleons have momenta
greater than 250~MeV/c, the reaction is dominated by events where two nucleons
each have less than 20\% of the transferred energy and the third `leading'
nucleon has the remainder.  Final state interactions of the leading nucleon
are suppressed by requiring that it has perpendicular momentum with respect to
$\vec q$ of less than 300 MeV/c. In these cases the two other nucleons (the
pn or pp pair) are found to be predominantly back-to-back and to have very
little total momentum in the direction of the momentum transfer. The relative
pair momentum is then almost isotropic with respect to $\vec{q}$, suggesting
that the nucleon-nucleon pair is
correlated and is a spectator to the virtual photon absorption. In this case,
the measured relative and total pair momenta, $\vec p_{rel} = (\vec p_1 - \vec
p_2)/2$ and $\vec p_{tot} = \vec p_1 + \vec p_2$ are closely related to the
initial momenta of the correlated pair. As shown in Fig.~\ref{src:spectator},
the pair relative momentum peaks at about 300--400 MeV/c and the pair total
momentum peaks at about 300 MeV/c. The pp and pn momentum distributions
are very similar. The advantage of this approach is that there are no
contributions from meson exchange currents or isobar configurations since the
virtual photon does not interact with the correlated pair. However, because
the continuum interaction of the correlated pair is very strong and reduces
the calculated cross section by a factor of about ten, this reaction is very
difficult to calculate precisely. Diagrammatic
calculations~\cite{Laget:1987jr, Audit:1996tq, Laget:1987zz}, are in
qualitative agreement with the measured momentum distributions, but only after
including the effects of the continuum interaction.

\begin{figure}[htb]
\begin{center}
\includegraphics[width=0.75\textwidth]{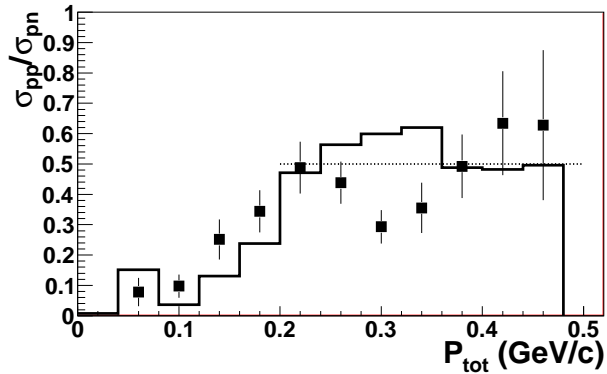}
\caption{\label{src:pptopn}
The ratio of cross sections with pp or pn SRC spectators (with relative
momentum from 300--500~MeV/c) as a function of the pair total
momentum~\cite{Baghdasaryan:2010nv} for $^3$He. The histogram is the Golak
one-body calculation~\cite{Golak:1994vq}, and the dotted line shows the
expectation based on simple pair-counting, assuming isospin-independent
pairing. The ratios (data and calculation) are multiplied by 1.5 to
approximately account for the difference in the ep and en elastic
cross sections associated with np and pp spectators. At low pair momentum,
they see strong pn dominance, consistent with the triple-coincidence
measurement on Carbon~\cite{Subedi:2008zz}, while at high total momentum, the
find equal contributions from pp and pn pairs.
Figure adapted from Ref.~\cite{Baghdasaryan:2010nv}
}
\end{center}
\end{figure}

The measurements were also able to examine the contribution from pn and pp
pairs as a function of the total pair momentum~\cite{Baghdasaryan:2010nv},
as shown in Fig.~\ref{src:pptopn}.
They observed a smaller difference in the fraction of high-momentum pn to
pp pairs, which were observed in a ratio of approximately 4:1.  However,
when taken as a function of the total momentum of the pair, the ratio becomes
much larger at small total momentum, in agreement with the observation of strong
np dominance in the triple-coincidence measurement on $^{12}$C.  They observe
a much smaller difference for the case where there is a large total momentum
of the pair, which for $^3$He corresponds to the case where all three nucleons
appear to have large momenta in the final state.  This is taken as further
indication of tensor dominance, as T=1 pairs are most likely to be in a spin-0
state, which yields a minimum in the distribution for large relative momentum
and small total pair momentum.  With increasing pair momentum, this minimum
is filled in, increasing the $pn/pp$ ratio~\cite{Baghdasaryan:2010nv,
Alvioli:2011aa}.

\section{Summary and Outlook}

We have reviewed the physics importance of studying short-range, high-momentum
nucleon configurations in nuclei.  We then presented in some detail the
necessary experimental requirements for measurements that can provide reliable
information on these SRCs, and discussed the main results of a new generation
of high energy and high momentum transfer measurements aimed at verifying the
presence and probing the nature of these correlations.

We emphasized the special role of high $Q^2$ studies of the deuteron
in advancing the field of the short-range nucleon correlations by means
of studying reaction mechanisms, off-shell nucleon effects and relativistic
dynamics of strongly bound nucleon systems.  Going beyond the deuteron, we
categorized the experimental program by the reaction mechanism, discussing
programs of inclusive, semi-inclusive and triple-coincidence measurements. We
reviewed the first high $Q^2$ experiments in these categories and demonstrated
how these experiments advanced our understanding the structure of the short
range correlations in the nuclei.  Further experimental details on the
Jefferson Lab program can be found in Ref.~\cite{Higinbotham:2009hi}, with
future plans, including some which go beyond hadronic degrees of freedom to
look at non-nucleonic components or medium modification within SRCs, presented
in Ref.~\cite{Sargsian:2002wc}.

The kinematic domain probed and the knowledge obtained from
these experiments can be summarized as follows:

\begin{itemize}

\item The deuteron momentum distribution has been probed up to $500$~MeV/c
internal momenta and $Q^2$ up to about 3.5~GeV$^2$. The observation is that MEC
and IC contributions can be well controlled by making measurements at large
$Q^2$ and $x>1$. In this region, FSI can be relatively well calculated within
the eikonal approximation, reproducing data taken in kinematic regimes where
FSI contributions dominate.  This allows for further studies in regions where
A(e,e$^\prime$p) measurements are most sensitive to the reaction mechanism,
in particular off-shell effects and relativistic dynamics, to provide the last
data needed to fully validate our understanding of the reaction in the
deuteron, thus allowing us to extend detailed and quantitative studies to
heavier nuclei.

\item For inclusive scattering from light-to-heavy nuclei, measurements at
high $Q^2$ and $x>1$ probe the momentum distribution of the nucleons in
nuclei. A clear signature of scaling associated with interaction from 2N SRCs
is observed in the $\sigma_{eA}/\sigma_{eD}$ ratios. These ratios allow us to
estimate the contribution of 2N SRCs in heavy nuclei relative to the
contributions in the deuteron. Moving to similar studies for $x>2$, initial
measurements have seen some indication of the dominance of 3N correlations
in these kinematics, although a recent measurement~\cite{e08014} as well as
more quantitative studies are required to verify and understand this.

\item Semi-inclusive knock-out A(e,e$^\prime$N) reactions established the clear
correlation signature between missing-momentum and missing-energy of the
reaction. For the
first time, evidence has been acquired that in the SRC domain the final state
interaction is localized within the short-range configuration.

\item Triple-coincidence experiments gave new insight into the dynamics of 2N
SRCs. They probed relative internal momenta up to $600$~MeV/c and gave
unambiguous evidence that the dynamics of 2N SRC in this domain are dominantly
controlled by the tensor component of the NN force.  An extension to higher
internal momenta was recently completed~\cite{e07006} and the data are
currently being analyzed.

\item Theoretical analysis of all these experiments allows us to conclude that
for up to $600$~MeV/c relative momenta, nucleons represent the relevant
degrees of freedom in the dynamics of two nucleon short-range correlations.

\end{itemize}

The knowledge of SRCs obtained so far sets up the main directions which in our
opinion the new generation of high $Q^2$ experiments should follow:

\begin{itemize}

\item Extension of the deuteron studies beyond 500~MeV/c. One may expect
especially interesting new physics beyond 750~MeV/c, when the
threshold of inelastic excitation in the iso-singlet NN system is passed and
the role of the non-nucleonic degrees of freedom may be increasingly
important. This region also will be defined by the core of the NN interaction,
by far the least well understood part of the NN interaction. A D(e,e$^\prime$p)
measurement is planned for Jefferson Lab~\cite{e1210003} with the goal of
precisely mapping out the high-momentum tail up to 1~GeV/c in the deuteron by
making measurements at $x>1$ and $Q^2=4.25$~GeV$^2$.

\item The next generation of inclusive measurements will pursue several
important issues. One is the detailed studies of $x>2$ region in probing the
onset of 3N correlations.  Jefferson Lab experiment E08-014~\cite{e08014}
recently completed data taking which should be able to clearly map out the
onset of scaling in the target ratios for $x>2$.  This experiment will also
compare scattering from $^{40}$Ca and $^{48}$Ca to provide another test of the
dominance of iso-singlet (np) pairs in SRCs, free from the large FSIs
associated with the triple-coincidence measurements.  In addition,
measurements planned for the 12~GeV upgrade will extend the 2N and 3N SRC
measurements to many additional light and heavy nuclei~\cite{e1206105}, as
well as comparing $^3$H and $^3$He~\cite{e1211112} to better study the
isospin structure of SRCs.

\item For knock-out and triple-coincidence experiments the challenge will be
to go beyond $600$~MeV/c (beyond the domain of the tensor-forces). It is clear
from Figs.~\ref{pp_to_pn} and \ref{theory-ratios} that for larger nucleon
momenta, $>$~600~MeV/c, the ratio of neutron-proton to proton-proton pairs
should become smaller as the reaction become dominated by the short-range
repulsive core of the nucleon-nucleon interaction. The recently completed
triple-coincidence experiment E07-006~\cite{e07006} will extend the previous
measurements beyond 600~MeV/c, and should be able to test this prediction.

\item Recent measurements of the EMC effect in light
nuclei~\cite{Seely:2009gt} yield a non-trivial A dependence in the
modification of the nuclear structure function in deep-inelastic scattering.
These data suggest that the nuclear structure function depends on the local,
microscopic structure of the nucleus. As the question of local, high-density
structure is the primary focus of SRC measurements, it was natural to compare
the two measurements, and a quantitative correlation between the size of the
EMC effect and the strength of SRCs in nuclei was
observed~\cite{Weinstein:2010rt}.  Extensions of both SRC~\cite{e1206105} and
EMC effect~\cite{E1210008} measurements after the Jefferson Lab 12 GeV upgrade,
including several light nuclei, some of which have significant cluster
structure, will help extend these comparisons which may relate both the
structure functions and the SRCs to the microscopic nuclear structure.

\item Finally, moving to the realm of scattering from quarks in nuclei provides
access to short-range structure in a region where non-hadronic degrees of
freedom may become important~\cite{Sargsian:2002wc, Arrington:2003qt,
Arrington:2006pn}.  Recent measurements suggest that even with 6~GeV
electrons, we may be able to reliably extract the distribution of
``super-fast'' quarks at $x>1$. The distribution of these super-fast quarks is
sensitive to the short-range structure of the nucleus which drives the
high-momentum components of the nuclear wave function.  While the present
data~\cite{Fomin:2010ei} suggest that the high-momentum contributions can be
explained by SRCs, without the need for any non-hadronic contributions, the
range of this data is still somewhat limited.  Measurements planned for 11~GeV
will move into a region with dramatically increased sensitivity to
non-hadronic components in nuclei~\cite{e1206105, Sargsian:2002wc,
Arrington:2006pn}.

\end{itemize}

\section*{Acknowledgments}

This work was supported by the U.S. Department of Energy, Office of Nuclear
Physics, under contracts DE-AC02-06CH11357, DE-AC05-06OR23177, and
DE-FG02-01ER-41172. We thank our colleagues who assisted in the preparation of
this work, including Werner Boeglin, Donal Day, Nadia Fomin, Roy Holt, Ushma
Kriplani, Eli Piasetzky, Mark Strikman, and Larry Weinstein for useful
comments and discussions, including preliminary results from recent
measurements, and providing figures.

\bibliography{references}{}

\begin{thebibliography}{100}
\expandafter\ifx\csname url\endcsname\relax
  \def\url#1{\texttt{#1}}\fi
\expandafter\ifx\csname urlprefix\endcsname\relax\def\urlprefix{URL }\fi
\expandafter\ifx\csname href\endcsname\relax
  \def\href#1#2{#2} \def\path#1{#1}\fi

\bibitem{Moniz:1971mt}
E.~J. Moniz, et~al., {Nuclear fermi momenta from quasielastic electron
  scattering}, Phys. Rev. Lett. 26 (1971) 445--448.

\bibitem{Bethe:1971xm}
H.~A. Bethe, {Theory of nuclear matter}, Ann. Rev. Nucl. Part. Sci. 21 (1971)
  93--244.

\bibitem{Mayer:1955yy}
M.~G. Mayer, J.~H.~D. Jensen, {Elementary Theory of Nuclear Shell Structure },
  {Elementary Theory of Nuclear Shell Structure }New York, USA: John Wiley and
  Sons, Inc.

\bibitem{Koltun:1972kh}
D.~Koltun, {Total Binding Energies of Nuclei, and Particle-Removal
  Experiments}, Phys.Rev.Lett. 28 (1972) 182--185.

\bibitem{Bernheim:1974aa}
M.~Bernheim, A.~Bussi\`ere, A.~Gillebert, J.~Mougey, P.~X. Ho, M.~Priou,
  D.~Royer, I.~Sick, G.~J. Wagner, {C12(e,e'p) Results as a Critical Test of an
  Energy Sum Rule}, Phys. Rev.Lett 32 (1974) 898--901.

\bibitem{Faessler:1975zz}
A.~Faessler, S.~Krewald, G.~J. Wagner, {Is there evidence of three-body forces
  from violation of the Koltun energy sum rule?}, Phys.Rev. C11 (1975)
  2069--2072.

\bibitem{Royer:1975zz}
D.~Royer, J.~Mougey, M.~Priou, M.~Bernheim, A.~Bussiere, et~al., {Influence of
  short range correlations on (e,e'p) cross sections}, Phys.Rev. C12 (1975)
  327--329.

\bibitem{Kelly:1996hd}
J.~J. Kelly, {Nucleon knockout by intermediate-energy electrons}, Adv. Nucl.
  Phys. 23 (1996) 75--294.

\bibitem{Lapikas:1003zz}
L.~Lapikas, {Quasi-elastic electron scattering off nuclei}, Nucl. Phys. A553
  (1993) 297--308.

\bibitem{CiofidegliAtti:1995qe}
C.~Ciofi~degli Atti, S.~Simula, {Realistic model of the nucleon spectral
  function in few and many nucleon systems}, Phys. Rev. C53 (1996) 1689.

\bibitem{Lacombe:1980dr}
M.~Lacombe, et~al., {Parametrization of the Paris n n Potential}, Phys. Rev.
  C21 (1980) 861--873.

\bibitem{degliAtti198414}
C.~Ciofi~degli Atti, E.~Pace, G.~Salmè, {Analysis of the proton momentum
  distribution in 3He}, Physics Letters B 141 (1984) 14 -- 18.

\bibitem{Schiavilla:1985gb}
R.~Schiavilla, V.~R. Pandharipande, R.~B. Wiringa, {Momentum distributions in a
  A = 3 and 4 nuclei}, Nucl. Phys. A449 (1986) 219--242.

\bibitem{Benhar1986135}
O.~Benhar, C.~Ciofi Degli~Atti, S.~Liuti, G.~Salmè, Realistic many-body wave
  functions and nucleon momentum distributions in finite nuclei, Physics
  Letters B 177 (1986) 135 -- 140.

\bibitem{Pieper:1992gr}
S.~C. Pieper, R.~B. Wiringa, V.~R. Pandharipande, {Variational calculation of
  the ground state of {$^{16}$}O}, Phys. Rev. C46 (1992) 1741--1756.

\bibitem{Ji:1989nr}
X.-D. Ji, J.~Engel, {High momentum nucleons in finite nuclei and y-scaling},
  Phys. Rev. C40 (1989) 497--501.

\bibitem{Jaminon1986445}
M.~Jaminon, C.~Mahaux, H.~Ngô, Effect of correlations on the momentum
  distribution of protons in $^{208}${P}b, Nuclear Physics A 452 (1986) 445 --
  461.

\bibitem{Fantoni1984473}
S.~Fantoni, V.~R. Pandharipande, Momentum distribution of nucleons in nuclear
  matter, Nucl. Phys. A 427 (1984) 473 -- 492.

\bibitem{CiofidegliAtti:1990rw}
C.~Ciofi~degli Atti, E.~Pace, G.~Salme, {y-scaling analysis of quasielastic
  electron scattering and nucleon momentum distributions in few body systems,
  complex nuclei and nuclear matter}, Phys. Rev. C43 (1991) 1155--1176.

\bibitem{Bussiere:1981mv}
M.~Bernheim, et~al., {Momentum distribution of nucleons in the deuteron from
  the d(e,e'p)n reaction}, Nucl. Phys. A365 (1981) 349--370.

\bibitem{TurckChieze:1984fy}
S.~Turck-Chieze, et~al., {Exclusive deuteron electrodisintegration at high
  neutron recoil momentum}, Phys. Lett. B142 (1984) 145--148.

\bibitem{Jans:1982aw}
E.~Jans, et~al., {Quasifree (e, e' p) reaction on He-3}, Phys. Rev. Lett. 49
  (1982) 974--978.

\bibitem{Jans:1987px}
E.~Jans, et~al., {The quasifree $^3$He (e,e'p) reaction}, Nucl. Phys. A475
  (1987) 687--719.

\bibitem{Marchand:1987hd}
C.~Marchand, M.~Bernheim, A.~Gerard, J.~Laget, A.~Magnon, et~al., {High proton
  momenta and short range nucleon-nucleon correlations in a He-3 (e,e'p)
  experiment}, Phys.Rev.Lett. 60 (1988) 1703.

\bibitem{VanDenBrand:1988pv}
J.~F.~J. Van Den~Brand, et~al., {Electrodisintegration of He-4 studied with the
  reaction $^4$He(e,e'p)$^3$H}, Phys. Rev. Lett. 60 (1988) 2006--2009.

\bibitem{LeGoff:1994cg}
J.~Le~Goff, M.~Bernheim, M.~Brussel, G.~Capitani, J.~Danel, et~al., {Short
  range interaction of nucleons inside the nucleus via He-4 (e,e'p) R
  reaction}, Phys.Rev. C50 (1994) 2278--2287.

\bibitem{Frullani:1984nn}
S.~Frullani, J.~Mougey, {Single Particle Properties of Nuclei Through (e, e' p)
  Reactions}, Adv. Nucl. Phys. 14 (1984) 1--283.

\bibitem{Muther:2000qx}
H.~Muther, A.~Polls, {Two-body correlations in nuclear systems},
  Prog.Part.Nucl.Phys. 45 (2000) 243--334.

\bibitem{Fabrocini:1999mz}
A.~Fabrocini, F.~Aria~de Saavedra, G.~Co, {Ground state correlations in
  $^{16}$O and $^{40}$Ca}, Phys.Rev. C61 (2000) 044302.

\bibitem{Dickhoff:2004xx}
W.~Dickhoff, C.~Barbieri, {Self consistent Green's function method for nuclei
  and nuclear matter}, Prog.Part.Nucl.Phys. 52 (2004) 377--496.

\bibitem{Alvioli:2005cz}
M.~Alvioli, C.~Ciofi~degli Atti, H.~Morita, {Ground-state energies, densities
  and momentum distributions in closed-shell nuclei calculated within a cluster
  expansion approach and realistic interactions}, Phys.Rev. C72 (2005) 054310.

\bibitem{Alvioli:2007zz}
M.~Alvioli, C.~Ciofi~degli Atti, H.~Morita, {Proton-neutron and proton-proton
  correlations in medium-weight nuclei and the role of the tensor force}, Phys.
  Rev. Lett. 100 (2008) 162503.

\bibitem{Suzuki:2008cy}
Y.~Suzuki, W.~Horiuchi, {Significance and properties of internucleon
  correlation functions}, Nucl.Phys. A818 (2009) 188--207.

\bibitem{Vanhalst:2011es}
M.~Vanhalst, W.~Cosyn, J.~Ryckebusch, {Counting the amount of correlated pairs
  in a nucleus}, Phys.Rev. C84 (2011) 031302.

\bibitem{Feldmeier:2011qy}
H.~Feldmeier, W.~Horiuchi, T.~Neff, Y.~Suzuki, {Universality of short-range
  nucleon-nucleon correlations}, Phys. Rev. C 84, 054003.

\bibitem{Forest:1996kp}
J.~L. Forest, et~al., {Femtometer Toroidal Structures in Nuclei}, Phys. Rev.
  C54 (1996) 646--667.

\bibitem{Schiavilla:2006xx}
R.~Schiavilla, R.~B. Wiringa, S.~C. Pieper, J.~Carlson, {Tensor Forces and the
  Ground-State Structure of Nuclei}, Phys. Rev. Lett. 98 (2007) 132501.

\bibitem{Wiringa:2008dn}
R.~B. Wiringa, R.~Schiavilla, S.~C. Pieper, J.~Carlson, {Dependence of
  two-nucleon momentum densities on total pair momentum}, Phys. Rev. C78 (2008)
  021001.

\bibitem{Muther:1993yg}
H.~Muther, W.~Dickhoff, {Single-particle spectral function of O-16}, Phys.Rev.
  C49 (1994) R17--R20.

\bibitem{Levinger:1951vp}
J.~Levinger, {The High-energy nuclear photoeffect}, Phys.Rev. 84 (1951) 43--51.

\bibitem{Gottfried:1958vp}
K.~Gottfried, {On the determination of the nuclear pair correlation function
  from the high energy photo-effect}, Nucl.Phys. 5 (1958) 557--587.

\bibitem{brown1976nucleon}
G.~Brown, A.~Jackson, \href{http://books.google.com/books?id=9uDvAAAAMAAJ}{The
  nucleon-nucleon interaction}, North-Holland, 1976.
\newline\urlprefix\url{http://books.google.com/books?id=9uDvAAAAMAAJ}

\bibitem{Frankfurt:1981mk}
L.~L. Frankfurt, M.~I. Strikman, {High-Energy Phenomena, Short Range Nuclear
  Structure and QCD}, Phys. Rept. 76 (1981) 215--347.

\bibitem{Frankfurt:2008zv}
L.~Frankfurt, M.~Sargsian, M.~Strikman, {Recent observation of short range
  nucleon correlations in nuclei and their implications for the structure of
  nuclei and neutron stars}, Int. J. Mod. Phys. A23 (2008) 2991--3055.

\bibitem{Nogga:2002qp}
A.~Nogga, et~al., {The three-nucleon bound state using realistic potential
  models}, Phys. Rev. C67 (2003) 034004.

\bibitem{Sargsian:2001ax}
M.~M. Sargsian, {Selected topics in high energy semi-exclusive electro- nuclear
  reactions}, Int. J. Mod. Phys. E10 (2001) 405--458.

\bibitem{VanOrden:1995eg}
J.~Van~Orden, N.~Devine, F.~Gross, {Elastic electron scattering from the
  deuteron using the gross equation}, Phys.Rev.Lett. 75 (1995) 4369--4372.

\bibitem{Brodsky:2003ip}
S.~Brodsky, L.~Frankfurt, R.~A. Gilman, J.~Hiller, G.~Miller, et~al., {Hard
  photodisintegration of a proton pair in He-3}, Phys.Lett. B578 (2004) 69--77.

\bibitem{Sargsian:2002wc}
M.~M. Sargsian, et~al., {Hadrons in the nuclear medium.}, J. Phys. G29 (2003)
  R1.

\bibitem{Stoler:1993yk}
P.~Stoler, {Baryon form-factors at high Q$^2$ and the transition to
  perturbative QCD}, Phys. Rept. 226 (1993) 103--171.

\bibitem{Ungaro:2006df}
M.~Ungaro, et~al., {Measurement of the N,Delta+(1232) transition at high
  momentum transfer by pi0 electroproduction}, Phys.Rev.Lett. 97 (2006) 112003.

\bibitem{Kogut:1969xa}
J.~B. Kogut, D.~E. Soper, {Quantum Electrodynamics in the Infinite Momentum
  Frame}, Phys. Rev. D1 (1970) 2901--2913.

\bibitem{Bjorken:1970ah}
J.~D. Bjorken, J.~B. Kogut, D.~E. Soper, {Quantum Electrodynamics at Infinite
  Momentum: Scattering from an External Field}, Phys. Rev. D3 (1971) 1382.

\bibitem{Feynman:1973xc}
R.~Feynman, {Photon-hadron interactions}.

\bibitem{Day:1987az}
D.~B. Day, et~al., {y-scaling in electron nucleus scattering}, Phys. Rev. Lett.
  59 (1987) 427--430.

\bibitem{Frankfurt:1988nt}
L.~L. Frankfurt, M.~I. Strikman, {Hard Nuclear Processes and Microscopic
  Nuclear Structure}, Phys. Rept. 160 (1988) 235--427.

\bibitem{Day:1990mf}
D.~B. Day, J.~S. McCarthy, T.~W. Donnelly, I.~Sick, {Scaling in inclusive
  electron - nucleus scattering}, Ann. Rev. Nucl. Part. Sci. 40 (1990)
  357--410.

\bibitem{Frankfurt:1993sp}
L.~L. Frankfurt, M.~I. Strikman, D.~B. Day, M.~Sargsian, {Evidence for short
  range correlations from high Q$^2$ (e,e') reactions}, Phys. Rev. C48 (1993)
  2451--2461.

\bibitem{Uchiyama:1989vr}
T.~Uchiyama, A.~Dieperink, O.~Scholten, {Final state interactions and y-scaling
  in inclusive electron scattering}, Phys.Lett. B233 (1989) 31--36.

\bibitem{Benhar:1995te}
O.~Benhar, S.~Liuti, {Can a highly virtual nucleon experience final state
  interactions in electron nucleus scattering?}, Phys.Lett. B389 (1996)
  649--654.

\bibitem{CiofidegliAtti:2009qc}
C.~Ciofi~degli Atti, C.~B. Mezzetti, {Obtaining information on Short Range
  Correlations from inclusive electron scattering}, Phys. Rev. C79 (2009)
  051302.

\bibitem{Mezzetti:2009ch}
C.~B. Mezzetti, C.~Ciofi~degli Atti, {A Novel analysis of the effects of short
  range correlations in inclusive lepton scattering off nuclei. }\href
  {http://arxiv.org/abs/0906.5564} {\path{arXiv:0906.5564}}.

\bibitem{Arrington:1998ps}
J.~Arrington, et~al., {Inclusive electron nucleus scattering at large momentum
  transfer}, Phys. Rev. Lett. 82 (1999) 2056--2059.

\bibitem{Fomin:2011ng}
N.~Fomin, J.~Arrington, R.~Asaturyan, F.~Benmokhtar, W.~Boeglin, et~al., {New
  measurements of high-momentum nucleons and short-range structures in
  nuclei.}, Phys. Rev. Lett. 108 (2012) 092502.

\bibitem{CiofidegliAtti:1994ys}
C.~Ciofi~degli Atti, S.~Simula, {Nucleon-nucleon correlations and final state
  interactions in inclusive quasielastic electron scattering off nuclei at
  x$>$1}, Phys.Lett. B325 (1994) 276--282.

\bibitem{Gurvitz:1986dn}
S.~Gurvitz, A.~Rinat, {y-scaling in inclusive scattering}, Phys.Rev. C35 (1987)
  696.

\bibitem{Gurvitz:1988ft}
S.~Gurvitz, A.~Rinat, R.~Rosenfelder, {y-scaling and hard core potentials},
  Phys.Rev. C40 (1989) 1363--1375.

\bibitem{Gurvitz:2001qm}
S.~Gurvitz, A.~Rinat, {Relativistic approaches to structure functions of
  nuclei}, Phys.Rev. C65 (2002) 024310.

\bibitem{Benhar:1991af}
O.~Benhar, A.~Fabrocini, S.~Fantoni, G.~Miller, V.~Pandharipande, et~al.,
  {Scattering of GeV electrons by nuclear matter}, Phys.Rev. C44 (1991)
  2328--2342.

\bibitem{Benhar:1993ja}
O.~Benhar, V.~Pandharipande, {Scattering of GeV electrons by light nuclei},
  Phys.Rev. C47 (1993) 2218--2227.

\bibitem{Benhar:1995xa}
O.~Benhar, A.~Fabrocini, S.~Fantoni, S.~Pieper, V.~Pandharipande, et~al.,
  {Higher order effects in inclusive electron nucleus scattering}, Phys.Lett.
  B359 (1995) 8--12.

\bibitem{Benhar:1995ph}
O.~Benhar, A.~Fabrocini, S.~Fantoni, I.~Sick, {Inclusive cross-section ratios
  at x$>$1}, Phys.Lett. B343 (1995) 47--52.

\bibitem{Makins:1994mm}
N.~Makins, R.~Ent, M.~Chapman, J.~Hansen, K.~Lee, et~al., {Momentum transfer
  dependence of nuclear transparency from the quasielastic C-12 (e,e'p)
  reaction}, Phys.Rev.Lett. 72 (1994) 1986--1989.

\bibitem{O'Neill:1994mg}
T.~G. O'Neill, et~al., {A-dependence of nuclear transparency in quasielastic
  A(e,e'p) at high Q$^2$}, Phys. Lett. B351 (1995) 87--92.

\bibitem{Abbott:1997bc}
D.~Abbott, et~al., {Quasifree (e,e'p) reactions and proton propagation in
  nuclei}, Phys. Rev. Lett. 80 (1998) 5072--5076.

\bibitem{Garrow:2001di}
K.~Garrow, et~al., {Nuclear transparency from quasielastic A(e,e'p) reactions
  up to Q$^2$ = 8.1-(GeV/c)$^2$}, Phys. Rev. C66 (2002) 044613.

\bibitem{Day:1993md}
D.~B. Day, et~al., {Inclusive electron nucleus scattering at high momentum
  transfer}, Phys. Rev. C48 (1993) 1849--1863.

\bibitem{Frankfurt:2009vv}
L.~Frankfurt, M.~Sargsian, M.~Strikman, {Future directions for probing two and
  three nucleon short-range correlations at high energies}, AIP Conf.Proc. 1056
  (2008) 322--329.

\bibitem{Abramovsky:1973fm}
V.~Abramovsky, V.~Gribov, O.~Kancheli, {Character of inclusive spectra and
  fluctuations produced in inelastic processes by multi-pomeron exchange},
  Yad.Fiz. 18 (1973) 595--616.

\bibitem{Bertocchi:1976bq}
L.~Bertocchi, D.~Treleani, {Glauber Theory, Unitarity, and the AGK
  Cancellation}, J.Phys. G3 (1977) 147.

\bibitem{CS}
W.~Cosyn, M.~Sargsian, work in progress.

\bibitem{Allasia:1986kg}
D.~Allasia, C.~Angelini, A.~Baldini, F.~Bobisut, A.~Borg, et~al., {Search for a
  delta (1236) - delta (1236) structure of the deuteron}, Phys.Lett. B174
  (1986) 450--452.

\bibitem{Blomqvist:1998fr}
K.~I. Blomqvist, et~al., {Large recoil momenta in the D(e,e'p)n reaction},
  Phys. Lett. B424 (1998) 33--38.

\bibitem{Arenhovel:1995be}
H.~Arenhovel, W.~Leidemann, E.~L. Tomusiak, {Nucleon polarization in exclusive
  deuteron electrodisintegration with polarized electrons and a polarized
  target}, Phys. Rev. C52 (1995) 1232--1253.

\bibitem{Passchier:2001uc}
I.~Passchier, et~al., {Spin-momentum correlations in quasi-elastic electron
  scattering from deuterium}, Phys. Rev. Lett. 88 (2002) 102302.

\bibitem{Ritz:1997cq}
F.~Ritz, H.~Arenhovel, T.~Wilbois, {Relativistic effects and the role of heavy
  meson exchange in deuteron photodisintegration}, Few Body Syst. 24 (1998)
  123--138.

\bibitem{Leidemann:1992fp}
W.~Leidemann, E.~L. Tomusiak, H.~Arenhoevel, {Polarization observables in
  deuteron electrodisintegration}, Few Body Syst. Suppl. 6 (1992) 236.

\bibitem{Schutz:1976he}
W.~P. Schutz, et~al., {Inelastic electron Deuteron Scattering in the Threshold
  Region at High Momentum Transfer}, Phys. Rev. Lett. 38 (1977) 259.

\bibitem{Rock:1982gf}
S.~Rock, et~al., {Measurement of Elastic electron - Neutron Cross-Sections Up
  to Q$^2$ = 10-(GeV/c)$^2$}, Phys. Rev. Lett. 49 (1982) 1139.

\bibitem{Lung:1992bu}
A.~Lung, et~al., {Measurements of the electric and magnetic form-factors of the
  neutron from Q$^2$ = 1.75 (GeV/c)$^2$ to 4 (GeV/c)$^2$}, Phys. Rev. Lett. 70
  (1993) 718--721.

\bibitem{Arrington:1998hz}
J.~R. Arrington, {Inclusive electron scattering from nuclei at $x > 1$ and high
  $Q^2$.~}Ph. D. Thesis, California Institute of Technology (1998).
\newblock \href {http://arxiv.org/abs/nucl-ex/0608013}
  {\path{arXiv:nucl-ex/0608013}}.

\bibitem{Arrington:2001ni}
J.~Arrington, et~al., {$x$ and $\xi$ scaling of the nuclear structure function
  at large $x$}, Phys. Rev. C64 (2001) 014602.

\bibitem{Fomin:2008iq}
N.~Fomin, {Inclusive electron scattering from nuclei in the quasielastic region
  at large momentum transfer}~Ph.D. Thesis, University of Virginia (2007).
\newblock \href {http://arxiv.org/abs/0812.2144} {\path{arXiv:0812.2144}}.

\bibitem{Fomin:2010ei}
N.~Fomin, et~al., {Scaling of the $F_2$ structure function in nuclei and quark
  distributions at $x>1$}, Phys. Rev. Lett. 105 (2010) 212502.

\bibitem{West:1974ua}
G.~B. West, {Electron Scattering from Atoms, Nuclei and Nucleons}, Phys.Rept.
  18 (1975) 263--323.

\bibitem{Sick:1980ey}
I.~Sick, D.~Day, J.~Mccarthy, {Nuclear high momentum components and y-scaling
  in electron scattering}, Phys.Rev.Lett. 45 (1980) 871--874.

\bibitem{Pace:1982xi}
E.~Pace, G.~Salme, {Nuclear scaling function and quasielastic electron
  scattering by nuclei}, Phys.Lett. B110 (1982) 411.

\bibitem{Benhar:2006wy}
O.~Benhar, D.~Day, I.~Sick, {Inclusive quasi-elastic electron-nucleus
  scattering}, Rev. Mod. Phys. 80 (2008) 189--224.

\bibitem{Arrington:2003tw}
J.~Arrington, {Nucleon momentum distributions from a modified scaling analysis
  of inclusive electron nucleus scattering.~~}\href
  {http://arxiv.org/abs/nucl-ex/0306016} {\path{arXiv:nucl-ex/0306016}}.

\bibitem{Sargsian:2009hf}
M.~M. Sargsian, {Large Q2 Electrodisintegration of the Deuteron in Virtual
  Nucleon Approximation}, Phys. Rev. C82 (2010) 014612.

\bibitem{Jeschonnek:2000nh}
S.~Jeschonnek, {Unfactorized versus factorized calculations for H-2(e,e' p)
  reactions at GeV energies}, Phys. Rev. C63 (2001) 034609.

\bibitem{CiofidegliAtti:2000xj}
C.~Ciofi~degli Atti, L.~P. Kaptari, D.~Treleani, {On the effects of the final
  state interaction in the electro-disintegration of the deuteron at
  intermediate and high energies}, Phys. Rev. C63 (2001) 044601.

\bibitem{Laget:2004sm}
J.~M. Laget, {The electro-disintegration of few body systems revisited}, Phys.
  Lett. B609 (2005) 49--56.

\bibitem{Jeschonnek:2008zg}
S.~Jeschonnek, J.~W. Van~Orden, {A new calculation for D(e,e'p)n at GeV
  energies}, Phys. Rev. C78 (2008) 014007.

\bibitem{Jeschonnek:2009tq}
S.~Jeschonnek, J.~W. Van~Orden, {Target Polarization for $^2 \vec H(e,e'p)n$ at
  GeV energies}, Phys. Rev. C80 (2009) 054001.

\bibitem{Jeschonnek:2009ds}
S.~Jeschonnek, J.~W. Van~Orden, {Ejectile Polarization for $^2 H(e,e'\vec p)n$
  at GeV energies}, Phys. Rev. C81 (2010) 014008.

\bibitem{Frankfurt:1994kt}
L.~L. Frankfurt, W.~R. Greenberg, G.~A. Miller, M.~M. Sargsian, M.~I. Strikman,
  {Color transparency effects in electron deuteron interactions at intermediate
  Q$^2$}, Z. Phys. A352 (1995) 97--113.

\bibitem{Frankfurt:1996xx}
L.~L. Frankfurt, M.~M. Sargsian, M.~I. Strikman, {Feynman graphs and
  generalized eikonal approach to high energy knock-out processes}, Phys. Rev.
  C56 (1997) 1124--1137.

\bibitem{Ulmer:2002jn}
P.~E. Ulmer, et~al., {H-2(e,e'p)n reaction at high recoil momenta}, Phys. Rev.
  Lett. 89 (2002) 062301.

\bibitem{Bulten95}
H.~Bulten, P.~Anthony, R.~Arnold, J.~Arrington, E.~Beise, et~al., {Exclusive
  electron scattering from deuterium at high momentum transfer}, Phys.Rev.Lett.
  74 (1995) 4775--4778.

\bibitem{e94004}
M.~Jones, P.~Ulmer, et~al., In-plane separations and high momentum structure in
  d(e,e'p)n, Jefferson Lab Experiment Proposal E94-004 (1994).

\bibitem{Tjon92}
J.~Tjon, {Relativistic analysis of electron deuteron scattering}, Few Body
  Syst.Suppl. 5 (1992) 5--16.

\bibitem{Beck92}
G.~Beck, H.~Arenhovel, {Relativistic corrections in quasi-free
  electro-disintegration of the deuteron}, Few Body Syst. 13 (1992) 165--188.

\bibitem{Egiyan:2007qj}
K.~S. Egiyan, et~al., {Experimental study of exclusive H-2(e,e'p)n reaction
  mechanisms at high Q$^2$}, Phys. Rev. Lett. 98 (2007) 262502.

\bibitem{Wiringa:1994wb}
R.~B. Wiringa, V.~G.~J. Stoks, R.~Schiavilla, {An Accurate nucleon-nucleon
  potential with charge independence breaking}, Phys. Rev. C51 (1995) 38--51.

\bibitem{Boeglin:2011mt}
W.~Boeglin, et~al., {Probing the high momentum component of the deuteron at
  high $Q^2$}, Phys.Rev.Lett. 107 (2011) 262501.

\bibitem{e01020}
W.~Boeglin, et~al., Jefferson Lab Experiment Proposal E01-020 (2001).

\bibitem{Luminita}
J.~Coman, { $^2H(e,e'p)$ Studies of the Deuteron at High $Q^2$'}, PhD. Thesis,
  FIU (2008).

\bibitem{Rock:1981aa}
S.~Rock, et~al., {Inelastic electron scattering from He-3 AND He-4 in the
  threshold region at high momentum transfer}, Phys. Rev. C26 (1982) 1592.

\bibitem{Bosted:1982gd}
P.~E. Bosted, R.~G. Arnold, S.~Rock, Z.~M. Szalata, {Nuclear scaling in
  inelastic electron scattering from d, He-3 and He-4}, Phys. Rev. Lett. 49
  (1982) 1380.

\bibitem{Meziani:1992xr}
Z.~E. Meziani, et~al., {High momentum transfer R(T,L) inclusive response
  functions for He-3, He-4. SLAC-NE-9 experiment}, Phys. Rev. Lett. 69 (1992)
  41--44.

\bibitem{Filippone:1992iz}
B.~W. Filippone, et~al., {Nuclear structure functions at $x > 1$}, Phys. Rev.
  C45 (1992) 1582--1585.

\bibitem{Bosted:1992fy}
P.~E. Bosted, et~al., {Measurements of $\nu W_2$ and $R = \sigma_L / \sigma_T$
  from inelastic electron - aluminum scattering near $x = 1$}, Phys. Rev. C46
  (1992) 2505--2515.

\bibitem{Arrington:1995hs}
J.~Arrington, et~al., {Inclusive electron scattering from nuclei at $x \approx
  1$}, Phys. Rev. C53 (1996) 2248--2251.

\bibitem{Egiyan:2003vg}
K.~S. Egiyan, et~al., {Observation of Nuclear Scaling in the A(e,e') Reaction
  at $x_B>$1}, Phys. Rev. C68 (2003) 014313.

\bibitem{Egiyan:2005hs}
K.~S. Egiyan, et~al., {Measurement of 2- and 3-Nucleon Short Range Correlation
  Probabilities in Nuclei}, Phys. Rev. Lett. 96 (2006) 082501.

\bibitem{Benhar:2006er}
O.~Benhar, D.~Day, I.~Sick, {An Archive for quasi-elastic electron-nucleus
  scattering data.~}\href {http://arxiv.org/abs/nucl-ex/0603032}
  {\path{arXiv:nucl-ex/0603032}}.

\bibitem{Arrington:2006pn}
J.~Arrington, {Hadrons in the nuclear medium - quarks, nucleons, or a bit of
  both?~~}\href {http://arxiv.org/abs/nucl-ex/0602007}
  {\path{arXiv:nucl-ex/0602007}}.

\bibitem{CiofidegliAtti:1997km}
C.~Ciofi~degli Atti, G.~B. West, {Old and new facets of y-scaling: The
  universal features of nuclear structure functions and nucleon momentum
  distributions.~~}\href {http://arxiv.org/abs/nucl-th/9702009}
  {\path{arXiv:nucl-th/9702009}}.

\bibitem{CiofidegliAtti:1999is}
C.~Ciofi~degli Atti, G.~B. West, {A new approach to y-scaling and the universal
  features of scaling functions and nucleon momentum distributions}, Phys.
  Lett. B458 (1999) 447--453.

\bibitem{e08014}
J.~Arrington, D.~Day, D.~W. Higinbotham, P.~Solvignon, et~al., {Three-nucleon
  short range correlations studies in inclusive scattering for $0.8 < Q^2 <
  2.8~({\rm GeV}/c)^2$}, Jefferson Lab Experiment Proposal E08-014 (2008).

\bibitem{Seely:2009gt}
J.~Seely, et~al., {New measurements of the EMC effect in very light nuclei},
  Phys. Rev. Lett. 103 (2009) 202301.

\bibitem{e1206105}
J.~Arrington, D.~Day, N.~Fomin, P.~Solvignon, et~al., Inclusive scattering from
  nuclei at $x>1$ in the quasielastic and deeply inelastic regimes, Jefferson
  Lab Experiment Proposal E12-06-105 (2006).

\bibitem{e1211112}
J.~Arrington, D.~Day, D.~W. Higinbotham, P.~Solvignon, et~al., {Precision
  measurement of the isospin dependence in the 2N and 3N short range
  correlation region}, Jefferson Lab Experiment Proposal E12-11-112 (2011).

\bibitem{vanLeeuwe:2001uc}
J.~van Leeuwe, W.~Hesselink, E.~Jans, W.~Kasdorp, J.~Laget, et~al., {The
  He-4(e,e'p) cross-section at high missing energies}, Phys.Lett. B523 (2001)
  6--12.

\bibitem{vanLeeuwe:1997wq}
J.~van Leeuwe, H.~Blok, J.~van~den Brand, H.~Bulten, G.~Dodge, et~al., {The
  $^4$He (e,e'p) cross-section at large missing energy}, Nucl.Phys. A631 (1998)
  593--596.

\bibitem{Sargsian:2005ru}
M.~M. Sargsian, T.~V. Abrahamyan, M.~I. Strikman, L.~L. Frankfurt, {Exclusive
  electro-disintegration of He-3 at high Q$^2$. II: Decay function formalism},
  Phys. Rev. C71 (2005) 044615.

\bibitem{Sargsian:2004tz}
M.~Sargsian, T.~Abrahamyan, M.~Strikman, L.~Frankfurt, {Exclusive
  electrodisintegration of He-3 at high Q$^2$. I. Generalized eikonal
  approximation}, Phys.Rev. C71 (2005) 044614.

\bibitem{Benmokhtar:2004fs}
F.~Benmokhtar, et~al., {Measurement of the He-3(e,e'p)pn reaction at high
  missing energies and momenta}, Phys. Rev. Lett. 94 (2005) 082305.

\bibitem{Rvachev:2004yr}
M.~M. Rvachev, et~al., {The quasielastic He-3(e,e'p)d reaction at Q$^2$ = 1.5-
  GeV$^2$ for recoil momenta up to 1-GeV/c}, Phys. Rev. Lett. 94 (2005) 192302.

\bibitem{degliAtti:2005dh}
C.~Ciofi~degli Atti, L.~P. Kaptari, {Calculations of the exclusive processes
  2H(e,e'p)n, 3He(e,e'p)2H, and 3He(e,e'p)(pn) within a generalized eikonal
  approximation}, Phys. Rev. C71 (2005) 024005.

\bibitem{CiofidegliAtti:2005qt}
C.~Ciofi~degli Atti, L.~P. Kaptari, {On the interpretation of the processes
  He-3(e,e'p)H-2 and He-3(e,e'p)(p n) at high missing momenta}, Phys. Rev.
  Lett. 95 (2005) 052502.

\bibitem{CiofidelgiAtti:2007qu}
C.~Ciofi~delgi Atti, L.~P. Kaptari, {A non factorized calculation of the
  process 3He(e,e'p)2H at medium energies}, Phys. Rev. Lett. 100 (2008) 122301.

\bibitem{Alvioli:2009zy}
M.~Alvioli, C.~Ciofi~degli Atti, L.~P. Kaptari, {Calculation of the cross
  section and the transverse- longitudinal asymmetry of the process
  $^3$He(e,e'p)pn at medium energies within the unfactorized generalized
  Glauber approach}, Phys. Rev. C81 (2010) 021001.

\bibitem{Rohe:2005vc}
D.~Rohe, et~al., {Nuclear transparency from quasielastic 12C(e,e'p)}, Phys.
  Rev. C72 (2005) 054602.

\bibitem{Barbieri:2004up}
C.~Barbieri, D.~Rohe, I.~Sick, L.~Lapikas, {Effect of kinematics on final state
  interactions in (e,e'p) reactions}, Phys. Lett. B608 (2005) 47--52.

\bibitem{Rohe:2004dz}
D.~Rohe, et~al., {Correlated Strength in Nuclear Spectral Function}, Phys. Rev.
  Lett. 93 (2004) 182501.

\bibitem{Benhar:1989aw}
O.~Benhar, A.~Fabrocini, S.~Fantoni, {The nucleon spectral function in nuclear
  matter}, Nucl.Phys. A505 (1989) 267--299.

\bibitem{Muther:1995bk}
H.~Muther, G.~Knehr, A.~Polls, {Momentum distribution in nuclear matter and
  finite nuclei}, Phys.Rev. C52 (1995) 2955--2968.

\bibitem{Hehl:1995iq}
T.~Hehl, {Study of (gamma,N N) reactions at MAMI}, Prog. Part. Nucl. Phys. 34
  (1995) 385--386.

\bibitem{Kester:1995zz}
L.~J. H.~M. Kester, et~al., {Short-Range Nucleon-Nucleon Correlations
  Investigated with the Reaction {C-12}(e,e'pp)}, Phys. Rev. Lett. 74 (1995)
  1712--1715.

\bibitem{Blomqvist:1998gq}
K.~I. Blomqvist, et~al., {Investigation of short-range nucleon nucleon
  correlations using the reaction {C-12(e,e'pp)} in close to 4$\pi$ geometry},
  Phys. Lett. B421 (1998) 71--78.

\bibitem{Onderwater:1998zz}
C.~J.~G. Onderwater, et~al., {Signatures for Short-Range Correlations in O-16
  Observed in the Reaction O-16 (e,e'pp) C-14}, Phys. Rev. Lett. 81 (1998)
  2213--2216.

\bibitem{Groep:2000cy}
D.~L. Groep, et~al., {Investigation of the exclusive He-3(e,e'pp)n reaction},
  Phys. Rev. C63 (2001) 014005.

\bibitem{Ryckebusch:1993tf}
J.~Ryckebusch, M.~Vanderhaeghen, L.~Machenil, M.~Waroquier, {Effects of the
  final state interaction in (gamma,pn) and (gamma,pp) processes}, Nucl. Phys.
  A568 (1994) 828--854.

\bibitem{Ryckebusch:1994zz}
J.~Ryckebusch, L.~Machenil, M.~Vanderhaeghen, V.~Van~der Sluys, M.~Waroquier,
  {Multinucleon mechanisms in (gamma, N) and (gamma, NN) reactions}, Phys.Rev.
  C49 (1994) 2704--2715.

\bibitem{Giusti:2007fn}
C.~Giusti, F.~D. Pacati, M.~Schwamb, S.~Boffi, {Electromagnetic proton-neutron
  knockout off $^{16}$O: new achievements in theory}, Eur. Phys. J. A33 (2007)
  29--38.

\bibitem{Ryckebusch:1995ze}
J.~Ryckebusch, M.~Vanderhaeghen, K.~Heyde, M.~Waroquier, {Short range
  correlations in (e,e'p) and (e,e'pp) reactions on complex nuclei}, Phys.
  Lett. B350 (1995) 1--7.

\bibitem{Ryckebusch:1989nn}
J.~Ryckebusch, K.~Heyde, D.~Van~Neck, M.~Waroquier, {Aspects of the final state
  interaction and long range correlations in quasielastic (e,e'p) and (e,e'n)
  reactions}, Nucl. Phys. A503 (1989) 694--722.

\bibitem{Frankfurt:1999ik}
L.~L. Frankfurt, G.~A. Miller, M.~M. Sargsian, M.~I. Strikman, {QCD
  rescattering and high-energy two-body photodisintegration of the deuteron},
  Phys.Rev.Lett. 84 (2000) 3045--3048.

\bibitem{Sargsian:2008zm}
M.~M. Sargsian, C.~Granados, {Hard Break-Up of Two-Nucleons from the He-3
  Nucleus}, Phys.Rev. C80 (2009) 014612.

\bibitem{Pomerantz:2009sb}
I.~Pomerantz, et~al., {Hard Photodisintegration of a Proton Pair}, Phys. Lett.
  B684 (2010) 106--109.

\bibitem{Aclander:1999fd}
J.~L. Aclander, J.~Alster, D.~Barton, G.~Bunce, A.~Carroll, et~al., {The large
  momentum transfer reaction {$^{12}$C(p,2p+n)} as a new method for measuring
  short range {N N} correlations in nuclei}, Phys.Lett. B453 (1999) 211--216.

\bibitem{Tang:2002ww}
A.~Tang, et~al., {np short-range correlations from (p,2p+n) measurements},
  Phys. Rev. Lett. 90 (2003) 042301.

\bibitem{Piasetzky:2006ai}
E.~Piasetzky, M.~Sargsian, L.~Frankfurt, M.~Strikman, J.~W. Watson, {Evidence
  for the Strong Dominance of Proton-Neutron Correlations in Nuclei}, Phys.
  Rev. Lett. 97 (2006) 162504.

\bibitem{Yaron:2002nv}
I.~Yaron, et~al., {Investigation of the high momentum component of nuclear wave
  function using hard quasielastic {A(p,2p)X} reactions}, Phys. Rev. C66 (2002)
  024601.

\bibitem{Shneor:2007tu}
R.~Shneor, et~al., {Investigation of Proton-Proton Short-Range Correlations via
  the 12C(e,e'pp) Reaction}, Phys. Rev. Lett. 99 (2007) 072501.

\bibitem{Subedi:2008zz}
R.~Subedi, et~al., {Probing Cold Dense Nuclear Matter}, Science 320 (2008)
  1476--1478.

\bibitem{Alcorn:2004sb}
J.~Alcorn, et~al., {Basic Instrumentation for Hall A at Jefferson Lab}, Nucl.
  Instrum. Meth. A522 (2004) 294--346.

\bibitem{deLange:1998au}
D.~J.~J. de~Lange, et~al., {The optical properties of the BigBite spectrometer
  at NIKHEF}, Nucl. Instrum. Meth. A412 (1998) 254--264.

\bibitem{Mardor:1992sb}
I.~Mardor, Y.~Mardor, E.~Piasetzky, J.~Alster, M.~M. Sargsian, {Effect of
  multiple scattering on the measurement of nuclear transparency}, Phys. Rev.
  C46 (1992) 761--767.

\bibitem{Higinbotham:2009zz}
D.~Higinbotham, E.~Piasetzky, M.~Strikman, {Protons and neutrons cosy up in
  nuclei and neutron stars}, CERN Cour. 49N1 (2009) 22--24.

\bibitem{RajSubedi:2007zz}
R.~Raj~Subedi, {Studying short-range correlations with the C-12(e, e' p n)
  reaction.}Ph.D. Thesis (Advisors: John Watson and Douglas Higinbotham).

\bibitem{Farrar:1988mf}
G.~Farrar, H.~Liu, L.~Frankfurt, M.~Strikman, {Study of Bound Nucleons by
  Quasi-Exclusive Scattering with Large Momentum Transfer}, Phys.Rev.Lett. 62
  (1989) 1095--1098.

\bibitem{Monaghan:2008zz}
P.~Monaghan, {Study of the $^{12}$C(e,e'p) reaction in a correlations dominant
  regime with $Q^2$ = 2.0 (GeV/c)$^2$ and $x_B>1$.~}Ph.D. Thesis.

\bibitem{Alvioli:2011aa}
M.~Alvioli, C.~Ciofi~degli Atti, L.~Kaptari, C.~Mezzetti, H.~Morita, et~al.,
  {Universality of nucleon-nucleon short-range correlations: two-nucleon
  momentum distributions in few-body systems}, Phys.Rev. C85 (2012) 021001.

\bibitem{Roth:2004ua}
R.~Roth, T.~Neff, H.~Hergert, H.~Feldmeier, {Nuclear Structure based on
  Correlated Realistic Nucleon- Nucleon Potentials}, Nucl. Phys. A745 (2004)
  3--33.

\bibitem{Laget:1987jr}
J.~M. Laget, {Three-body exchange currents. 1. the He-3 (gamma, 2p) n
  Reaction}, J. Phys. G14 (1988) 1445--1451.

\bibitem{Audit:1996tq}
G.~Audit, et~al., {Study of three nucleon mechanisms in the photodisintegration
  of He-3}, Nucl. Phys. A614 (1997) 461--471.

\bibitem{Laget:1987zz}
J.-M. Laget, {Role of correlations in the (e,e'2p) n reaction}, Phys. Rev. C35
  (1987) 832--835.

\bibitem{Niyazov:2003zr}
R.~A. Niyazov, et~al., {Two nucleon momentum distributions measured in
  He-3(e,e'pp)n}, Phys. Rev. Lett. 92 (2004) 052303.

\bibitem{Baghdasaryan:2010nv}
H.~Baghdasaryan, et~al., {Tensor Correlations Measured in 3He(e,e'pp)n}, Phys.
  Rev. Lett. 105 (2010) 222501.

\bibitem{Golak:1994vq}
J.~Golak, H.~Kamada, H.~Witala, W.~Gloeckle, S.~Ishikawa, {Electron induced p d
  and p p n breakup of He-3 with full inclusion of final-state interactions},
  Phys.Rev. C51 (1995) 1638--1647.

\bibitem{Higinbotham:2009hi}
D.~Higinbotham, E.~Piasetzky, S.~Wood, {Short-Distance Structure of Nuclei},
  J.Phys.Conf.Ser. 299 (2011) 012010.

\bibitem{e07006}
S.~Gilad, D.~W. Higinbotham, E.~Piasetzky, V.~Sulkosky, J.~Watson, et~al.,
  Studying short-range correlations in nuclei at the repulsive core limit via
  the triple coincidence (e,e'pn) reaction, Jefferson Lab Experiment Proposal
  E07-006 (2007).

\bibitem{e1210003}
W.~Boeglin, M.~K. Jones, et~al., Deuteron electro-disintegration at very high
  missing momenta, Jefferson Lab Experiment Proposal E12-10-003 (2010).

\bibitem{Weinstein:2010rt}
L.~Weinstein, E.~Piasetzky, D.~Higinbotham, J.~Gomez, O.~Hen, et~al., {Short
  Range Correlations and the EMC Effect}, Phys.Rev.Lett. 106 (2011) 052301.

\bibitem{E1210008}
J.~Arrington, A.~Daniel, D.~G. Gaskell, et~al., Detailed studies of the nuclear
  dependence of F$_2$ in light nuclei, Jefferson Lab Experiment Proposal
  E12-10-008 (2010).

\bibitem{Arrington:2003qt}
J.~Arrington, {Do ordinary nuclei contain exotic states of matter?}, Heavy Ion
  Phys. 21 (2004) 295--300.
\newblock \href {http://arxiv.org/abs/hep-ph/0304213}
  {\path{arXiv:hep-ph/0304213}}.

\end{thebibliography}
\bibliographystyle{elsarticle-num}

\end{document}